\documentclass[12pt,a4paper]{article}
\usepackage{jheplikemod,ifpdf,tocloft,rotating,amsmath,textcomp,mathrsfs,mathtools,xcolor}
\usepackage{tikz}
\usetikzlibrary{calc,decorations.markings,shapes.geometric}

\definecolor{mhvblue2}{rgb}{0.3,0.3,0.575}

\definecolor{mhvblue}{rgb}{0.6,0.6,0.7765}
\definecolor{nmhvred}{rgb}{0.6765,0.15,0.3}
\definecolor{ampgrey}{rgb}{0.9,0.9,0.9}
\definecolor{unord}{rgb}{0,0,0}
\definecolor{ord}{rgb}{0,0,0.575}
\definecolor{anchorLeg}{rgb}{0.575,0.0,0.225}
\definecolor{labelcolor}{rgb}{0,0,0}

\newcounter{legSteps}
\newcounter{offset}

\def\figScale{0.9}
\def\legSpread{4}
\def\extLegLen{0.9*0.32*\figScale}
\def\edgeLen{1*\figScale}

\pgfmathsetmacro{\pLen}{\edgeLen/(2*sin(72/2))}
\def\legLen{\edgeLen*0.45}

\def\labelDist{\legLen*1.5}
\def\lineThickness{(1pt)}
\def\dotSize{(\figScale*12pt)}
\def\ampSize{(1*\figScale*12pt)}
\def\eph{0.4}
\def\pageW{15.175/\figScale}

\tikzset{fullamp/.style={coordinate,minimum size=0.7*\ampSize,ball color=black!20,circle,text=white,inner sep=0}}
\tikzset{fullmhv/.style={coordinate,minimum size=0.8*\ampSize,ball color=mhvblue,circle,text=white,inner sep=0}}
\tikzset{fullnmhv/.style={coordinate,minimum size=0.8*\ampSize,ball color=nmhvred,circle,text=white,inner sep=0}}
\tikzset{fullmhvBar/.style={coordinate,minimum size=0.8*\ampSize,ball color=white,circle,text=white,inner sep=0}}
\tikzset{ordAmp/.style={fill=ampgrey,circle,draw=black,line width=\lineThickness,minimum size=0.6*\ampSize,text=white,inner sep=0}}
\tikzset{mhv/.style={fill=mhvblue,circle,draw=black,line width=\lineThickness,minimum size=0.8*\ampSize,text=white,inner sep=0}}
\tikzset{mhvBar/.style={fill=white,circle,draw=black,line width=\lineThickness,minimum size=0.8*\ampSize,text=white,inner sep=0}}
\tikzset{fgraphEdge/.style={anchorLeg,line width=\lineThickness,line cap=round}}
\tikzset{fgraphExt/.style={ord,line width=\lineThickness,line cap=round}}
\tikzset{fgraphOpt/.style={ord,dotted,line width=\lineThickness,line cap=round}}
\tikzset{fdot/.style={fill=anchorLeg,circle,minimum size=0.35*\ampSize,inner sep=0}}
\tikzset{bdot/.style={fill=black,circle,minimum size=0.35*\ampSize,inner sep=0}}
\tikzset{ext/.style={black,line width=\lineThickness,line cap=round}}
\tikzset{under/.style={white,line width=4*\lineThickness,line cap=round}}
\tikzset{optExt/.style={black,dotted,line width=\lineThickness,line cap=round,rounded corners=10pt}}
\tikzset{optExtSc/.style={black,line width=\lineThickness,line cap=round,rounded corners=10pt}}
\tikzset{dashed/.style={black!70,dotted,line width=\lineThickness,line cap=round,rounded corners=10pt}}
\tikzset{ddot/.style={fill=black,circle,minimum size=0.35*\dotSize,inner sep=0}}
\tikzset{int/.style={black,line width=\lineThickness,line cap=round,rounded corners=1.5pt}}

\tikzset{intInfR/.style={nmhvred,line width=\lineThickness,line cap=round,rounded corners=1.5pt}}
\tikzset{blueDot/.style={fill=mhvblue,circle,draw=black,line width=\lineThickness,minimum size=0.5*\ampSize,text=white,inner sep=0}}
\tikzset{whiteDot/.style={fill=white,circle,draw=black,line width=\lineThickness,minimum size=0.5*\ampSize,text=white,inner sep=0}}
\tikzset{blackDot/.style={fill=black,circle,minimum size=0.5*\ampSize,inner sep=0}}
\tikzset{compositeDot/.style={fill=none,draw=black,line width=\lineThickness,circle,minimum size=0.75*\ampSize,inner sep=0}}

\tikzset{directedEdge/.style={draw=none,decoration={markings,mark connection node=connode,mark=at position 0.5 with {\node[transform shape, scale=0.205*\figScale,shape=dart,aspect=0.5,fill=black,draw] (connode) {};}},postaction={decorate}}}

\tikzset{directedEdgeBend/.style={rounded corners=10pt,draw=none,decoration={markings,mark connection node=connode,mark=at position 0.5 with {\node[transform shape, scale=0.205,shape=dart,aspect=0.5,fill=black,draw] (connode) {};}},postaction={decorate}}}

\newcommand{\legOpt}[3]{\draw[optExt] #1--($#1+(#2:\legLen*1.15)$);\node at ($#1+(#2:\labelDist*1.15)$)[]{{\footnotesize #3}};}

\newcommand{\leg}[3]{\draw[ext] #1--($#1+(#2:\legLen)$);\node at ($#1+(#2:\labelDist)$)[]{{\footnotesize #3}};}

\newcommand{\edgeA}{\text{{\footnotesize$a$}}}
\newcommand{\edgeB}{\text{{\footnotesize$b$}}}
\newcommand{\edgeC}{\text{{\footnotesize$c$}}}
\newcommand{\edgeD}{\text{{\footnotesize$d$}}}
\newcommand{\edgeE}{\text{{\footnotesize$e$}}}
\newcommand{\edgeF}{\text{{\footnotesize$f$}}}
\newcommand{\edgeG}{\text{{\footnotesize$g$}}}
\newcommand{\edgeH}{\text{{\footnotesize$h$}}}

\def\boundingDraw{red}
\def\boundingDraw{none}

\definecolor{legColour}{rgb}{0.35,0.35,0.35}
\definecolor{ndotColor}{rgb}{0.65,0.25,0.25}
\def\markStroke{0.65}

\newcommand{\tikzBox}[2][0.5]{\begin{tikzpicture}[scale=1,baseline=-3.05,rotate=0]\useasboundingbox ($\figScale*(-#1,-#1)$) rectangle ($\figScale*(#1,#1)$);#2\end{tikzpicture}}

\tikzset{ndot/.style={transform shape,scale=0.35*\figScale,aspect=0.65,draw=ndotColor,line width=\markStroke*\figScale,shape=circle,fill=none}}

\tikzset{markedEdgeR/.style={draw=none,decoration={markings,mark connection node=connode,mark=at position 0.5 with {\node[ndot] (connode) {};}},postaction={decorate}}}

\tikzset{ndotR/.style={transform shape,scale=0.35*\figScale,aspect=0.65,draw=ndotColor,line width=\markStroke*\figScale,shape=circle,fill=white}}

\newcommand{\legAlt}[2]{
\fill[legColour] #1--($#1+(#2+\legSpread*3:\extLegLen)$)--($#1+(#2-\legSpread*3:\extLegLen)$);
\node at #1 [ddot]{};
}

\newcommand{\legAltLabelled}[3]{
\fill[legColour] #1--($#1+(#2+\legSpread*3:\extLegLen*1.4)$)--($#1+(#2-\legSpread*3:\extLegLen*1.4)$);
\node at #1 [ddot]{};
\node at ($#1+(#2:\labelDist)$)[]{{\normalsize #3}};
}

\newcommand{\oneLoopGraphElement}[2][0]{\def\rotn{-360/#2}\def\edgeLen{0.75*\figScale}
\ifthenelse{#2=3}{\coordinate (v0) at (0,-\edgeLen/9)}{\coordinate (v0) at (0,0)};
\ifthenelse{#2=1}{\draw[int]($(v0)+(\edgeLen/2*0.3,0)$) arc (0:360:\edgeLen/2*0.3 and \edgeLen/1.125*0.3)coordinate[pos=0.25](e1);\legAlt{($(v0)-(0,\edgeLen/1.125*0.3)$)}{-90}}{\ifthenelse{#2=2}{
\draw[int]($(v0)+(\edgeLen/1.125*0.45,0)$) arc (0:360:\edgeLen/1.125*0.45 and \edgeLen/2*0.45)coordinate[pos=0.5](a1)coordinate[pos=0](a0)coordinate[pos=0.25](e1)coordinate[pos=0.75](e2);\legAlt{(a1)}{180};\legAlt{(a0)}{0};
}{
\foreach\a in {0,...,#2}{\coordinate (a\a) at ($(v0)+(\a*\rotn-90-\rotn/2:\edgeLen/2)$);};
\foreach\a[remember=\a as \la] in {0,...,#2}{\ifthenelse{\a=0}{}{\draw[int](a\la)--(a\a)coordinate (e\a) at ($(a\la)!0.5!(a\a)$);}};
\foreach\a in {1,...,#2}{\legAlt{(a\a)}{\a*\rotn-90-\rotn/2};};
}};
}

\newcommand{\oneLoopGraphElementLS}[2][0]{\def\extLegLen{1.1*0.32*\figScale}
\def\rotn{-360/#2}\def\edgeLen{0.75*\figScale}
\ifthenelse{#2=3}{\coordinate (v0) at (0,-\edgeLen/9)}{\coordinate (v0) at (0,0)};
\ifthenelse{#2=1}{\draw[int]($(v0)+(\edgeLen/2*0.3,0)$) arc (0:360:\edgeLen/2*0.3 and \edgeLen/1.125*0.3)coordinate[pos=0.25](e1);\legAlt{($(v0)-(0,\edgeLen/1.125*0.3)$)}{-90}}{\ifthenelse{#2=2}{
\draw[int]($(v0)+(\edgeLen/1.125*0.45,0)$) arc (0:360:\edgeLen/1.125*0.45 and \edgeLen/2*0.45)coordinate[pos=0.5](a1)coordinate[pos=0](a0)coordinate[pos=0.25](e1)coordinate[pos=0.75](e2);\legAlt{(a1)}{180};\legAlt{(a0)}{0};
}{
\foreach\a in {0,...,#2}{\coordinate (a\a) at ($(v0)+(\a*\rotn-90-\rotn/2:\edgeLen/2)$);};
\foreach\a[remember=\a as \la] in {0,...,#2}{\ifthenelse{\a=0}{}{\draw[int](a\la)--(a\a)coordinate (e\a) at ($(a\la)!0.5!(a\a)$);}};
\foreach\a in {1,...,#2}{\legAlt{(a\a)}{\a*\rotn-90-\rotn/2};};
}};\def\extLegLen{0.9*0.32*\figScale}
}

\newcommand{\intDots}[1]{
\foreach\n in {1,...,#1}{\node at (v\n) [bdot] {};};}

\newcommand{\lsVerts}[7]{
\ifthenelse{#1=1}{\node at (v1) [fullmhvBar] {};}{\ifthenelse{#1=2}{\node at (v1) [fullmhv] {};}{\ifthenelse{#1=3}{\node at (v1) [fullnmhv] {};}{\ifthenelse{#1=0}{\node at (v1) [bdot] {};}{}}}}
\ifthenelse{#2=1}{\node at (v2) [fullmhvBar] {};}{\ifthenelse{#2=2}{\node at (v2) [fullmhv] {};}{\ifthenelse{#2=3}{\node at (v2) [fullnmhv] {};}{\ifthenelse{#2=0}{\node at (v2) [bdot] {};}{}}}}
\ifthenelse{#3=1}{\node at (v3) [fullmhvBar] {};}{\ifthenelse{#3=2}{\node at (v3) [fullmhv] {};}{\ifthenelse{#3=3}{\node at (v3) [fullnmhv] {};}{\ifthenelse{#3=0}{\node at (v3) [bdot] {};}{}}}}
\ifthenelse{#4=1}{\node at (v4) [fullmhvBar] {};}{\ifthenelse{#4=2}{\node at (v4) [fullmhv] {};}{\ifthenelse{#4=3}{\node at (v4) [fullnmhv] {};}{\ifthenelse{#4=0}{\node at (v4) [bdot] {};}{}}}}
\ifthenelse{#5=1}{\node at (v5) [fullmhvBar] {};}{\ifthenelse{#5=2}{\node at (v5) [fullmhv] {};}{\ifthenelse{#5=3}{\node at (v5) [fullnmhv] {};}{\ifthenelse{#5=0}{\node at (v5) [bdot] {};}{}}}}
\ifthenelse{#6=1}{\node at (v6) [fullmhvBar] {};}{\ifthenelse{#6=2}{\node at (v6) [fullmhv] {};}{\ifthenelse{#6=3}{\node at (v6) [fullnmhv] {};}{\ifthenelse{#6=0}{\node at (v6) [bdot] {};}{}}}}
\ifthenelse{#7=1}{\node at (v7) [fullmhvBar] {};}{\ifthenelse{#7=2}{\node at (v7) [fullmhv] {};}{\ifthenelse{#7=3}{\node at (v7) [fullnmhv] {};}{\ifthenelse{#7=0}{\node at (v7) [bdot] {};}{}}}}}

\newcommand{\dPent}[1]{\begin{tikzpicture}[scale=\figScale,baseline=-3.05]\useasboundingbox ($(-\pageW/9,-1.4)$) rectangle ($(\pageW/9,1.4)$);\draw[int,line width=0.1,red,draw=\boundingDraw] ($(-\pageW/9,-1.4)$) rectangle ($(\pageW/9,1.4)$);\coordinate(v7)at($\figScale*(0,0)$);\coordinate(v6)at($(v7)+(-90:\figScale*0.65)$);\coordinate(v3)at($(v7)+(90:\figScale*0.65)$);\coordinate(v2)at($(v3)+(180:\figScale*1.25)$);\coordinate(v4)at($(v3)+(0:\figScale*1.25)$);\coordinate(v1)at($(v6)+(180:\figScale*1.25)$);\coordinate(v5)at($(v6)+(0:\figScale*1.25)$);
#1\end{tikzpicture}}

\newcommand{\dBox}[1]{\begin{tikzpicture}[scale=\figScale,baseline=-3.05]\useasboundingbox ($(-\pageW/9,-1.4)$) rectangle ($(\pageW/9,1.4)$);\draw[int,line width=0.1,red,draw=\boundingDraw] ($(-\pageW/9,-1.4)$) rectangle ($(\pageW/9,1.4)$);\coordinate(v7)at($\figScale*(0,0)$);\coordinate(v6)at($(v7)+(-90:\figScale*0.65)$);\coordinate(v3)at($(v7)+(90:\figScale*0.65)$);\coordinate(v2)at($(v3)+(180:\figScale*1.25)$);\coordinate(v4)at($(v3)+(0:\figScale*1.25)$);\coordinate(v1)at($(v6)+(180:\figScale*1.25)$);\coordinate(v5)at($(v6)+(0:\figScale*1.25)$);
#1\end{tikzpicture}}

\newcommand{\pT}[1]{\begin{tikzpicture}[scale=\figScale,baseline=-3.05]\useasboundingbox ($(-\pageW/9,-1.5)$) rectangle ($(\pageW/9,1.5)$);\draw[int,line width=0.1,red,draw=\boundingDraw] ($(-\pageW/9,-1.5)$) rectangle ($(\pageW/9,1.5)$);\coordinate(v2)at($\figScale*(-1.25,0)$);\coordinate(v1)at($(v2)+(-54:\figScale*1.119)$);\coordinate(v3)at($(v2)+(54:\figScale*1.119)$);\coordinate(v4)at($\figScale*(0.3,0.65)$);\coordinate(v5)at($\figScale*(1.2,-0)$);\coordinate(v6)at($\figScale*(0.3,-0.65)$);
#1\end{tikzpicture}}

\newcommand{\npPbox}[1]{\begin{tikzpicture}[scale=\figScale,baseline=-3.05]\useasboundingbox ($(-\pageW/9,-1.5)$) rectangle ($(\pageW/9,1.5)$);\draw[int,line width=0.1,red,draw=\boundingDraw] ($(-\pageW/9,-1.5)$) rectangle ($(\pageW/9,1.5)$);\coordinate(v7)at($\figScale*(-1.85,0)$);\coordinate(v1)at($(v7)+(-54:\figScale*1.119)$);\coordinate(v2)at($(v7)+(54:\figScale*1.119)$);\coordinate(v3)at($\figScale*(0.35,0.65)$);\coordinate(v4)at($\figScale*(1.0,-0)$);\coordinate(v5)at($\figScale*(0.35,-0.65)$);\coordinate(v6)at($\figScale*(-0.2,0)$);
#1\end{tikzpicture}}

\newcommand{\kT}[1]{\begin{tikzpicture}[scale=\figScale,baseline=-3.05]\useasboundingbox ($(-\pageW/9,-1.2)$) rectangle ($(\pageW/9,1.2)$);\draw[int,line width=0.1,red,draw=\boundingDraw] ($(-\pageW/9,-1.2)$) rectangle ($(\pageW/9,1.2)$);
\coordinate(v3)at(-0.0,0);\coordinate(v1)at($\figScale*(-0.65-0.2,-0.65)$);\coordinate(v2)at($\figScale*(-0.65-0.2,0.65)$);\coordinate(v4)at($\figScale*(0.65+0.2,0.65)$);\coordinate(v5)at($\figScale*(0.65+0.2,-0.65)$);
#1\end{tikzpicture}}

\newcommand{\bT}[1]{\begin{tikzpicture}[scale=\figScale,baseline=-3.05]\useasboundingbox ($(-\pageW/9,-1.2)$) rectangle ($(\pageW/9,1.2)$);\draw[int,line width=0.1,red,draw=\boundingDraw] ($(-\pageW/9,-1.2)$) rectangle ($(\pageW/9,1.2)$);\coordinate(v7)at($\figScale*(0.2,0)$);\coordinate(v5)at($(v7)+(-90:\figScale*0.65)$);\coordinate(v3)at($(v7)+(90:\figScale*0.65)$);\coordinate(v2)at($(v3)+(180:\figScale*1.25)$);\coordinate(v4)at($(v7)+(0:\figScale*0.9)$);\coordinate(v1)at($(v5)+(180:\figScale*1.25)$);
#1\end{tikzpicture}}

\newcommand{\dT}[1]{\begin{tikzpicture}[scale=\figScale,baseline=-3.05]\useasboundingbox ($(-\pageW/9,-1.2)$) rectangle ($(\pageW/9,1.2)$);\draw[int,line width=0.1,red,draw=\boundingDraw] ($(-\pageW/9,-1.2)$) rectangle ($(\pageW/9,1.2)$);\coordinate(v1)at($(180:\figScale*1.05)$);\coordinate(v2)at($(90:\figScale*0.65)$);\coordinate(v3)at($(0:\figScale*1.05)$);\coordinate(v4)at($(-90:\figScale*0.65)$);
#1\end{tikzpicture}}

\newcommand{\tardi}[1]{\begin{tikzpicture}[scale=\figScale,baseline=-3.05]\useasboundingbox ($(-\pageW/9,-1.35)$) rectangle ($(\pageW/9,1.35)$);\draw[int,line width=0.1,red,draw=\boundingDraw] ($(-\pageW/9,-1.35)$) rectangle ($(\pageW/9,1.35)$);\coordinate(v1)at($(0,0.2)+(180:\figScale*1.15)$);\coordinate(v2)at($(0,0.2)+(90:\figScale*0.75)$);\coordinate(v3)at($(0,0.2)+(0:\figScale*1.15)$);\coordinate(v4)at($(0,0.2)+(-90:\figScale*0.75)$);\coordinate(v5)at (0,0.2);
#1\end{tikzpicture}}

\newcommand{\dPentPlainEdges}{\draw[int](v6)--(v1);\draw[int](v1)--(v2);\draw[int](v2)--(v3);\draw[int](v3)--(v4);\draw[int](v4)--(v5);\draw[int](v5)--(v6);\draw[int](v6)--(v7);\draw[int](v7)--(v3);}
\newcommand{\dPentScalarEdges}{\dPentPlainEdges\draw[directedEdge](v6)--(v1);\node[inner sep=3pt,anchor=north] at (connode) [] {\edgeA};\draw[directedEdge](v1)--(v2);\node[inner sep=3pt,anchor=east] at (connode) [] {\edgeB};\draw[directedEdge](v2)--(v3);\node[inner sep=3pt,anchor=south] at (connode) [] {\edgeC};\draw[directedEdge](v3)--(v4);\node[inner sep=3pt,anchor=south] at (connode) [] {\edgeD};\draw[directedEdge](v4)--(v5);\node[inner sep=3pt,anchor=west] at (connode) [] {\edgeE};\draw[directedEdge](v5)--(v6);\node[inner sep=3pt,anchor=north] at (connode) [] {\edgeF};\draw[directedEdge](v6)--(v7);\node[inner sep=2pt,anchor=west] at (connode) [] {\edgeG};\draw[directedEdge](v7)--(v3);\node[inner sep=2pt,anchor=west] at (connode) [] {\edgeH};}

\newcommand{\dBoxPlainEdges}{\draw[int](v6)--(v1);\draw[int](v1)--(v2);\draw[int](v2)--(v3);\draw[int](v3)--(v4);\draw[int](v4)--(v5);\draw[int](v5)--(v6);\draw[int](v6)--(v3);}
\newcommand{\dBoxScalarEdges}{\dBoxPlainEdges\draw[directedEdge](v6)--(v1);\node[inner sep=3pt,anchor=north] at (connode) [] {\edgeA};\draw[directedEdge](v1)--(v2);\node[inner sep=3pt,anchor=east] at (connode) [] {\edgeB};\draw[directedEdge](v2)--(v3);\node[inner sep=3pt,anchor=south] at (connode) [] {\edgeC};\draw[directedEdge](v3)--(v4);\node[inner sep=3pt,anchor=south] at (connode) [] {\edgeD};\draw[directedEdge](v4)--(v5);\node[inner sep=3pt,anchor=west] at (connode) [] {\edgeE};\draw[directedEdge](v5)--(v6);\node[inner sep=3pt,anchor=north] at (connode) [] {\edgeF};\draw[directedEdge](v6)--(v3);\node[inner sep=2pt,anchor=west] at (connode) [] {\edgeG};}

\newcommand{\pTArrows}{
\draw[directedEdge](v6)--(v1);\node[inner sep=1pt,anchor=north west] at (connode) [] {\edgeA};\draw[directedEdge](v1)--(v2);\node[inner sep=1pt,anchor=north east] at (connode) [] {\edgeB};\draw[directedEdge](v2)--(v3);\node[inner sep=1pt,anchor=south east] at (connode) [] {\edgeC};\draw[directedEdge](v3)--(v4);\node[inner sep=1pt,anchor=south west] at (connode) [] {\edgeD};\draw[directedEdge](v4)--(v5);\node[inner sep=1pt,anchor=south west] at (connode) [] {\edgeE};\draw[directedEdge](v5)--(v6);\node[inner sep=1pt,anchor=north west] at (connode) [] {\edgeF};\draw[directedEdge](v6)--(v4);\node[inner sep=1pt,anchor=east] at (connode) [] {\edgeG};}

\newcommand{\kTScalarEdges}{\draw[int](v3)--(v1);\draw[int](v1)--(v2);\draw[int](v2)--(v3);\draw[int](v3)--(v4);\draw[int](v4)--(v5);\draw[int](v5)--(v3);
\draw[directedEdge](v3)--(v1);\node[inner sep=2.5pt,anchor=north] at (connode) [] {\edgeA};\draw[directedEdge](v1)--(v2);\node[inner sep=2pt,anchor=east] at (connode) [] {\edgeB};\draw[directedEdge](v2)--(v3);\node[inner sep=2.5pt,anchor=south] at (connode) [] {\edgeC};\draw[directedEdge](v3)--(v4);\node[inner sep=2.5pt,anchor=south] at (connode) [] {\edgeD};\draw[directedEdge](v4)--(v5);\node[inner sep=1.5pt,anchor=south west] at (connode) [] {\edgeE};\draw[directedEdge](v5)--(v3);\node[inner sep=0.5pt,anchor=north] at (connode) [] {\edgeF\;\,\,};}

\newcommand{\bTScalarEdges}{\draw[int](v5)--(v1);\draw[int](v1)--(v2);\draw[int](v2)--(v3);\draw[int](v3)--(v4);\draw[int](v4)--(v5);\draw[int](v5)--(v3);
\draw[directedEdge](v5)--(v1);\node[inner sep=1pt,anchor=north] at (connode) [] {\edgeA};\draw[directedEdge](v1)--(v2);\node[inner sep=2.5pt,anchor=east] at (connode) [] {\edgeB};\draw[directedEdge](v2)--(v3);\node[inner sep=1pt,anchor=south west] at (connode) [] {\edgeC};\draw[directedEdge](v3)--(v4);\node[inner sep=1pt,anchor=south west] at (connode) [] {\edgeD};\draw[directedEdge](v4)--(v5);\node[inner sep=1pt,anchor=north west] at (connode) [] {\edgeE};\draw[directedEdge](v5)--(v3);\node[inner sep=1pt,anchor=east] at (connode) [] {\edgeF};}

\newcommand{\dTScalarEdges}{\draw[int](v4)--(v1);\draw[int](v1)--(v2);\draw[int](v2)--(v3);\draw[int](v3)--(v4);\draw[int](v4)--(v2);
\draw[directedEdge](v4)--(v1);\node[inner sep=1pt,anchor=north east] at (connode) [] {\edgeA};\draw[directedEdge](v1)--(v2);\node[inner sep=1.5pt,anchor=south east] at (connode) [] {\edgeB};\draw[directedEdge](v2)--(v3);\node[inner sep=1pt,anchor=south west] at (connode) [] {\edgeC};\draw[directedEdge](v3)--(v4);\node[inner sep=1pt,anchor=north west] at (connode) [] {\edgeD};\draw[directedEdge](v4)--(v2);\node[inner sep=2pt,anchor=east] at (connode) [] {\edgeE};}

\newcommand{\tardiScalarEdges}{\draw[int](v4)--(v1);\draw[int](v1)--(v2);\draw[int](v2)--(v3);\draw[int](v3)--(v4);\draw[int](v4)--(v5);\draw[int](v5)--(v2);
\draw[directedEdge](v4)--(v1);\node[inner sep=1pt,anchor=north east] at (connode) [] {\edgeA};\draw[directedEdge](v1)--(v2);\node[inner sep=1.5pt,anchor=south east] at (connode) [] {\edgeB};\draw[directedEdge](v2)--(v3);\node[inner sep=1pt,anchor=south west] at (connode) [] {\edgeC};\draw[directedEdge](v3)--(v4);\node[inner sep=1pt,anchor=north west] at (connode) [] {\edgeD};\draw[directedEdge](v4)--(v5);\node[inner sep=1.2pt,anchor=west] at (connode) [] {\edgeE};\draw[directedEdge](v5)--(v2);\node[inner sep=1.2pt,anchor=west] at (connode) [] {\edgeF};}

\newcommand{\kBoxLegs}[7]{
\setcounter{legSteps}{0}
\def\zeroAngle{-90}\def\spread{45}\setcounter{offset}{-1}\addtocounter{offset}{#1}
\ifthenelse{#1=0}{}{\ifthenelse{#1=1}{\stepcounter{legSteps}\leg{(v1)}{\zeroAngle}{\arabic{legSteps}};}{
\foreach\n in {1,...,#1}{\def\eph{\arabic{offset}}\def\angle{\zeroAngle-2*\n*\spread/\eph+#1*\spread/\eph+\spread/\eph}\stepcounter{legSteps}\leg{(v1)}{\angle}{\arabic{legSteps}}}}}

\def\zeroAngle{180}\def\spread{45}\setcounter{offset}{-1}\addtocounter{offset}{#2}
\ifthenelse{#2=0}{}{\ifthenelse{#2=1}{\stepcounter{legSteps}\leg{(v2)}{\zeroAngle}{\arabic{legSteps}};}{
\foreach\n in {1,...,#2}{\def\eph{\arabic{offset}}\def\angle{\zeroAngle-2*\n*\spread/\eph+#2*\spread/\eph+\spread/\eph}\stepcounter{legSteps}\leg{(v2)}{\angle}{\arabic{legSteps}}}}}

\def\zeroAngle{90}\def\spread{45}\setcounter{offset}{-1}\addtocounter{offset}{#3}
\ifthenelse{#3=0}{}{\ifthenelse{#3=1}{\stepcounter{legSteps}\leg{(v3)}{\zeroAngle}{\arabic{legSteps}};}{
\foreach\n in {1,...,#3}{\def\eph{\arabic{offset}}\def\angle{\zeroAngle-2*\n*\spread/\eph+#3*\spread/\eph+\spread/\eph}\stepcounter{legSteps}\leg{(v3)}{\angle}{\arabic{legSteps}}}}}

\def\zeroAngle{90}\def\spread{15}\setcounter{offset}{-1}\addtocounter{offset}{#4}
\ifthenelse{#4=0}{}{\ifthenelse{#4=1}{\stepcounter{legSteps}\leg{(v4)}{\zeroAngle}{\arabic{legSteps}};}{
\foreach\n in {1,...,#4}{\def\eph{\arabic{offset}}\def\angle{\zeroAngle-2*\n*\spread/\eph+#4*\spread/\eph+\spread/\eph}\stepcounter{legSteps}\leg{(v4)}{\angle}{\arabic{legSteps}}}}}

\def\zeroAngle{90}\def\spread{45}\setcounter{offset}{-1}\addtocounter{offset}{#5}
\ifthenelse{#5=0}{}{\ifthenelse{#5=1}{\stepcounter{legSteps}\leg{(v5)}{\zeroAngle}{\arabic{legSteps}};}{
\foreach\n in {1,...,#5}{\def\eph{\arabic{offset}}\def\angle{\zeroAngle-2*\n*\spread/\eph+#5*\spread/\eph+\spread/\eph}\stepcounter{legSteps}\leg{(v5)}{\angle}{\arabic{legSteps}}}}}

\def\zeroAngle{0}\def\spread{45}\setcounter{offset}{-1}\addtocounter{offset}{#6}
\ifthenelse{#6=0}{}{\ifthenelse{#6=1}{\stepcounter{legSteps}\leg{(v6)}{\zeroAngle}{\arabic{legSteps}};}{
\foreach\n in {1,...,#6}{\def\eph{\arabic{offset}}\def\angle{\zeroAngle-2*\n*\spread/\eph+#6*\spread/\eph+\spread/\eph}\stepcounter{legSteps}\leg{(v6)}{\angle}{\arabic{legSteps}}}}}

\def\zeroAngle{-90}\def\spread{45}\setcounter{offset}{-1}\addtocounter{offset}{#7}
\ifthenelse{#7=0}{}{\ifthenelse{#7=1}{\stepcounter{legSteps}\leg{(v7)}{\zeroAngle}{\arabic{legSteps}};}{
\foreach\n in {1,...,#7}{\def\eph{\arabic{offset}}\def\angle{\zeroAngle-2*\n*\spread/\eph+#7*\spread/\eph+\spread/\eph}\stepcounter{legSteps}\leg{(v7)}{\angle}{\arabic{legSteps}}}}}
}

\newcommand{\kTboxLegs}[6]{
\setcounter{legSteps}{0}
\def\zeroAngle{-135}\def\spread{45}\setcounter{offset}{-1}\addtocounter{offset}{#1}
\ifthenelse{#1=0}{}{\ifthenelse{#1=1}{\stepcounter{legSteps}\leg{(v1)}{\zeroAngle}{\arabic{legSteps}};}{
\foreach\n in {1,...,#1}{\def\eph{\arabic{offset}}\def\angle{\zeroAngle-2*\n*\spread/\eph+#1*\spread/\eph+\spread/\eph}\stepcounter{legSteps}\leg{(v1)}{\angle}{\arabic{legSteps}}}}}

\def\zeroAngle{135}\def\spread{45}\setcounter{offset}{-1}\addtocounter{offset}{#2}
\ifthenelse{#2=0}{}{\ifthenelse{#2=1}{\stepcounter{legSteps}\leg{(v2)}{\zeroAngle}{\arabic{legSteps}};}{
\foreach\n in {1,...,#2}{\def\eph{\arabic{offset}}\def\angle{\zeroAngle-2*\n*\spread/\eph+#2*\spread/\eph+\spread/\eph}\stepcounter{legSteps}\leg{(v2)}{\angle}{\arabic{legSteps}}}}}

\def\zeroAngle{90}\def\spread{15}\setcounter{offset}{-1}\addtocounter{offset}{#3}
\ifthenelse{#3=0}{}{\ifthenelse{#3=1}{\stepcounter{legSteps}\leg{(v3)}{\zeroAngle}{\arabic{legSteps}};}{
\foreach\n in {1,...,#3}{\def\eph{\arabic{offset}}\def\angle{\zeroAngle-2*\n*\spread/\eph+#3*\spread/\eph+\spread/\eph}\stepcounter{legSteps}\leg{(v3)}{\angle}{\arabic{legSteps}}}}}

\def\zeroAngle{90}\def\spread{45}\setcounter{offset}{-1}\addtocounter{offset}{#4}
\ifthenelse{#4=0}{}{\ifthenelse{#4=1}{\stepcounter{legSteps}\leg{(v4)}{\zeroAngle}{\arabic{legSteps}};}{
\foreach\n in {1,...,#4}{\def\eph{\arabic{offset}}\def\angle{\zeroAngle-2*\n*\spread/\eph+#4*\spread/\eph+\spread/\eph}\stepcounter{legSteps}\leg{(v4)}{\angle}{\arabic{legSteps}}}}}

\def\zeroAngle{0}\def\spread{45}\setcounter{offset}{-1}\addtocounter{offset}{#5}
\ifthenelse{#5=0}{}{\ifthenelse{#5=1}{\stepcounter{legSteps}\leg{(v5)}{\zeroAngle}{\arabic{legSteps}};}{
\foreach\n in {1,...,#5}{\def\eph{\arabic{offset}}\def\angle{\zeroAngle-2*\n*\spread/\eph+#5*\spread/\eph+\spread/\eph}\stepcounter{legSteps}\leg{(v5)}{\angle}{\arabic{legSteps}}}}}

\def\zeroAngle{-90}\def\spread{45}\setcounter{offset}{-1}\addtocounter{offset}{#6}
\ifthenelse{#6=0}{}{\ifthenelse{#6=1}{\stepcounter{legSteps}\leg{(v6)}{\zeroAngle}{\arabic{legSteps}};}{
\foreach\n in {1,...,#6}{\def\eph{\arabic{offset}}\def\angle{\zeroAngle-2*\n*\spread/\eph+#6*\spread/\eph+\spread/\eph}\stepcounter{legSteps}\leg{(v6)}{\angle}{\arabic{legSteps}}}}}
}

\newcommand{\kTLegs}[5]{
\setcounter{legSteps}{0}
\def\zeroAngle{-135}\def\spread{45}\setcounter{offset}{-1}\addtocounter{offset}{#1}
\ifthenelse{#1=0}{}{\ifthenelse{#1=1}{\stepcounter{legSteps}\leg{(v1)}{\zeroAngle}{\arabic{legSteps}};}{
\foreach\n in {1,...,#1}{\def\eph{\arabic{offset}}\def\angle{\zeroAngle-2*\n*\spread/\eph+#1*\spread/\eph+\spread/\eph}\stepcounter{legSteps}\leg{(v1)}{\angle}{\arabic{legSteps}}}}}

\def\zeroAngle{135}\def\spread{45}\setcounter{offset}{-1}\addtocounter{offset}{#2}
\ifthenelse{#2=0}{}{\ifthenelse{#2=1}{\stepcounter{legSteps}\leg{(v2)}{\zeroAngle}{\arabic{legSteps}};}{
\foreach\n in {1,...,#2}{\def\eph{\arabic{offset}}\def\angle{\zeroAngle-2*\n*\spread/\eph+#2*\spread/\eph+\spread/\eph}\stepcounter{legSteps}\leg{(v2)}{\angle}{\arabic{legSteps}}}}}

\def\zeroAngle{90}\def\spread{15}\setcounter{offset}{-1}\addtocounter{offset}{#3}
\ifthenelse{#3=0}{}{\ifthenelse{#3=1}{\stepcounter{legSteps}\leg{(v3)}{\zeroAngle}{\arabic{legSteps}};}{
\foreach\n in {1,...,#3}{\def\eph{\arabic{offset}}\def\angle{\zeroAngle-2*\n*\spread/\eph+#3*\spread/\eph+\spread/\eph}\stepcounter{legSteps}\leg{(v3)}{\angle}{\arabic{legSteps}}}}}

\def\zeroAngle{45}\def\spread{45}\setcounter{offset}{-1}\addtocounter{offset}{#4}
\ifthenelse{#4=0}{}{\ifthenelse{#4=1}{\stepcounter{legSteps}\leg{(v4)}{\zeroAngle}{\arabic{legSteps}};}{
\foreach\n in {1,...,#4}{\def\eph{\arabic{offset}}\def\angle{\zeroAngle-2*\n*\spread/\eph+#4*\spread/\eph+\spread/\eph}\stepcounter{legSteps}\leg{(v4)}{\angle}{\arabic{legSteps}}}}}

\def\zeroAngle{-45}\def\spread{45}\setcounter{offset}{-1}\addtocounter{offset}{#5}
\ifthenelse{#5=0}{}{\ifthenelse{#5=1}{\stepcounter{legSteps}\leg{(v5)}{\zeroAngle}{\arabic{legSteps}};}{
\foreach\n in {1,...,#5}{\def\eph{\arabic{offset}}\def\angle{\zeroAngle-2*\n*\spread/\eph+#5*\spread/\eph+\spread/\eph}\stepcounter{legSteps}\leg{(v5)}{\angle}{\arabic{legSteps}}}}}
}

\newcommand{\pBoxLegs}[7]{
\setcounter{legSteps}{0}
\def\zeroAngle{-108}\def\spread{45}\setcounter{offset}{-1}\addtocounter{offset}{#1}
\ifthenelse{#1=0}{}{\ifthenelse{#1=1}{\stepcounter{legSteps}\leg{(v1)}{\zeroAngle}{\arabic{legSteps}};}{
\foreach\n in {1,...,#1}{\def\eph{\arabic{offset}}\def\angle{\zeroAngle-2*\n*\spread/\eph+#1*\spread/\eph+\spread/\eph}\stepcounter{legSteps}\leg{(v1)}{\angle}{\arabic{legSteps}}}}}

\def\zeroAngle{180}\def\spread{45}\setcounter{offset}{-1}\addtocounter{offset}{#2}
\ifthenelse{#2=0}{}{\ifthenelse{#2=1}{\stepcounter{legSteps}\leg{(v2)}{\zeroAngle}{\arabic{legSteps}};}{
\foreach\n in {1,...,#2}{\def\eph{\arabic{offset}}\def\angle{\zeroAngle-2*\n*\spread/\eph+#2*\spread/\eph+\spread/\eph}\stepcounter{legSteps}\leg{(v2)}{\angle}{\arabic{legSteps}}}}}

\def\zeroAngle{108}\def\spread{45}\setcounter{offset}{-1}\addtocounter{offset}{#3}
\ifthenelse{#3=0}{}{\ifthenelse{#3=1}{\stepcounter{legSteps}\leg{(v3)}{\zeroAngle}{\arabic{legSteps}};}{
\foreach\n in {1,...,#3}{\def\eph{\arabic{offset}}\def\angle{\zeroAngle-2*\n*\spread/\eph+#3*\spread/\eph+\spread/\eph}\stepcounter{legSteps}\leg{(v3)}{\angle}{\arabic{legSteps}}}}}

\def\zeroAngle{81}\def\spread{20}\setcounter{offset}{-1}\addtocounter{offset}{#4}
\ifthenelse{#4=0}{}{\ifthenelse{#4=1}{\stepcounter{legSteps}\leg{(v4)}{\zeroAngle}{\arabic{legSteps}};}{
\foreach\n in {1,...,#4}{\def\eph{\arabic{offset}}\def\angle{\zeroAngle-2*\n*\spread/\eph+#4*\spread/\eph+\spread/\eph}\stepcounter{legSteps}\leg{(v4)}{\angle}{\arabic{legSteps}}}}}

\def\zeroAngle{45}\def\spread{45}\setcounter{offset}{-1}\addtocounter{offset}{#5}
\ifthenelse{#5=0}{}{\ifthenelse{#5=1}{\stepcounter{legSteps}\leg{(v5)}{\zeroAngle}{\arabic{legSteps}};}{
\foreach\n in {1,...,#5}{\def\eph{\arabic{offset}}\def\angle{\zeroAngle-2*\n*\spread/\eph+#5*\spread/\eph+\spread/\eph}\stepcounter{legSteps}\leg{(v5)}{\angle}{\arabic{legSteps}}}}}

\def\zeroAngle{-45}\def\spread{45}\setcounter{offset}{-1}\addtocounter{offset}{#6}
\ifthenelse{#6=0}{}{\ifthenelse{#6=1}{\stepcounter{legSteps}\leg{(v6)}{\zeroAngle}{\arabic{legSteps}};}{
\foreach\n in {1,...,#6}{\def\eph{\arabic{offset}}\def\angle{\zeroAngle-2*\n*\spread/\eph+#6*\spread/\eph+\spread/\eph}\stepcounter{legSteps}\leg{(v6)}{\angle}{\arabic{legSteps}}}}}

\def\zeroAngle{-81}\def\spread{20}\setcounter{offset}{-1}\addtocounter{offset}{#7}
\ifthenelse{#7=0}{}{\ifthenelse{#7=1}{\stepcounter{legSteps}\leg{(v7)}{\zeroAngle}{\arabic{legSteps}};}{
\foreach\n in {1,...,#7}{\def\eph{\arabic{offset}}\def\angle{\zeroAngle-2*\n*\spread/\eph+#7*\spread/\eph+\spread/\eph}\stepcounter{legSteps}\leg{(v7)}{\angle}{\arabic{legSteps}}}}}
}

\newcommand{\hBoxLegs}[7]{
\setcounter{legSteps}{0}
\def\zeroAngle{-115}\def\spread{45}\setcounter{offset}{-1}\addtocounter{offset}{#1}
\ifthenelse{#1=0}{}{\ifthenelse{#1=1}{\stepcounter{legSteps}\leg{(v1)}{\zeroAngle}{\arabic{legSteps}};}{
\foreach\n in {1,...,#1}{\def\eph{\arabic{offset}}\def\angle{\zeroAngle-2*\n*\spread/\eph+#1*\spread/\eph+\spread/\eph}\stepcounter{legSteps}\leg{(v1)}{\angle}{\arabic{legSteps}}}}}

\def\zeroAngle{180}\def\spread{45}\setcounter{offset}{-1}\addtocounter{offset}{#2}
\ifthenelse{#2=0}{}{\ifthenelse{#2=1}{\stepcounter{legSteps}\leg{(v2)}{\zeroAngle}{\arabic{legSteps}};}{
\foreach\n in {1,...,#2}{\def\eph{\arabic{offset}}\def\angle{\zeroAngle-2*\n*\spread/\eph+#2*\spread/\eph+\spread/\eph}\stepcounter{legSteps}\leg{(v2)}{\angle}{\arabic{legSteps}}}}}

\def\zeroAngle{115}\def\spread{45}\setcounter{offset}{-1}\addtocounter{offset}{#3}
\ifthenelse{#3=0}{}{\ifthenelse{#3=1}{\stepcounter{legSteps}\leg{(v3)}{\zeroAngle}{\arabic{legSteps}};}{
\foreach\n in {1,...,#3}{\def\eph{\arabic{offset}}\def\angle{\zeroAngle-2*\n*\spread/\eph+#3*\spread/\eph+\spread/\eph}\stepcounter{legSteps}\leg{(v3)}{\angle}{\arabic{legSteps}}}}}

\def\zeroAngle{61.5}\def\spread{30}\setcounter{offset}{-1}\addtocounter{offset}{#4}
\ifthenelse{#4=0}{}{\ifthenelse{#4=1}{\stepcounter{legSteps}\leg{(v4)}{\zeroAngle}{\arabic{legSteps}};}{
\foreach\n in {1,...,#4}{\def\eph{\arabic{offset}}\def\angle{\zeroAngle-2*\n*\spread/\eph+#4*\spread/\eph+\spread/\eph}\stepcounter{legSteps}\leg{(v4)}{\angle}{\arabic{legSteps}}}}}

\def\zeroAngle{0}\def\spread{45}\setcounter{offset}{-1}\addtocounter{offset}{#5}
\ifthenelse{#5=0}{}{\ifthenelse{#5=1}{\stepcounter{legSteps}\leg{(v5)}{\zeroAngle}{\arabic{legSteps}};}{
\foreach\n in {1,...,#5}{\def\eph{\arabic{offset}}\def\angle{\zeroAngle-2*\n*\spread/\eph+#5*\spread/\eph+\spread/\eph}\stepcounter{legSteps}\leg{(v5)}{\angle}{\arabic{legSteps}}}}}

\def\zeroAngle{-61.5}\def\spread{30}\setcounter{offset}{-1}\addtocounter{offset}{#6}
\ifthenelse{#6=0}{}{\ifthenelse{#6=1}{\stepcounter{legSteps}\leg{(v6)}{\zeroAngle}{\arabic{legSteps}};}{
\foreach\n in {1,...,#6}{\def\eph{\arabic{offset}}\def\angle{\zeroAngle-2*\n*\spread/\eph+#6*\spread/\eph+\spread/\eph}\stepcounter{legSteps}\leg{(v6)}{\angle}{\arabic{legSteps}}}}}

\def\zeroAngle{180}\def\spread{25}\setcounter{offset}{-1}\addtocounter{offset}{#7}
\ifthenelse{#7=0}{}{\ifthenelse{#7=1}{\stepcounter{legSteps}\leg{(v7)}{\zeroAngle}{\arabic{legSteps}};}{
\foreach\n in {1,...,#7}{\def\eph{\arabic{offset}}\def\angle{\zeroAngle-2*\n*\spread/\eph+#7*\spread/\eph+\spread/\eph}\stepcounter{legSteps}\leg{(v7)}{\angle}{\arabic{legSteps}}}}}
}

\newcommand{\npPboxLegs}[6]{
\setcounter{legSteps}{0}
\def\zeroAngle{-135}\def\spread{45}\setcounter{offset}{-1}\addtocounter{offset}{#1}
\ifthenelse{#1=0}{}{\ifthenelse{#1=1}{\stepcounter{legSteps}\leg{(v1)}{\zeroAngle}{{\color{labelcolor}\arabic{legSteps}}};}{
\foreach\n in {1,...,#1}{\def\eph{\arabic{offset}}\def\angle{\zeroAngle-2*\n*\spread/\eph+#1*\spread/\eph+\spread/\eph}\stepcounter{legSteps}\leg{(v1)}{\angle}{{\color{labelcolor}\arabic{legSteps}}}}}}

\def\zeroAngle{135}\def\spread{45}\setcounter{offset}{-1}\addtocounter{offset}{#2}
\ifthenelse{#2=0}{}{\ifthenelse{#2=1}{\stepcounter{legSteps}\leg{(v2)}{\zeroAngle}{{\color{labelcolor}\arabic{legSteps}}};}{
\foreach\n in {1,...,#2}{\def\eph{\arabic{offset}}\def\angle{\zeroAngle-2*\n*\spread/\eph+#2*\spread/\eph+\spread/\eph}\stepcounter{legSteps}\leg{(v2)}{\angle}{{\color{labelcolor}\arabic{legSteps}}}}}}

\def\zeroAngle{61.5}\def\spread{30}\setcounter{offset}{-1}\addtocounter{offset}{#3}
\ifthenelse{#3=0}{}{\ifthenelse{#3=1}{\stepcounter{legSteps}\leg{(v3)}{\zeroAngle}{{\color{labelcolor}\arabic{legSteps}}};}{
\foreach\n in {1,...,#3}{\def\eph{\arabic{offset}}\def\angle{\zeroAngle-2*\n*\spread/\eph+#3*\spread/\eph+\spread/\eph}\stepcounter{legSteps}\leg{(v3)}{\angle}{{\color{labelcolor}\arabic{legSteps}}}}}}

\def\zeroAngle{0}\def\spread{45}\setcounter{offset}{-1}\addtocounter{offset}{#4}
\ifthenelse{#4=0}{}{\ifthenelse{#4=1}{\stepcounter{legSteps}\leg{(v4)}{\zeroAngle}{{\color{labelcolor}\arabic{legSteps}}};}{
\foreach\n in {1,...,#4}{\def\eph{\arabic{offset}}\def\angle{\zeroAngle-2*\n*\spread/\eph+#4*\spread/\eph+\spread/\eph}\stepcounter{legSteps}\leg{(v4)}{\angle}{{\color{labelcolor}\arabic{legSteps}}}}}}

\def\zeroAngle{-61.5}\def\spread{30}\setcounter{offset}{-1}\addtocounter{offset}{#5}
\ifthenelse{#5=0}{}{\ifthenelse{#5=1}{\stepcounter{legSteps}\leg{(v5)}{\zeroAngle}{{\color{labelcolor}\arabic{legSteps}}};}{
\foreach\n in {1,...,#5}{\def\eph{\arabic{offset}}\def\angle{\zeroAngle-2*\n*\spread/\eph+#5*\spread/\eph+\spread/\eph}\stepcounter{legSteps}\leg{(v5)}{\angle}{{\color{labelcolor}\arabic{legSteps}}}}}}

\def\zeroAngle{180}\def\spread{25}\setcounter{offset}{-1}\addtocounter{offset}{#6}
\ifthenelse{#6=0}{}{\ifthenelse{#6=1}{\stepcounter{legSteps}\leg{(v6)}{\zeroAngle}{{\color{labelcolor}\arabic{legSteps}}};}{
\foreach\n in {1,...,#6}{\def\eph{\arabic{offset}}\def\angle{\zeroAngle-2*\n*\spread/\eph+#6*\spread/\eph+\spread/\eph}\stepcounter{legSteps}\leg{(v6)}{\angle}{{\color{labelcolor}\arabic{legSteps}}}}}}
}

\newcommand{\pTLegs}[6]{
\setcounter{legSteps}{0}
\def\zeroAngle{-115}\def\spread{45}\setcounter{offset}{-1}\addtocounter{offset}{#1}
\ifthenelse{#1=0}{}{\ifthenelse{#1=1}{\stepcounter{legSteps}\leg{(v1)}{\zeroAngle}{\arabic{legSteps}};}{
\foreach\n in {1,...,#1}{\def\eph{\arabic{offset}}\def\angle{\zeroAngle-2*\n*\spread/\eph+#1*\spread/\eph+\spread/\eph}\stepcounter{legSteps}\leg{(v1)}{\angle}{\arabic{legSteps}}}}}

\def\zeroAngle{180}\def\spread{45}\setcounter{offset}{-1}\addtocounter{offset}{#2}
\ifthenelse{#2=0}{}{\ifthenelse{#2=1}{\stepcounter{legSteps}\leg{(v2)}{\zeroAngle}{\arabic{legSteps}};}{
\foreach\n in {1,...,#2}{\def\eph{\arabic{offset}}\def\angle{\zeroAngle-2*\n*\spread/\eph+#2*\spread/\eph+\spread/\eph}\stepcounter{legSteps}\leg{(v2)}{\angle}{\arabic{legSteps}}}}}

\def\zeroAngle{115}\def\spread{45}\setcounter{offset}{-1}\addtocounter{offset}{#3}
\ifthenelse{#3=0}{}{\ifthenelse{#3=1}{\stepcounter{legSteps}\leg{(v3)}{\zeroAngle}{\arabic{legSteps}};}{
\foreach\n in {1,...,#3}{\def\eph{\arabic{offset}}\def\angle{\zeroAngle-2*\n*\spread/\eph+#3*\spread/\eph+\spread/\eph}\stepcounter{legSteps}\leg{(v3)}{\angle}{\arabic{legSteps}}}}}

\def\zeroAngle{61.5}\def\spread{30}\setcounter{offset}{-1}\addtocounter{offset}{#4}
\ifthenelse{#4=0}{}{\ifthenelse{#4=1}{\stepcounter{legSteps}\leg{(v4)}{\zeroAngle}{\arabic{legSteps}};}{
\foreach\n in {1,...,#4}{\def\eph{\arabic{offset}}\def\angle{\zeroAngle-2*\n*\spread/\eph+#4*\spread/\eph+\spread/\eph}\stepcounter{legSteps}\leg{(v4)}{\angle}{\arabic{legSteps}}}}}

\def\zeroAngle{0}\def\spread{45}\setcounter{offset}{-1}\addtocounter{offset}{#5}
\ifthenelse{#5=0}{}{\ifthenelse{#5=1}{\stepcounter{legSteps}\leg{(v5)}{\zeroAngle}{\arabic{legSteps}};}{
\foreach\n in {1,...,#5}{\def\eph{\arabic{offset}}\def\angle{\zeroAngle-2*\n*\spread/\eph+#5*\spread/\eph+\spread/\eph}\stepcounter{legSteps}\leg{(v5)}{\angle}{\arabic{legSteps}}}}}

\def\zeroAngle{-61.5}\def\spread{30}\setcounter{offset}{-1}\addtocounter{offset}{#6}
\ifthenelse{#6=0}{}{\ifthenelse{#6=1}{\stepcounter{legSteps}\leg{(v6)}{\zeroAngle}{\arabic{legSteps}};}{
\foreach\n in {1,...,#6}{\def\eph{\arabic{offset}}\def\angle{\zeroAngle-2*\n*\spread/\eph+#6*\spread/\eph+\spread/\eph}\stepcounter{legSteps}\leg{(v6)}{\angle}{\arabic{legSteps}}}}}
}

\newcommand{\dPentLegs}[7]{
\setcounter{legSteps}{0}
\def\zeroAngle{-90-45}\def\spread{45}\setcounter{offset}{-1}\addtocounter{offset}{#1}
\ifthenelse{#1=0}{}{\ifthenelse{#1=1}{\stepcounter{legSteps}\leg{(v1)}{\zeroAngle}{\arabic{legSteps}};}{
\foreach\n in {1,...,#1}{\def\eph{\arabic{offset}}\def\angle{\zeroAngle-2*\n*\spread/\eph+#1*\spread/\eph+\spread/\eph}\stepcounter{legSteps}\leg{(v1)}{\angle}{\arabic{legSteps}}}}}

\def\zeroAngle{90+45}\def\spread{45}\setcounter{offset}{-1}\addtocounter{offset}{#2}
\ifthenelse{#2=0}{}{\ifthenelse{#2=1}{\stepcounter{legSteps}\leg{(v2)}{\zeroAngle}{\arabic{legSteps}};}{
\foreach\n in {1,...,#2}{\def\eph{\arabic{offset}}\def\angle{\zeroAngle-2*\n*\spread/\eph+#2*\spread/\eph+\spread/\eph}\stepcounter{legSteps}\leg{(v2)}{\angle}{\arabic{legSteps}}}}}

\def\zeroAngle{90}\def\spread{25}\setcounter{offset}{-1}\addtocounter{offset}{#3}
\ifthenelse{#3=0}{}{\ifthenelse{#3=1}{\stepcounter{legSteps}\leg{(v3)}{\zeroAngle}{\arabic{legSteps}};}{
\foreach\n in {1,...,#3}{\def\eph{\arabic{offset}}\def\angle{\zeroAngle-2*\n*\spread/\eph+#3*\spread/\eph+\spread/\eph}\stepcounter{legSteps}\leg{(v3)}{\angle}{\arabic{legSteps}}}}}

\def\zeroAngle{45}\def\spread{45}\setcounter{offset}{-1}\addtocounter{offset}{#4}
\ifthenelse{#4=0}{}{\ifthenelse{#4=1}{\stepcounter{legSteps}\leg{(v4)}{\zeroAngle}{\arabic{legSteps}};}{
\foreach\n in {1,...,#4}{\def\eph{\arabic{offset}}\def\angle{\zeroAngle-2*\n*\spread/\eph+#4*\spread/\eph+\spread/\eph}\stepcounter{legSteps}\leg{(v4)}{\angle}{\arabic{legSteps}}}}}

\def\zeroAngle{-45}\def\spread{45}\setcounter{offset}{-1}\addtocounter{offset}{#5}
\ifthenelse{#5=0}{}{\ifthenelse{#5=1}{\stepcounter{legSteps}\leg{(v5)}{\zeroAngle}{\arabic{legSteps}};}{
\foreach\n in {1,...,#5}{\def\eph{\arabic{offset}}\def\angle{\zeroAngle-2*\n*\spread/\eph+#5*\spread/\eph+\spread/\eph}\stepcounter{legSteps}\leg{(v5)}{\angle}{\arabic{legSteps}}}}}

\def\zeroAngle{-90}\def\spread{25}\setcounter{offset}{-1}\addtocounter{offset}{#6}
\ifthenelse{#6=0}{}{\ifthenelse{#6=1}{\stepcounter{legSteps}\leg{(v6)}{\zeroAngle}{\arabic{legSteps}};}{
\foreach\n in {1,...,#6}{\def\eph{\arabic{offset}}\def\angle{\zeroAngle-2*\n*\spread/\eph+#6*\spread/\eph+\spread/\eph}\stepcounter{legSteps}\leg{(v6)}{\angle}{\arabic{legSteps}}}}}

\def\zeroAngle{180}\def\spread{20}\setcounter{offset}{-1}\addtocounter{offset}{#7}
\ifthenelse{#7=0}{}{\ifthenelse{#7=1}{\stepcounter{legSteps}\leg{(v7)}{\zeroAngle}{\arabic{legSteps}};}{
\foreach\n in {1,...,#7}{\def\eph{\arabic{offset}}\def\angle{\zeroAngle-2*\n*\spread/\eph+#7*\spread/\eph+\spread/\eph}\stepcounter{legSteps}\leg{(v7)}{\angle}{\arabic{legSteps}}}}}
}

\newcommand{\dBoxLegs}[6]{
\setcounter{legSteps}{0}
\def\zeroAngle{-90-45}\def\spread{45}\setcounter{offset}{-1}\addtocounter{offset}{#1}
\ifthenelse{#1=0}{}{\ifthenelse{#1=1}{\stepcounter{legSteps}\leg{(v1)}{\zeroAngle}{\arabic{legSteps}};}{
\foreach\n in {1,...,#1}{\def\eph{\arabic{offset}}\def\angle{\zeroAngle-2*\n*\spread/\eph+#1*\spread/\eph+\spread/\eph}\stepcounter{legSteps}\leg{(v1)}{\angle}{\arabic{legSteps}}}}}

\def\zeroAngle{90+45}\def\spread{45}\setcounter{offset}{-1}\addtocounter{offset}{#2}
\ifthenelse{#2=0}{}{\ifthenelse{#2=1}{\stepcounter{legSteps}\leg{(v2)}{\zeroAngle}{\arabic{legSteps}};}{
\foreach\n in {1,...,#2}{\def\eph{\arabic{offset}}\def\angle{\zeroAngle-2*\n*\spread/\eph+#2*\spread/\eph+\spread/\eph}\stepcounter{legSteps}\leg{(v2)}{\angle}{\arabic{legSteps}}}}}

\def\zeroAngle{90}\def\spread{25}\setcounter{offset}{-1}\addtocounter{offset}{#3}
\ifthenelse{#3=0}{}{\ifthenelse{#3=1}{\stepcounter{legSteps}\leg{(v3)}{\zeroAngle}{\arabic{legSteps}};}{
\foreach\n in {1,...,#3}{\def\eph{\arabic{offset}}\def\angle{\zeroAngle-2*\n*\spread/\eph+#3*\spread/\eph+\spread/\eph}\stepcounter{legSteps}\leg{(v3)}{\angle}{\arabic{legSteps}}}}}

\def\zeroAngle{45}\def\spread{45}\setcounter{offset}{-1}\addtocounter{offset}{#4}
\ifthenelse{#4=0}{}{\ifthenelse{#4=1}{\stepcounter{legSteps}\leg{(v4)}{\zeroAngle}{\arabic{legSteps}};}{
\foreach\n in {1,...,#4}{\def\eph{\arabic{offset}}\def\angle{\zeroAngle-2*\n*\spread/\eph+#4*\spread/\eph+\spread/\eph}\stepcounter{legSteps}\leg{(v4)}{\angle}{\arabic{legSteps}}}}}

\def\zeroAngle{-45}\def\spread{45}\setcounter{offset}{-1}\addtocounter{offset}{#5}
\ifthenelse{#5=0}{}{\ifthenelse{#5=1}{\stepcounter{legSteps}\leg{(v5)}{\zeroAngle}{\arabic{legSteps}};}{
\foreach\n in {1,...,#5}{\def\eph{\arabic{offset}}\def\angle{\zeroAngle-2*\n*\spread/\eph+#5*\spread/\eph+\spread/\eph}\stepcounter{legSteps}\leg{(v5)}{\angle}{\arabic{legSteps}}}}}

\def\zeroAngle{-90}\def\spread{25}\setcounter{offset}{-1}\addtocounter{offset}{#6}
\ifthenelse{#6=0}{}{\ifthenelse{#6=1}{\stepcounter{legSteps}\leg{(v6)}{\zeroAngle}{\arabic{legSteps}};}{
\foreach\n in {1,...,#6}{\def\eph{\arabic{offset}}\def\angle{\zeroAngle-2*\n*\spread/\eph+#6*\spread/\eph+\spread/\eph}\stepcounter{legSteps}\leg{(v6)}{\angle}{\arabic{legSteps}}}}}
}

\newcommand{\bTLegs}[5]{
\setcounter{legSteps}{0}
\def\zeroAngle{-90-45}\def\spread{45}\setcounter{offset}{-1}\addtocounter{offset}{#1}
\ifthenelse{#1=0}{}{\ifthenelse{#1=1}{\stepcounter{legSteps}\leg{(v1)}{\zeroAngle}{{\color{labelcolor}\arabic{legSteps}}};}{
\foreach\n in {1,...,#1}{\def\eph{\arabic{offset}}\def\angle{\zeroAngle-2*\n*\spread/\eph+#1*\spread/\eph+\spread/\eph}\stepcounter{legSteps}\leg{(v1)}{\angle}{{\color{labelcolor}\arabic{legSteps}}}}}}

\def\zeroAngle{90+45}\def\spread{45}\setcounter{offset}{-1}\addtocounter{offset}{#2}
\ifthenelse{#2=0}{}{\ifthenelse{#2=1}{\stepcounter{legSteps}\leg{(v2)}{\zeroAngle}{{\color{labelcolor}\arabic{legSteps}}};}{
\foreach\n in {1,...,#2}{\def\eph{\arabic{offset}}\def\angle{\zeroAngle-2*\n*\spread/\eph+#2*\spread/\eph+\spread/\eph}\stepcounter{legSteps}\leg{(v2)}{\angle}{{\color{labelcolor}\arabic{legSteps}}}}}}

\def\zeroAngle{70}\def\spread{25}\setcounter{offset}{-1}\addtocounter{offset}{#3}
\ifthenelse{#3=0}{}{\ifthenelse{#3=1}{\stepcounter{legSteps}\leg{(v3)}{\zeroAngle}{{\color{labelcolor}\arabic{legSteps}}};}{
\foreach\n in {1,...,#3}{\def\eph{\arabic{offset}}\def\angle{\zeroAngle-2*\n*\spread/\eph+#3*\spread/\eph+\spread/\eph}\stepcounter{legSteps}\leg{(v3)}{\angle}{{\color{labelcolor}\arabic{legSteps}}}}}}

\def\zeroAngle{0}\def\spread{45}\setcounter{offset}{-1}\addtocounter{offset}{#4}
\ifthenelse{#4=0}{}{\ifthenelse{#4=1}{\stepcounter{legSteps}\leg{(v4)}{\zeroAngle}{{\color{labelcolor}\arabic{legSteps}}};}{
\foreach\n in {1,...,#4}{\def\eph{\arabic{offset}}\def\angle{\zeroAngle-2*\n*\spread/\eph+#4*\spread/\eph+\spread/\eph}\stepcounter{legSteps}\leg{(v4)}{\angle}{{\color{labelcolor}\arabic{legSteps}}}}}}

\def\zeroAngle{-70}\def\spread{25}\setcounter{offset}{-1}\addtocounter{offset}{#5}
\ifthenelse{#5=0}{}{\ifthenelse{#5=1}{\stepcounter{legSteps}\leg{(v5)}{\zeroAngle}{{\color{labelcolor}\arabic{legSteps}}};}{
\foreach\n in {1,...,#5}{\def\eph{\arabic{offset}}\def\angle{\zeroAngle-2*\n*\spread/\eph+#5*\spread/\eph+\spread/\eph}\stepcounter{legSteps}\leg{(v5)}{\angle}{{\color{labelcolor}\arabic{legSteps}}}}}}
}

\newcommand{\dTLegs}[4]{
\setcounter{legSteps}{0}
\def\zeroAngle{180}\def\spread{45}\setcounter{offset}{-1}\addtocounter{offset}{#1}
\ifthenelse{#1=0}{}{\ifthenelse{#1=1}{\stepcounter{legSteps}\leg{(v1)}{\zeroAngle}{{\color{labelcolor}\arabic{legSteps}}};}{
\foreach\n in {1,...,#1}{\def\eph{\arabic{offset}}\def\angle{\zeroAngle-2*\n*\spread/\eph+#1*\spread/\eph+\spread/\eph}\stepcounter{legSteps}\leg{(v1)}{\angle}{{\color{labelcolor}\arabic{legSteps}}}}}}

\def\zeroAngle{90}\def\spread{55}\setcounter{offset}{-1}\addtocounter{offset}{#2}
\ifthenelse{#2=0}{}{\ifthenelse{#2=1}{\stepcounter{legSteps}\leg{(v2)}{\zeroAngle}{{\color{labelcolor}\arabic{legSteps}}};}{
\foreach\n in {1,...,#2}{\def\eph{\arabic{offset}}\def\angle{\zeroAngle-2*\n*\spread/\eph+#2*\spread/\eph+\spread/\eph}\stepcounter{legSteps}\leg{(v2)}{\angle}{{\color{labelcolor}\arabic{legSteps}}}}}}

\def\zeroAngle{0}\def\spread{60}\setcounter{offset}{-1}\addtocounter{offset}{#3}
\ifthenelse{#3=0}{}{\ifthenelse{#3=1}{\stepcounter{legSteps}\leg{(v3)}{\zeroAngle}{{\color{labelcolor}\arabic{legSteps}}};}{
\foreach\n in {1,...,#3}{\def\eph{\arabic{offset}}\def\angle{\zeroAngle-2*\n*\spread/\eph+#3*\spread/\eph+\spread/\eph}\stepcounter{legSteps}\leg{(v3)}{\angle}{{\color{labelcolor}\arabic{legSteps}}}}}}

\def\zeroAngle{-90}\def\spread{55}\setcounter{offset}{-1}\addtocounter{offset}{#4}
\ifthenelse{#4=0}{}{\ifthenelse{#4=1}{\stepcounter{legSteps}\leg{(v4)}{\zeroAngle}{{\color{labelcolor}\arabic{legSteps}}};}{
\foreach\n in {1,...,#4}{\def\eph{\arabic{offset}}\def\angle{\zeroAngle-2*\n*\spread/\eph+#4*\spread/\eph+\spread/\eph}\stepcounter{legSteps}\leg{(v4)}{\angle}{{\color{labelcolor}\arabic{legSteps}}}}}}
}

\newcommand{\tardiLegs}[5]{
\setcounter{legSteps}{0}
\def\zeroAngle{180}\def\spread{45}\setcounter{offset}{-1}\addtocounter{offset}{#1}
\ifthenelse{#1=0}{}{\ifthenelse{#1=1}{\stepcounter{legSteps}\leg{(v1)}{\zeroAngle}{{\color{labelcolor}\arabic{legSteps}}};}{
\foreach\n in {1,...,#1}{\def\eph{\arabic{offset}}\def\angle{\zeroAngle-2*\n*\spread/\eph+#1*\spread/\eph+\spread/\eph}\stepcounter{legSteps}\leg{(v1)}{\angle}{{\color{labelcolor}\arabic{legSteps}}}}}}

\def\zeroAngle{90}\def\spread{55}\setcounter{offset}{-1}\addtocounter{offset}{#2}
\ifthenelse{#2=0}{}{\ifthenelse{#2=1}{\stepcounter{legSteps}\leg{(v2)}{\zeroAngle}{{\color{labelcolor}\arabic{legSteps}}};}{
\foreach\n in {1,...,#2}{\def\eph{\arabic{offset}}\def\angle{\zeroAngle-2*\n*\spread/\eph+#2*\spread/\eph+\spread/\eph}\stepcounter{legSteps}\leg{(v2)}{\angle}{{\color{labelcolor}\arabic{legSteps}}}}}}

\def\zeroAngle{0}\def\spread{45}\setcounter{offset}{-1}\addtocounter{offset}{#3}
\ifthenelse{#3=0}{}{\ifthenelse{#3=1}{\stepcounter{legSteps}\leg{(v3)}{\zeroAngle}{{\color{labelcolor}\arabic{legSteps}}};}{
\foreach\n in {1,...,#3}{\def\eph{\arabic{offset}}\def\angle{\zeroAngle-2*\n*\spread/\eph+#3*\spread/\eph+\spread/\eph}\stepcounter{legSteps}\leg{(v3)}{\angle}{{\color{labelcolor}\arabic{legSteps}}}}}}

\def\zeroAngle{-90}\def\spread{55}\setcounter{offset}{-1}\addtocounter{offset}{#4}
\ifthenelse{#4=0}{}{\ifthenelse{#4=1}{\stepcounter{legSteps}\leg{(v4)}{\zeroAngle}{{\color{labelcolor}\arabic{legSteps}}};}{
\foreach\n in {1,...,#4}{\def\eph{\arabic{offset}}\def\angle{\zeroAngle-2*\n*\spread/\eph+#4*\spread/\eph+\spread/\eph}\stepcounter{legSteps}\leg{(v4)}{\angle}{{\color{labelcolor}\arabic{legSteps}}}}}}

\def\zeroAngle{180}\def\spread{35}\setcounter{offset}{-1}\addtocounter{offset}{#5}
\ifthenelse{#5=0}{}{\ifthenelse{#5=1}{\stepcounter{legSteps}\leg{(v5)}{\zeroAngle}{{\color{labelcolor}\arabic{legSteps}}};}{\ifthenelse{#5=2}{\stepcounter{legSteps}\leg{(v5)}{180}{{\color{labelcolor}\arabic{legSteps}}};\stepcounter{legSteps}\leg{(v5)}{0}{{\color{labelcolor}\arabic{legSteps}}};}{
\foreach\n in {1,...,#5}{\def\eph{\arabic{offset}}\def\angle{\zeroAngle-2*\n*\spread/\eph+#5*\spread/\eph+\spread/\eph}\stepcounter{legSteps}\leg{(v5)}{\angle}{{\color{labelcolor}\arabic{legSteps}}}}}}}

}

\renewcommand{\triangle}[1]{\begin{tikzpicture}[scale=\figScale,baseline=-3.05]\useasboundingbox ($(-\pageW/12,-1.)$) rectangle ($(\pageW/12,1.)$);\draw[int,line width=0.1,red,draw=\boundingDraw] ($(-\pageW/12,-1.)$) rectangle ($(\pageW/12,1.)$);\coordinate(v1)at($(210:\figScale*0.5)$);\coordinate(v2)at($(90:\figScale*0.5)$);\coordinate(v3)at($(-30:\figScale*0.5)$);
#1\end{tikzpicture}}
\newcommand{\boxOL}[1]{\begin{tikzpicture}[scale=\figScale,baseline=-3.05]\useasboundingbox ($(-\pageW/12,-1.)$) rectangle ($(\pageW/12,1.)$);\draw[int,line width=0.1,red,draw=\boundingDraw] ($(-\pageW/12,-1.)$) rectangle ($(\pageW/12,1.)$);\coordinate(v1)at($(225:\figScale*0.6)$);\coordinate(v2)at($(135:\figScale*0.6)$);\coordinate(v3)at($(45:\figScale*0.6)$);\coordinate(v4)at($(-45:\figScale*0.6)$);
#1\end{tikzpicture}}

\newcommand{\contourVerts}[7]{
\ifthenelse{#1=1}{\node at (v1) [whiteDot] {};}{\ifthenelse{#1=2}{\node at (v1) [blueDot] {};}{\ifthenelse{#1=4}{\node at (v1) [blackDot] {};}{\ifthenelse{#1=3}{\node at (v1) [bdot] {};\node at (v1) [compositeDot] {};
}{\ifthenelse{#1=0}{\node at (v1) [bdot] {};}{}}}}}

\ifthenelse{#2=1}{\node at (v2) [whiteDot] {};}{\ifthenelse{#2=2}{\node at (v2) [blueDot] {};}{\ifthenelse{#2=4}{\node at (v2) [blackDot] {};}{\ifthenelse{#2=3}{\node at (v2) [bdot] {};\node at (v2) [compositeDot] {};
}{\ifthenelse{#2=0}{\node at (v2) [bdot] {};}{}}}}}

\ifthenelse{#3=1}{\node at (v3) [whiteDot] {};}{\ifthenelse{#3=2}{\node at (v3) [blueDot] {};}{\ifthenelse{#3=4}{\node at (v3) [blackDot] {};}{\ifthenelse{#3=3}{\node at (v3) [bdot] {};\node at (v3) [compositeDot] {};
}{\ifthenelse{#3=0}{\node at (v3) [bdot] {};}{}}}}}

\ifthenelse{#4=1}{\node at (v4) [whiteDot] {};}{\ifthenelse{#4=2}{\node at (v4) [blueDot] {};}{\ifthenelse{#4=4}{\node at (v4) [blackDot] {};}{\ifthenelse{#4=3}{\node at (v4) [bdot] {};\node at (v4) [compositeDot] {};
}{\ifthenelse{#4=0}{\node at (v4) [bdot] {};}{}}}}}

\ifthenelse{#5=1}{\node at (v5) [whiteDot] {};}{\ifthenelse{#5=2}{\node at (v5) [blueDot] {};}{\ifthenelse{#5=4}{\node at (v5) [blackDot] {};}{\ifthenelse{#5=3}{\node at (v5) [bdot] {};\node at (v5) [compositeDot] {};
}{\ifthenelse{#5=0}{\node at (v5) [bdot] {};}{}}}}}
\ifthenelse{#6=1}{\node at (v6) [whiteDot] {};}{\ifthenelse{#6=2}{\node at (v6) [blueDot] {};}{\ifthenelse{#6=4}{\node at (v6) [blackDot] {};}{\ifthenelse{#6=3}{\node at (v6) [bdot] {};\node at (v6) [compositeDot] {};
}{\ifthenelse{#6=0}{\node at (v6) [bdot] {};}{}}}}}
\ifthenelse{#7=1}{\node at (v7) [whiteDot] {};}{\ifthenelse{#7=2}{\node at (v7) [blueDot] {};}{\ifthenelse{#7=4}{\node at (v7) [blackDot] {};}{\ifthenelse{#7=3}{\node at (v7) [bdot] {};\node at (v7) [compositeDot] {};
}{\ifthenelse{#7=0}{\node at (v7) [bdot] {};}{}}}}}
}

\newcommand{\dBoxInfA}{\begin{scope}\path[clip](v1)--(v2)--(v3)--(v6)--(v1);
\node at (v1) [fill=nmhvred,circle,minimum size=0.95*\ampSize,inner sep=0] {};
\node at (v2) [fill=nmhvred,circle,minimum size=0.95*\ampSize,inner sep=0] {};
\node at (v3) [fill=nmhvred,circle,minimum size=0.95*\ampSize,inner sep=0] {};
\node at (v6) [fill=nmhvred,circle,minimum size=0.95*\ampSize,inner sep=0] {};
\draw[nmhvred,line width=5*\lineThickness,line cap=round,rounded corners=1.5pt](v6)--(v1)--(v2)--(v3)--(v6);
\draw[white,line width=3*\lineThickness,line cap=round,rounded corners=1.5pt](v6)--(v1)--(v2)--(v3)--(v6);
\node at (v1) [fill=white,circle,minimum size=0.75*\ampSize,inner sep=0] {};
\node at (v2) [fill=white,circle,minimum size=0.75*\ampSize,inner sep=0] {};
\node at (v3) [fill=white,circle,minimum size=0.75*\ampSize,inner sep=0] {};
\node at (v6) [fill=white,circle,minimum size=0.75*\ampSize,inner sep=0] {};
\node at ($(v1)!0.55!(v6)+(90:0.2)$) [] {\tiny{${\color{nmhvred}\!\!\!\infty}$}};\end{scope}}

\newcommand{\bTInfA}{\begin{scope}\path[clip](v1)--(v2)--(v3)--(v5)--(v1);
\node at (v1) [fill=nmhvred,circle,minimum size=0.95*\ampSize,inner sep=0] {};
\node at (v2) [fill=nmhvred,circle,minimum size=0.95*\ampSize,inner sep=0] {};
\node at (v3) [fill=nmhvred,circle,minimum size=0.95*\ampSize,inner sep=0] {};
\node at (v5) [fill=nmhvred,circle,minimum size=0.95*\ampSize,inner sep=0] {};
\draw[nmhvred,line width=5*\lineThickness,line cap=round,rounded corners=1.5pt](v5)--(v1)--(v2)--(v3)--(v5);
\draw[white,line width=3*\lineThickness,line cap=round,rounded corners=1.5pt](v5)--(v1)--(v2)--(v3)--(v5);
\node at (v1) [fill=white,circle,minimum size=0.75*\ampSize,inner sep=0] {};
\node at (v2) [fill=white,circle,minimum size=0.75*\ampSize,inner sep=0] {};
\node at (v3) [fill=white,circle,minimum size=0.75*\ampSize,inner sep=0] {};
\node at (v5) [fill=white,circle,minimum size=0.75*\ampSize,inner sep=0] {};
\node at ($(v1)!0.55!(v5)+(90:0.2)$) [] {\tiny{${\color{nmhvred}\!\!\!\infty}$}};\end{scope}}

\newcommand{\dTInfA}{\begin{scope}\path[clip](v1)--(v2)--(v4)--(v1);
\node at (v1) [fill=nmhvred,circle,minimum size=0.95*\ampSize,inner sep=0] {};
\node at (v2) [fill=nmhvred,circle,minimum size=0.95*\ampSize,inner sep=0] {};
\node at (v4) [fill=nmhvred,circle,minimum size=0.95*\ampSize,inner sep=0] {};
\draw[nmhvred,line width=5*\lineThickness,line cap=round,rounded corners=1.5pt](v4)--(v1)--(v2)--(v4);
\draw[white,line width=3*\lineThickness,line cap=round,rounded corners=1.5pt](v4)--(v1)--(v2)--(v4);
\node at (v1) [fill=white,circle,minimum size=0.75*\ampSize,inner sep=0] {};
\node at (v2) [fill=white,circle,minimum size=0.75*\ampSize,inner sep=0] {};
\node at (v4) [fill=white,circle,minimum size=0.75*\ampSize,inner sep=0] {};
\node at ($(v4)!0.5!(v2)+(180:0.2)$) [] {\tiny{${\color{nmhvred}\infty\,}$}};\end{scope}}

\newcommand{\dBoxInfB}{\begin{scope}\path[clip] (v3)--(v4)--(v5)--(v6)--(v3);\node at (v3) [fill=blue,circle,minimum size=0.95*\ampSize,inner sep=0] {};
\node at (v4) [fill=blue,circle,minimum size=0.95*\ampSize,inner sep=0] {};
\node at (v5) [fill=blue,circle,minimum size=0.95*\ampSize,inner sep=0] {};
\node at (v6) [fill=blue,circle,minimum size=0.95*\ampSize,inner sep=0] {};
\draw[blue,line width=5*\lineThickness,line cap=round,rounded corners=1.5pt](v3)--(v4)--(v5)--(v6)--(v3);
\draw[white,line width=3*\lineThickness,line cap=round,rounded corners=1.5pt](v3)--(v4)--(v5)--(v6)--(v3);
\node at (v3) [fill=white,circle,minimum size=0.75*\ampSize,inner sep=0] {};
\node at (v4) [fill=white,circle,minimum size=0.75*\ampSize,inner sep=0] {};
\node at (v5) [fill=white,circle,minimum size=0.75*\ampSize,inner sep=0] {};
\node at (v6) [fill=white,circle,minimum size=0.75*\ampSize,inner sep=0] {};
\node at ($(v5)!0.45!(v6)+(90:0.2)$) [] {\tiny{${\color{blue}\!\!\!\infty}$}};\end{scope}}

\newcommand{\bTInfB}{\begin{scope}\path[clip](v3)--(v4)--(v5)--(v3);
\node at (v3) [fill=blue,circle,minimum size=0.95*\ampSize,inner sep=0] {};
\node at (v4) [fill=blue,circle,minimum size=0.95*\ampSize,inner sep=0] {};
\node at (v5) [fill=blue,circle,minimum size=0.95*\ampSize,inner sep=0] {};
\draw[blue,line width=5*\lineThickness,line cap=round,rounded corners=1.5pt](v3)--(v4)--(v5)--(v3);
\draw[white,line width=3*\lineThickness,line cap=round,rounded corners=1.5pt](v3)--(v4)--(v5)--(v3);
\node at (v3) [fill=white,circle,minimum size=0.75*\ampSize,inner sep=0] {};
\node at (v4) [fill=white,circle,minimum size=0.75*\ampSize,inner sep=0] {};
\node at (v5) [fill=white,circle,minimum size=0.75*\ampSize,inner sep=0] {};
\node at ($(v5)!0.5!(v3)+(0:0.2)$) [] {\tiny{${\color{blue}\,\,\,\infty}$}};\end{scope}}

\newcommand{\dTInfB}{\begin{scope}\path[clip](v2)--(v3)--(v4)--(v2);
\node at (v2) [fill=blue,circle,minimum size=0.95*\ampSize,inner sep=0] {};
\node at (v3) [fill=blue,circle,minimum size=0.95*\ampSize,inner sep=0] {};
\node at (v4) [fill=blue,circle,minimum size=0.95*\ampSize,inner sep=0] {};
\draw[blue,line width=5*\lineThickness,line cap=round,rounded corners=1.5pt](v2)--(v3)--(v4)--(v2);
\draw[white,line width=3*\lineThickness,line cap=round,rounded corners=1.5pt](v2)--(v3)--(v4)--(v2);
\node at (v2) [fill=white,circle,minimum size=0.75*\ampSize,inner sep=0] {};
\node at (v3) [fill=white,circle,minimum size=0.75*\ampSize,inner sep=0] {};
\node at (v4) [fill=white,circle,minimum size=0.75*\ampSize,inner sep=0] {};
\node at ($(v4)!0.5!(v2)+(0:0.2)$) [] {\tiny{${\color{blue}\,\infty}$}};\end{scope}}

\newcommand{\bTInfAB}{\bTInfA\bTInfB}

\newcommand{\dTInfAB}{\dTInfA\dTInfB}

\let\olditemize\itemize\renewcommand{\itemize}{\vspace{-2pt}\olditemize\setlength{\itemsep}{1pt}\setlength{\parskip}{0pt}\setlength{\parsep}{-0pt}}
\let\oldenumerate\enumerate\renewcommand{\enumerate}{\vspace{-4pt}\oldenumerate\setlength{\itemsep}{1pt}\setlength{\parskip}{0pt}\setlength{\parsep}{0pt}}
\makeatletter\renewcommand\section{\addtocontents{toc}{\protect\addvspace{-2.25\p@}}\@startsection {section}{1}{\z@}{-0.0ex \@plus .2ex \@minus 0.2ex}{1ex \@plus.1ex\@minus .5ex}{\normalfont\large\bfseries}}
\renewcommand\subsection{\addtocontents{toc}{\protect\addvspace{-2.5\p@}}\@startsection {subsection}{1}{\z@}{0.5ex \@plus .2ex \@minus 0.2ex}{0.75ex \@plus.1ex\@minus .5ex}{\normalfont\bfseries}}
\newcommand{\eq}[1]{\vspace{-0.5pt}\begin{equation}#1\vspace{-0.5pt}\end{equation}}
\newcommand{\eqs}[1]{\vspace{-0.5pt}\begin{equation}\begin{split}#1\end{split}\vspace{-0.5pt}\end{equation}}
\newcommand{\fwbox}[2]{\text{\makebox[#1][c]{$\hspace{-150pt}\displaystyle#2\hspace{-150pt}$}}}
\newcommand{\fwboxL}[2]{\text{\makebox[#1][l]{$#2$}}}
\newcommand{\fwboxR}[2]{\text{\makebox[#1][r]{$#2$}}}
\newcommand{\equivR}{\fwbox{14.5pt}{\hspace{-0pt}\fwboxR{0pt}{\raisebox{0.47pt}{\hspace{1.25pt}:\hspace{-4pt}}}=\fwboxL{0pt}{}}}
\newcommand{\equivL}{\fwbox{14.5pt}{\fwboxR{0pt}{}=\fwboxL{0pt}{\raisebox{0.47pt}{\hspace{-4pt}:\hspace{1.25pt}}}}}
\newcommand{\fig}[3]{\raisebox{#1}{\includegraphics[scale=#2]{#3}}}

\renewcommand{\phi}{\varphi}

\renewcommand{\hat}{\widehat}
\newcommand{\br}[1]{\left[\!\!\,\left[#1\right]\!\!\,\right]}
\renewcommand{\tilde}{\widetilde}

\newcommand{\x}[2]{{\color{black}(}\hspace{-0.85pt}{\color{black}#1}\hspace{-0.25pt}{\color{black}|}\hspace{-0.25pt}{\color{black}#2}\hspace{-0.85pt}{\color{black})}}
\newcommand{\dbar}{\fwboxL{7.2pt}{\raisebox{4.5pt}{\fwboxL{0pt}{\scalebox{1.5}[0.75]{\hspace{1.25pt}\text{-}}}}d}}

\definecolor{varcolor}{rgb}{0.08,0.44,0.2}
\definecolor{functioncolor}{rgb}{0.08,0.28,0.6}

\newcommand{\var}[1]{{\tt{\color{varcolor}{\sl#1}}}}
\newcommand{\fun}[1]{{\color{functioncolor}#1}}
\newcommand{\uscore}{\rule[-1.05pt]{7.5pt}{.75pt}}

\definecolor{rindou1}{rgb}{0.4431,0.2862,0.7960}
\definecolor{rindou2}{rgb}{0.0078,0.1215,0.4392}
\definecolor{lapis}{rgb}{0.0.0470,0.2941,0.5568}
\definecolor{emerald}{rgb}{0.31, 0.78, 0.47}
\definecolor{pinegreen}{rgb}{0.0, 0.47, 0.44}
\definecolor{jade}{rgb}{0.0, 0.66, 0.42}
\definecolor{teal}{rgb}{0.0, 0.5, 0.5}
\definecolor{totalCount}{rgb}{0,0,0.575}
\definecolor{topCount}{rgb}{0.575,0.0,0.225}
\definecolor{dim}{rgb}{0.55,0.55,0.55}
\definecolor{deemph}{rgb}{0.25,0.25,0.25}
\definecolor{hblue}{rgb}{0,0,0.575}
\definecolor{hred}{rgb}{0.575,0.0,0.225}
\definecolor{hgreen}{rgb}{0.0,0.4,0.2}
\definecolor{hteal}{rgb}{0.0,0.545,0.7451}
\renewcommand{\r}[1]{{\color{hred}#1}}
\renewcommand{\b}[1]{{\color{hblue}#1}}
\newcommand{\g}[1]{{\color{hgreen}#1}}

\title{\texorpdfstring{{\huge \mbox{Illustrations of Integrand-Basis\hspace{-10pt}}}\\[-6pt]{\huge\mbox{Building at Two Loops}}\\[-0pt]}{Illustrations of Integrand-Basis Building at Two Loops}}
\author[a,b]{\vspace{-24pt}Jacob~L.~Bourjaily,}\emailAdd{bourjaily@psu.edu}
\author[a]{Cameron~Langer,}\emailAdd{ckl5552@psu.edu}
\author[a]{Yaqi~Zhang}\emailAdd{yjz5289@psu.edu}
\affiliation[a]{Institute for Gravitation and the Cosmos, Department of Physics,\\Pennsylvania State University, University Park, PA 16802, USA}
\affiliation[b]{Niels Bohr International Academy and Discovery Center, Niels Bohr Institute,\\University of Copenhagen, Blegdamsvej 17, DK-2100, Copenhagen \O, Denmark}
\abstract{%
We outline the concrete steps involved in building prescriptive master integrand bases for scattering amplitudes beyond the planar limit. We highlight the role of contour choices in such bases, and illustrate the full process by constructing a complete, triangle power-counting basis at two loops for six particles. We show how collinear contour choices can be used to divide integrand bases into separately finite and divergent subspaces, and how double-poles can be used to further subdivide these spaces according to (transcendental) weight.\\~\\ Complete details of the basis constructed for six particles is provided in the ancillary files for this work's submission to the \texttt{arXiv}. 
}
\preprint{}

\begin{document}
\maketitle
\pagenumbering{roman}

\pagenumbering{roman}

\pagenumbering{arabic}
\vspace{0pt}%
\section{Introduction and Overview}\label{sec:introduction}\vspace{0pt}
Recent years have been witness to incredible advances in our understanding of perturbative quantum field theory; much of this has resulted from concrete applications of generalized unitarity: through constructions of specific target amplitudes, matching field theory cuts at the integrand-level. In particular, such investigations \cite{Bern:1994zx,Bern:1994cg,Britto:2004nc} led to the discovery of tree-level \cite{Britto:2004ap,Britto:2005fq} (and loop-level \cite{ArkaniHamed:2010kv}) on-shell recursion relations; the discovery of dual-conformal (and ultimately, Yangian-)invariance \mbox{\cite{Drummond:2006rz,Alday:2007hr,Drummond:2008vq,Drummond:2009fd}} of planar, maximally supersymmetric Yang-Mills theory; the connection between on-shell diagrams and subspaces of Grassmannian manifolds \cite{ArkaniHamed:2009sx,ArkaniHamed:2009dg,Kaplan:2009mh,Bourjaily:2010kw,ArkaniHamed:2012nw,Bourjaily:2012gy,Arkani-Hamed:2014bca}; and the amplituhedron \cite{ArkaniHamed:2010gg,Arkani-Hamed:2013jha,Arkani-Hamed:2013kca}---among much else. 

Perhaps not surprisingly, the vast majority of this progress has been made in the context of planar theories. For one thing, loop integrands in planar theories can be assigned (symmetrized) dual-momentum coordinates for loop momenta in which all Feynman diagrams depend universally. This allows for \emph{the} loop integrand to correspond to a specific, unambiguous rational function of internal and external (dual) momenta. Moreover, dual coordinates allow a preferential stratification---or organization---of loop integrand bases according to their UV behavior in these coordinates. 

Beyond the planar limit, the non-existence of a \emph{particular} rational function has dramatically hindered progress. This is somewhat surprising, because very little about generalized unitarity depends on the coordinates used to describe loop momenta or how momentum-conservation is solved to express an $L$-loop integrand in terms of some particular choice of loop momenta to be integrated (how the loop momenta are `routed' in the language of ref.~\cite{Bourjaily:2020qca}). Nevertheless, applications of generalized unitarity beyond the planar limit have been surprisingly sparse. Due to the importance of the possible UV divergence of maximally supersymmetric ($\mathcal{N}\!=\!8$) supergravity and the connections between this theory and ($\mathcal{N}\!=\!4$) supersymmetric Yang-Mills theory (`sYM'), the most impressive results that exist are for four-point amplitudes---which are known through five loops \cite{Bern:1997nh,Arkani-Hamed:2014via,Bern:2006kd,Bern:2007hh,Bern:2008pv,Bern:2009kd,Bern:2010tq,Bern:2017ucb}. Outside of this case, only relatively isolated examples of amplitudes are known. These include the five and six-point MHV amplitudes in sYM \cite{Carrasco:2011mn,Bourjaily:2019iqr} and a more recent, general formula for all-multiplicity MHV amplitudes at two loops  \cite{Bourjaily:2019gqu}. Compare this with planar sYM, for which local integrand representations of all-multiplicity, all N$^{k}$MHV amplitudes are known through three loops \cite{Bourjaily:2013mma,Bourjaily:2015jna,Bourjaily:2017wjl}, and isolated examples of integrands at lower-multiplicity are known as high as ten loops \cite{Bourjaily:2011hi,Bourjaily:2015bpz,Bourjaily:2016evz}. 

Interestingly, and in contrast with the work on planar amplitudes, none of the previous applications of generalized unitarity to non-planar amplitudes built upon a \emph{complete} basis of integrands in which to express the result. Rather, all such results were obtained by starting from a reasonably good guess for a \emph{sufficient} subspace of integrands and amending this guess only as far as needed to match the amplitudes in question.\\

In this work, we would like to take some of the guess-work and cleverness out of the construction of non-planar amplitudes by illustrating the construction of a \emph{complete} and \emph{prescriptive} basis for all (not necessarily planar) integrands with triangle power-counting, (denoted `$\mathfrak{B}_3^{(4)}$' in \cite{Bourjaily:2020qca}) involving six particles at two loops, following the general strategy described in ref.~\cite{Bourjaily:2020qca}. To be clear, although ref.~\cite{Bourjaily:2020qca} described how to \emph{define} and \emph{enumerate} the space of such integrands beyond the planar limit, it did not discuss how \emph{particular} choices of basis elements should be chosen. 

A \emph{prescriptive} basis of integrands $\{\mathcal{I}_i\}$---viewed as a basis for cohomology on the space of differential forms on loop momenta---is one which is the cohomological dual of some choice of integration contours $\{\Omega_j\}$. That is, a basis of loop integrands $\{\mathcal{I}_i\}$ is called \emph{prescriptive} if there exists a set of cycles $\{\Omega_j\}$ such that 
\eq{\oint\limits_{\Omega_j}\!\mathcal{I}_i=\delta_{i,j}\,.\label{prescriptivity_condition}}
When this is the case, amplitudes may be expanded in this basis simply as 
\eq{\mathcal{A}=\sum_i\mathfrak{a}_i\mathcal{I}_i\,}
with coefficients $\mathfrak{a}_i$ being directly (possibly generalized \cite{Bourjaily:2020qca,Bourjaily:2021vyj}) `leading singularities'
\eq{\mathfrak{a}_i\equivR\oint\limits_{\Omega_i}\!\!\mathcal{A}}
---on-shell functions computed in terms of tree-amplitudes. 

Obviously, the particular integrands appearing in a prescriptive basis will depend strongly on the integration contours $\{\Omega_j\}$ to which they are dual. One motivation for our present work is to illustrate the scope of possible choices for these cycles, and how these choices affect the resulting integrand basis in a concrete case of relevance to amplitudes in sYM. For example, by choosing as many contours as possible to encompass all regions responsible for IR-divergence \cite{Bourjaily:2021ewt}, diagonalization of the basis implicit in (\ref{prescriptivity_condition}) should render all integrands \emph{not} manifestly dual to such contours IR-finite. Such a splitting of integrands in the basis should simplify the computation of finite quantities (such as the ratio function)---possibly even achieving local finiteness as described in \cite{Bourjaily:2021ewt}. It remains to be seen how these ideas generalize to beyond the planar limit, and it may be viewed as a conjecture---to be tested by direct integration---that the integrand basis we construct here has this property. 

Regardless of the integration contours chosen for the basis, it is empirically the case that prescriptive integrand bases turn out to be `simple' with respect to the difficult challenge of loop integration. Specifically, we mean that they are almost certainly `pure' (at least when maximal in weight) \cite{ArkaniHamed:2010gh,Drummond:2010cz}; that is, viewed as master integrals, they are found to satisfy nilpotent systems of differential equations \cite{Henn:2013pwa,Broedel:2018qkq}. This property may be complicated by the presence of integrals with double-poles (or equivalently, integrals that should be viewed as maximal in weight for a lower spacetime dimension of integration). The basis we construct certainly includes such integrands; and so it may be viewed as a conjecture---again, to be tested by direct integration---as to whether the full-weight subspace of integrands we construct are indeed `pure' after loop integration.

The existence of integrand basis elements with less than maximal transcendental weight---features widely believed to be absent from amplitudes in sYM---illustrates another motivation for this work: to explore and clarify the ways in which the basis of triangle power-counting integrands (`$\mathfrak{B}_3^{(4)}$') as defined in ref.~\cite{Bourjaily:2020qca} is \emph{too large} for amplitudes in sYM. As we will see, many integration contours can be chosen for which all amplitudes in sYM are known to vanish. In addition to integrands with double-poles, there are many which can be normalized on contours involving poles at infinity---for which amplitudes in sYM are known (at this loop order) or widely expected to vanish \cite{Arkani-Hamed:2014via,Bourjaily:2018omh}.

Finally, a practical motivation for this present work is the prospect of constructing a single integrand basis large enough to span all N$^{k}$MHV amplitudes in sYM at some fixed multiplicity. Although the six-particle MHV amplitudes at two loops are known in multiple forms \cite{Bourjaily:2019gqu,Bourjaily:2019iqr}, the six-point NMHV amplitude remains an important, and open target for investigation. By expressing both amplitudes in the \emph{same} basis of master integrands, there is some hope that it may simplify the work of loop integration for some analogue of the ratio function in the planar case. The analogous results in the planar limit proved important seeds for future discovery and development \cite{Bern:2008ap,Kosower:2010yk,DelDuca:2009au,Goncharov:2010jf}, and we hope this work may inspire some ambitious loop-integrators to take up this challenge once the corresponding amplitudes are known \cite{6ptAmps}.

\vspace{4pt}\subsection{Why Illustrate Basis-Building for \emph{Six}-Particles at Two Loops?}\vspace{0pt}
Let us briefly describe why we have chosen to use the case of non-planar integrands at two-loops involving \emph{six} external particles as the primary example discussed in this work. As discussed at length in \cite{Bourjaily:2019iqr}, six-particle amplitudes are the last multiplicity which can be represented in terms of \emph{manifestly} (and individually) polylogarithmic integrands. In the context of maximal sYM, this can be made obvious by considering the seven-point `tardigrade' topology, whose maximal cut surface has support on an elliptic sub-topology \cite{Bourjaily:2018yfy},
\vspace{2pt}\eq{\label{max_cut_elliptic}\underset{a^2=\cdots=f^2=0}{\mathrm{Res}}\left[\raisebox{-1pt}{\tardi{\tardiScalarEdges\tardiLegs{2}{1}{2}{1}{1}\intDots{5}}}\right]\propto\frac{\dbar x}{y}\,,\quad \text{where}\quad y^2\equivR Q(x)\,,}
where $Q(x)$ is an irreducible quartic in the remaining loop variable; as such, this integral involves the geometry of an elliptic curve. Because the integral in (\ref{max_cut_elliptic}) has NMHV coloring, \emph{any} representation of the 7-point NMHV amplitude \emph{necessarily} requires terms with support on this elliptic integral. To be clear, although the maximal cut (\ref{max_cut_elliptic}) has no further residues on which to define a contour, this presents no fundamental obstacle to defining a spanning set of contours with which to diagonalize numerators. Indeed, the recent works \cite{Bourjaily:2020hjv,Bourjaily:2021vyj} (see also \cite{Bourjaily:2017bsb,Bourjaily:2018ycu,Bourjaily:2018yfy,Bourjaily:2019hmc}) demonstrate that for \emph{any} topology with elliptic (or worse) structures a natural choice of contours may be furnished by some choice of homological cycles within the cut-surface's geometry. In the above example, these would simply be a choice of either the $a$- or $b$-cycle of the elliptic curve. 

Another reason we have chosen to focus on the case of six particles is primarily pragmatic: integrals with this multiplicity are (or just beyond) the current state-of-the-art in loop integration. Moreover, a six-point prescriptive basis (in which both the MHV and NMHV amplitudes were expressed) would allow for the direct search of some non-planar analogue of the ratio function beyond the planar limit.

\vspace{4pt}%
\subsection{Organization and Outline}\label{subsec:outline}\vspace{0pt}
This work is organized as follows. In section~\ref{sec:general_basis_building_review} we review the relevant aspects and benefits of the prescriptive approach to generalized unitarity, as well as the graph-theoretic definition of power-counting for multi-loop integrand bases introduced in \cite{Bourjaily:2020qca}. In section~\ref{sec:one_loop_review} we explore the relation between contour choices and the properties of the induced prescriptive integrand bases at one loop, and provide a pedagogical derivation of novel triangle power-counting bases related to the `chiral box expansion' of \cite{Bourjaily:2013mma}. Section~\ref{sec:six_point_basis} contains the main results of this work: a thorough outline of the steps involved in constructing prescriptive integrand bases, and the application of these ideas to the case of six particles at two loops. In section~\ref{sec:discussion_of_our_basis}, we discuss desirable features and possible applications of the aforementioned basis, both from the perspective of amplitude integrand construction and for loop integration. Finally, we conclude in section~\ref{sec:conclusions} with a summary and a discussion of natural directions for further research.\\

The complete details of the triangle power-counting basis constructed at two loops for six particles are made available as ancillary files to this work's submission on the \texttt{arXiv}. These are accessible directly from the abstract page for this work on the \texttt{arXiv} (linked-to below `Download' on the right-hand pane). The organization and content of these files is described in \mbox{appendix \ref{ancillary_files}}. In addition to including complete details of our contour choices and the basis of integrands that results, we have also prepared some \textsc{Mathematica} code for the manipulation of these data and for the evaluation of the expressions involved.

\newpage\vspace{0pt}%
\section{Review: Essential Elements of Integrand-Basis Building}\label{sec:general_basis_building_review}\vspace{0pt}
Even before we describe the choices involved in making a particular choice for the basis elements of set of loop integrands, it is important to first enumerate the space of integrands under consideration. Typically, this consists of a choice of Feynman integral topologies---defining the propagators of the integrands in the basis---and then some stratum of allowed loop-dependent numerator degrees of freedom. These loop-dependent numerators are usually organized according to some notion of power-counting---limiting the degree of polynomial loop-dependence of the numerators to be considered in the basis. 

In \cite{Bourjaily:2020qca} it was argued at length that the numerators of any basis are best organized not by Lorentz-invariant scalar products as is more typical, but rather in terms of the \emph{translates} of inverse propagators involving the momentum flowing through some choice of edges of the graph. This choice manifests both the identification of some of these polynomial degrees of freedom as `contact terms' and this organization is naturally translationally invariant---independent of how the origins of loop momenta are chosen, or how the loop momenta are `routed' through a Feynman graph---how momentum conservation is solved in order to specify some number $L$ independent loop momenta. Thus, this framework organizes loop-dependent numerators in a purely graph-theoretic way. 
 
As this formalism is still somewhat novel, we begin with a rapid but largely self-contained review of the essential ideas and notation introduced in \cite{Bourjaily:2020qca} which will be required later in this work.\\

At one loop, the loop-dependent part of the denominator of any Feynman integrand consists of a single, closed cycle of some number of Feynman propagators. A graph involving $p$ propagators (and any loop-dependent polynomial in the numerator) is called a `$p$-gon'. The denominator of a $p$-gon consists of a product of factors of the form $\x{\ell}{D_i}\equivR(\ell{-}D_i)^2$; provided the $D_i$ are indexed cyclically around the graph, consecutive $D_i$'s differ by the total (external) momentum flowing into a given vertex along the cycle. (Obviously, translation-invariance in $\ell$ allows us to set any $D_i$ to $\vec{0}$; but it is best if we leave this redundancy in place.)

To describe loop-dependent numerators, we choose (without any loss of generality) to use the same building blocks as the factors appearing in the denominator. Namely, we choose to write any loop-dependent numerator as a polynomial involving products of inverse propagators as monomials. Specifically, we may define
\eq{[\ell]\equivR\text{span}_{Q}\{\x{\ell}{Q}\}\quad\text{for}\quad Q\!\in\!\mathbb{R}^4\,.}
That is, $[\ell]$ consists of all polynomials in $\ell$ that can be written as single inverse propagators with loop-independent coefficients. (These coefficients may depend arbitrarily---typically algebraically---on the external momenta.) It is not hard to see that for four dimensions, $\mathrm{rank}\big([\ell]\big)\!=\!6$. Moreover, it is easy to see that this vector space is equivalent to $[\ell]\!\simeq\!\text{span}\{1,(\ell\!\cdot\!e_i),\ell^2\}$ where $e_i$ are some basis vectors for the space of external momenta, $\mathbb{R}^4$. Notice that $[\ell]\!=\![\ell{+}Q]$ for any $Q\!\in\!\mathbb{R}^4$; thus, this is a naturally translationally-invariant notion for a set of loop-dependent numerators.

Higher-order polynomials in $\ell$ can be constructed as products of inverse propagators. Specifically, we may define
\eq{[\ell]^q\equivR\text{span}_{\oplus Q_i}\left\{\prod_{i=1}^{q}\x{\ell}{Q_i}\right\}\quad\text{for}\quad Q_i\!\in\!\mathbb{R}^4\,.}
It not hard to see that $[\ell]^a\!\subset\![\ell]^b$ for any $a\!<\!b$. In particular, $[\ell]^0\!=\!\{1\}\!\subset\![\ell]$, as seen above. These spaces form symmetric, traceless products of $6$-dimensional representations of $SO_6$---a fact that is easy to see in the embedding formalism (see e.g.~\cite{Bourjaily:2019exo}). For this work, we will mostly be interested in spaces $[\ell]^0$, $[\ell]^1,[\ell]^2,[\ell]^3$, which have ranks 1, 6, 20, and 50, respectively. 

We can discuss the space of loop-dependent numerators assigned to a given set of Feynman propagators using the following graphical notation:
\eq{\fwbox{65pt}{\tikzBox{\draw[int](0,0)--(1,0);\draw[markedEdgeR](0,0)--(1,0);\node[anchor=north] at (0.5,0) {$\vec{\ell}$};}}\equivR\frac{[\ell]}{\ell^2}\simeq\text{span}\left\{\frac{1}{\ell^2},\frac{\ell^i}{\ell^2},1\right\}\,.\label{vector_space_of_decorated_edge}}
(Notice that this space is large enough to include the propagators of scalars, fermions, and vector bosons (in any gauge).) To illustrate how this notation can be used, consider the space of box integrands with triangle power-counting:
\def\labelDist{\legLen*1.9}
\eq{\label{eq:box_with_tri_pc}\frac{[\ell]}{\x{\ell}{D_1}\x{\ell}{D_2}\x{\ell}{D_3}\x{\ell}{D_4}}\leftrightarrow\tikzBox{\oneLoopGraphElement[0]{4}\draw[markedEdgeR](a2)--(a3);\legAltLabelled{(a1)}{-135}{$P_1$};\legAltLabelled{(a2)}{135}{$P_2\,$};\legAltLabelled{(a3)}{45}{$P_3$};\legAltLabelled{(a4)}{-45}{$P_4$};}\,\,.\def\labelDist{\legLen*1.5}}
(Momentum conservation requires, for example, that $D_2{-}D_1\!=\!P_1$; but to specify the $D$'s would require that we eliminate the translational-invariance of the loop momentum---something we'd like to avoid.) Because the space $[\ell]$ is invariant under translations, decorating any pair of propagators whose momenta differ by some sum of external momenta would result in the same vector space of loop-dependent numerators; in particular, the decoration on the top edge in (\ref{eq:box_with_tri_pc}) can be placed on any edge of the graph and would define precisely the same set of Feynman-like loop integrands.

Notice that any element of the space of integrands in (\ref{eq:box_with_tri_pc}) scales \emph{manifestly} like a scalar triangle integral as $\ell\!\to\!\infty$. Because of this, we describe this space of box  integrands in (\ref{eq:box_with_tri_pc}) as those with `triangle' power-counting.

It is natural to partition the space of loop-dependent numerators as much as possible into `contact-terms'---monomials which eliminate one (or more) of the propagators of the graph. In the case of the box in (\ref{eq:box_with_tri_pc}), the space of contact terms are obviously $\text{span}\{\x{\ell}{D_1},\x{\ell}{D_2},\x{\ell}{D_3},\x{\ell}{D_4}\}\!\subset\![\ell]$. The space of numerators \emph{not spanned by} contact terms is called the numerator's \emph{{\color{topCount}top-level}} degrees of freedom. This space can be spanned by $\{\x{\ell}{Q^i}\}$ for $i\!=\!1,2$ where $Q^i$ represents one of the solutions to the `quad-cut' equation $\x{Q^i}{D_1}\!=\x{Q^i}{D_2}\!=\!\x{Q^i}{D_3}\!=\!\x{Q^i}{D_4}\!=\!0$. 

Thus, any particular numerator for a box with `triangle' power-counting---any element of (\ref{eq:box_with_tri_pc})---could be expressed by
\eq{[\ell]\ni\mathfrak{n}(\ell)\equivL\underbrace{c^1(\ell | Q^1)+c^2(\ell | Q^2)}_{\text{top-level}}+\underbrace{c^3(\ell | P_1)+\cdots c^6(\ell | P_4)}_{\text{contact terms}}}
where the coefficients $\{c^i\}_{i=1,\ldots,6}$ are some $\ell$-independent `constants' (arbitrary functions of the external momenta). We would say that a box integral with triangle power-counting's six-dimensional space of numerators consists of {\color{topCount}2} `{\color{topCount}top-level}' degrees of freedom and 4 `contact-terms'. 

In four dimensions, it turns out that any pentagon (or higher) with triangle power-counting can be expanded into the boxes (\ref{eq:box_with_tri_pc}) and scalar triangle integrands. Thus, the complete basis of one-loop integrands with triangle power-counting would be given by 
\def\labelDist{\legLen*1.9}
\eq{\begin{split}\mathfrak{B}_3^{(4),L=1}&\equivR\text{span}_{\oplus D_i}\left\{\frac{1}{\x{\ell}{D_1}\x{\ell}{D_2}\x{\ell}{D_3}},\frac{[\ell]}{\x{\ell}{D_1}\x{\ell}{D_2}\x{\ell}{D_3}\x{\ell}{D_4}}\right\}\\[10pt]
&=\text{span}_{\oplus P_i}\Bigg\{\hspace{15pt}\tikzBox{\oneLoopGraphElement[0]{3}\legAltLabelled{(a1)}{-150}{$P_1$};\legAltLabelled{(a2)}{90}{$P_2$};\legAltLabelled{(a3)}{-30}{$P_3$};}\hspace{15pt},\hspace{15pt}\tikzBox{\oneLoopGraphElement[0]{4}\draw[markedEdgeR](a2)--(a3);\legAltLabelled{(a1)}{-135}{$P_1$};\legAltLabelled{(a2)}{135}{$P_2\,$};\legAltLabelled{(a3)}{45}{$P_3$};\legAltLabelled{(a4)}{-45}{$P_4$};}\hspace{15pt}\Bigg\}.\end{split}\def\labelDist{\legLen*1.5}\label{eq:one_loop_triangle_pc}}

To specify a \emph{particular} basis for this space, both top-level and contact-term degrees of freedom must be defined for every integrand in the basis. It may seem tempting to simply set all contact-terms to zero in the space of numerators, but this is not obviously in line with prescriptivity (\ref{prescriptivity_condition}); rather, in any prescriptive basis the contact terms of the boxes are uniquely fixed by the requirement that the box integrands in the basis \emph{vanish on the contours to which the triangles are dual}. Provided any contour used for diagonalization includes `cutting' of every propagator of the integrand, all triangle integrands will automatically vanish on any contour defining a box integrand. This is an illustration of the upper-triangular nature of bases with $\!<\!d$-gon power-counting in $d$ dimensions---something we will discuss in more detail below.\\

At two-loops (or higher), loop-dependent numerators are organized in a similar way. Before we discuss this, it is worthwhile to review the possible loop dependence in the denominator of any two-loop Feynman graph. Any such graph can be labelled by three numbers, $\{\r{a},\g{b},\b{c}\}$ which indicate the number of propagators that are related by translates involving only external momenta; we can see this graphically by describing the denominator of any two-loop integrand as
\eq{\fwboxR{0pt}{\Gamma_{[\r{a},\g{b},\b{c}]}\Leftrightarrow\,}\fig{-26pt}{0.3}{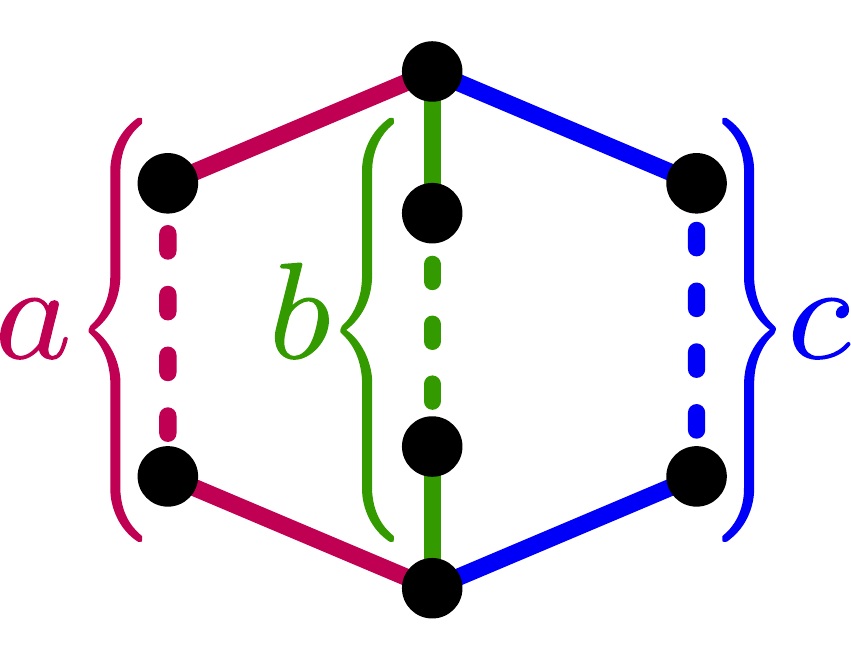}}

When discussing loop-dependent numerators for such integrands, we may use inverse propagators involving translates of the momenta flowing through the edges each type in the graph:
\begin{align}
\begin{split}
[\r{\ell_A}] &\equivR \text{span}_{Q}\!\left\{(\r{\ell_A}|Q)\right\}, \\
[\g{\ell_B}] &\equivR \text{span}_{R}\left\{(\g{\ell_B}|R)\right\}, \quad \text{where}\quad Q,R,S\in\mathbb{R}^4,\\
[\b{\ell_C}] &\equivR \text{span}_{S\,}\left\{(\b{\ell_C}|S)\right\},
\end{split}            
\end{align} 
as well as (sums of) products thereof. Of course, $[\g{\ell_B}]\!\simeq\![\r{\ell_A}{-}\b{\ell_C}]$; but as before, we choose to avoid making any choice of loop-momentum routing in order to eliminate one of the three labels. 

For any choice of numerators constructed in this way, there will be a natural decomposition into {\color{topCount}top-level} degrees of freedom and contact terms. But to specify a particular \emph{set} of such numerators (even before we discuss choosing \emph{particular} representative integrands to span that set) requires that we make some choice of \emph{which} numerators to include for each topology in the basis. Complicating matters is the fact that for non-planar graphs the na\"ive notion of `scaling like a $p$-gon' as $\ell\!\to\!\infty$ depends strongly on how the loop momenta are routed---which two cycles of unfixed, internal loop momenta are considered as independent. A proposal for a space of triangle power-counting integrands at any loop order for non-planar graphs was defined in ref.~\cite{Bourjaily:2020qca}, which we review presently.

\vspace{4pt}%
\subsection{Power-Counting of Integrands Beyond the Planar Limit}\label{subsec:general_power_counting}\vspace{0pt}
The essential idea behind the graph-theoretic definition of power-counting described in ref.~\cite{Bourjaily:2020qca} is that when spaces of loop-dependent numerators are constructed using (\ref{vector_space_of_decorated_edge}) it is clear how every integrand in this space behaves as the loop momentum through any edge goes to infinity: like a constant. That is, putting a decoration as in (\ref{vector_space_of_decorated_edge}) on any edge of a graph results in a space of integrands that scales exactly like the Feynman integrand where all decorated-edges have collapsed. 

Thus, while there may be no invariant notion of how a Feynman integrand scales at `infinity' without specifying a particular routing of loop momenta, there is an invariant sense in which one loop integrand can scale like one (or more) among its \emph{contact-terms}. Thus, the starting point for defining $\mathfrak{B}_p$---the space of integrands with `$p$-gon' power-counting---beyond one loop is a specification of which integrands in the basis should be defined as `scalars'---those integrands with loop-independent numerators, like which every other integrand in the basis scales at infinity. 

Given any choice of `scalar' Feynman graphs, a complete space of integrands which scale like these may be constructed by adding decorated edges (\ref{vector_space_of_decorated_edge}) to these. In any fixed spacetime dimension, there is an upper bound on the number of edges that can be added before the rank of that graph's top-level degrees of freedom vanishes.

The authors of \cite{Bourjaily:2020qca} defined the set of $p$-gon scalars as any graph with girth $p$ which would be lowered by the collapse of any edge. At two loops, these scalar $3$-gons are defined to be
\eq{\mathfrak{S}_3\equivR\left\{\hspace{-10pt}\kT{\kTScalarEdges\intDots{5}\legOpt{(v1)}{-140}{};\leg{(v1)}{-90}{};\leg{(v2)}{140}{};\legOpt{(v2)}{90}{};\leg{(v3)}{105}{};\legOpt{(v3)}{75}{};\leg{(v4)}{90}{};\legOpt{(v4)}{40}{};\leg{(v5)}{-40}{};\legOpt{(v5)}{-90}{};}\hspace{-5pt},\hspace{-5pt}\dT{\dTScalarEdges\intDots{4}\leg{(v1)}{210}{};\legOpt{(v1)}{150}{};\leg{(v2)}{150}{};\legOpt{(v2)}{30}{};\leg{(v3)}{30}{};\legOpt{(v3)}{-30}{};\leg{(v4)}{-30}{};\legOpt{(v4)}{210}{};}\hspace{-10pt}\right\}.\label{scalar_3_gons}}
From these, the basis $\mathfrak{B}_3$ would be generated by adding any number of decorated edges (\ref{vector_space_of_decorated_edge}) to either of these graphs (without increasing the loop number). Said another way, any graph that contains either of (\ref{scalar_3_gons}) as a contact term would be endowed with a vector-space of numerators given by the product of the translates of inverse propagators corresponding to each edge in the quotient of the parent relative to the daughter. For graphs where more than one edge-sets exists, the space of numerators is defined as the outer-sum of the vector-spaces generated by each.

\vspace{4pt}%
\subsection{Non-Planar, Triangle Power-Counting Basis at Two Loops}\label{subsec:review_of_size_of_basis}\vspace{0pt}

Tthe set of triangle power-counting integrands consists of all graphs which scale like one or the other of (\ref{scalar_3_gons}). In any fixed spacetime dimension (or equivalently, external multiplicity), the space of such integrands is in fact finite dimensional: all but a finite number of integrands with a bounded number of propagators is \emph{reducible}---meaning that its numerator space would entirely spanned by contact terms (its {\color{topCount}top-level} rank would be zero). 

In four dimensions, any integrand with more than eight propagators is reducible in $\mathfrak{B}_3$, and those integrands with precisely eight propagators will have precisely as many {\color{topCount}top-level} degrees of freedom as there are solutions to the maximal-cut equations (defining contours which `encircle' all of the propagators of the graph). The complete space of such integrands was described in some detail in \cite{Bourjaily:2020qca}, which is summarized in \mbox{table \ref{4d_3_gon_pc_graphs}}.\\
\begin{table}[h]\centering\caption{The starting point in our basis construction: the complete list of integrand topologies compatible with triangle power-counting, as defined in \cite{Bourjaily:2020qca}, reproduced from \cite{Bourjaily:2019iqr}. Included is the {\color{totalCount}total rank} of the vector space of numerators in $\mathfrak{B}_3$, as well as the decomposition of these ranks into {\color{topCount}top-level} and contact terms.}\vspace{-25pt} $$\fig{-120pt}{1.5}{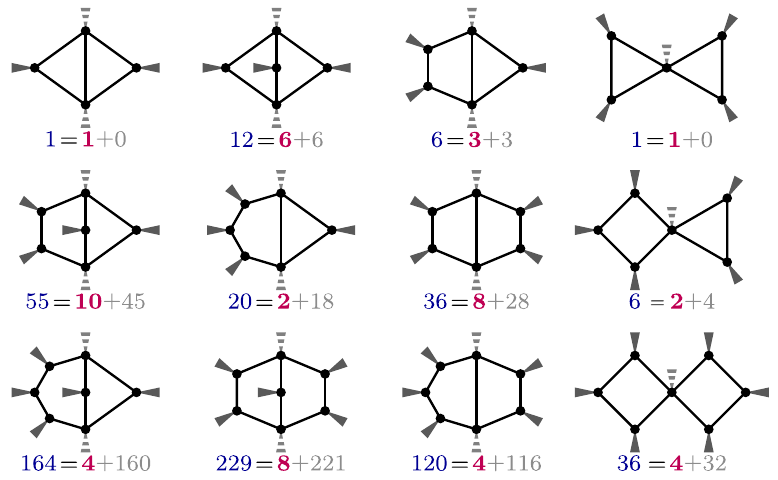}$$\vspace{-20pt}\label{4d_3_gon_pc_graphs}\end{table}

In the case of six external particles, the integrands in \mbox{table \ref{4d_3_gon_pc_graphs}} involve 94 distinct (graph non-isomorphic) instances of distinct leg distributions. Among these, we have chosen to disregard 7 of which include massless-triangle sub-topologies---for example,
\eq{\label{eq:massless_tri_eg}\pT{\draw[int](v1)--(v2);\draw[int](v2)--(v3);\draw[int](v3)--(v4);\draw[int](v4)--(v5);\draw[int](v5)--(v6);\draw[int](v6)--(v1);\draw[int](v6)--(v4);\pTLegs{1}{1}{3}{0}{1}{0}\pTArrows}.}
Such integrands are problematic from a unitarity point of view for the simple reason that when the internal propagators flowing into the massless triangle ($a$ and $d$ above) and any two within the triangle ($e$, $f$, and $g$ above) are cut, the third propagator of the triangle will automatically be on-shell (in four dimensions). Thus, the five propagators involved in a massless triangle do give rise to a co-dimension 5, transverse residue. This technical problem can be easily resolved by considering the propagators as massive, for example; but as such integrands form a closed subspace under edge contractions (no integrand \emph{without} a massless triangle has one as a contact term) and as their coefficients would always vanish in sYM, we have chosen to ignore them in our classification. 

Thus, including all non-graph-isomorphic leg distributions, and excluding any graphs with massless triangles, we find 87 distinct integrand topologies for six external particles at two loops. These are enumerated in full detail in the ancillary files to this work (see \mbox{appendix \ref{ancillary_files}}), but we summarize the basis that results in \mbox{table \ref{enumeration_of_6pt_basis}}. 
\begin{table}[h]
\caption{Enumeration of the topology distributions and rank-counts for graphs with six external legs. The index ranges are those used to identify integrands in the ancillary files.\label{enumeration_of_6pt_basis}}\vspace{2pt}\centering\hspace{-0pt}\begin{tabular}[c]{@{$\,$}c@{}rc@{$\,$}c@{$\,$}r@{$\,$}llll}\multicolumn{1}{c}{topology}&\#&\begin{tabular}{@{}c@{}}index\\[-5pt]range\end{tabular}&\multicolumn{1}{c}{$\mathfrak{N}_3$}&\multicolumn{2}{l}{$\fwboxL{0pt}{\hspace{-39.5pt}\text{\begin{tabular}{@{}r@{}l@{}}{\color{totalCount}total rank}\,=&\phantom{+}{\color{topCount}top-level rank}\\[-5pt]&+{\color{black}contact-terms}\end{tabular}}}$}\\
\hline\hline$\Gamma_{[\r{4},\g{0},\b{4}]}$&1&[1]&$\r{[\ell_A]}\b{[\ell_C]}$&{\color{totalCount}36}&=\,{\color{topCount}\phantom{0}4}{\color{dim}+\,32}\\
$\Gamma_{[\r{4},\g{1},\b{3}]}$&4&[2-5]&$\r{[\ell_A]^2}\b{[\ell_C]}\!\oplus\!\r{[\ell_A]}\g{[\ell_B]}$&{\color{totalCount}120}&=\,{\color{topCount}\phantom{0}4}{\color{dim}+\,116}\\
$\Gamma_{[\r{4},\g{2},\b{2}]}$&4&[6-9]&$\r{[\ell_A]^3}\!\!\oplus\!\r{[\ell_A]^2}\b{[\ell_C]}\!\oplus\!\r{[\ell_A]^2}\g{[\ell_B]}$&{\color{totalCount}164}&=\,{\color{topCount}\phantom{0}4}{\color{dim}+\,160}\\
$\Gamma_{[\r{3},\g{2},\b{3}]}$&3&[10-12]&$\g{[\ell_B]^2}\!\!\oplus\!\r{[\ell_A]^2}\b{[\ell_C]}\!\oplus\!\r{[\ell_A]^2}\g{[\ell_B]}\!\oplus\!\r{[\ell_A]}\g{[\ell_B]}\b{[\ell_C]}$&{\color{totalCount}229}&=\,{\color{topCount}\phantom{0}8}{\color{dim}+\,221}\\
$\Gamma_{[\r{3},\g{0},\b{4}]}$&4&[13-16]&$\b{[\ell_C]}$&{\color{totalCount}6}&=\,{\color{topCount}\phantom{0}2}{\color{dim}+\,4}\\
$\Gamma_{[\r{3},\g{1},\b{3}]}$&8&[17-24]&$\g{[\ell_B]}\!\oplus\!\r{[\ell_A]}\b{[\ell_C]}$&{\color{totalCount}36}&=\,{\color{topCount}\phantom{0}8}{\color{dim}+\,28}\\
$\Gamma_{[\r{4},\g{1},\b{2}]}$&9&[25-33]&$\r{[\ell_A]^2}$&{\color{totalCount}20}&=\,{\color{topCount}\phantom{0}2}{\color{dim}+\,18}\\
$\Gamma_{[\r{3},\g{2},\b{2}]}$&10&[34-43]&$\r{[\ell_A]^2}\!\!\oplus\!\r{[\ell_A]}\b{[\ell_C]}\!\oplus\!\r{[\ell_A]}\g{[\ell_B]}$&{\color{totalCount}55}&=\,{\color{topCount}10}{\color{dim}+\,45}\\
$\Gamma_{[\r{3},\g{1},\b{2}]}$&17&[44-60]&$\r{[\ell_A]}$&{\color{totalCount}6}&=\,{\color{topCount}\phantom{0}3}{\color{dim}+\,3}\\
$\Gamma_{[\r{2},\g{2},\b{2}]}$&9&[61-69]&$\r{[\ell_A]}\!\oplus\!\g{[\ell_B]}\oplus\b{[\ell_C]}$&{\color{totalCount}12}&=\,{\color{topCount}\phantom{0}6}{\color{dim}+\,6}\\
$\Gamma_{[\r{3},\g{0},\b{3}]}$&5&[70-74]&$1$&{\color{totalCount}1}&=\,{\color{topCount}\phantom{0}1}{\color{dim}+\,0}\\
$\Gamma_{[\r{2},\g{1},\b{2}]}$&13&[75-87]&$1$&{\color{totalCount}1}&=\,{\color{topCount}\phantom{0}1}{\color{dim}+\,0}\fwboxL{43pt}{~}\\\hline
\end{tabular}
\end{table}

In total, we have 87 six-particle, two-loop integrand topologies spanning a space of $373$ top-level degrees of freedom ($3129$ degrees of freedom in total). This is the number of contours which must be specified to define a prescriptive basis. All 3129 degrees of freedom will be fixed by these contours, together with their graph-isomorphic images as contact-term conditions according to (\ref{prescriptivity_condition}). 

We will return to the details of this integrand basis in \mbox{section \ref{sec:six_point_basis}}; but first, let us briefly return to one loop in order to better understand role of contour choices in the prescriptivity condition (\ref{prescriptivity_condition}).

\newpage\vspace{0pt}%
\section{Illustrating Implications of Contour-Choices at One Loop}\label{sec:one_loop_review}\vspace{0pt}
Generalized unitarity at one loop has a long and rich history \cite{Bern:1994zx,Bern:1994cg,Bern:1995db,Britto:2004nj,Britto:2004nc,Bidder:2005ri,Anastasiou:2006jv,Ossola:2006us}, and its many refinements have yielded many practical and theoretical discoveries \mbox{\cite{Britto:2004ap,Britto:2005fq,Drummond:2006rz,Alday:2007hr,Drummond:2008vq,ArkaniHamed:2012nw}}. Examples of pedagogical reviews of this material can be found in \cite{Dixon:1996wi,Bern:2011qt,Elvang:2013cua} as well as in recent, related works by some of the authors \cite{Bourjaily:2017wjl,Bourjaily:2019iqr,Bourjaily:2020qca}. Before proceeding to the main two-loop results of this paper, in this section we shall take a detour to re-analyze one-loop bases with box and triangle-power counting, with a particular focus on the relation between the contour choices and the properties of the associated bases. In addition to providing some simple examples of the kinds of contour prescriptions we use throughout this work, the main results of this section are two different and novel triangle-power-counting bases which are related to---but slightly different than---the `chiral box' expansion of \cite{Bourjaily:2013mma}.

\vspace{0pt}%
\subsection{Box Power-Counting Basis and the No-Triangle Property}\label{subsec:box_power_counting}\vspace{0pt}
The standard lore of one-loop generalized unitarity in maximally supersymmetric Yang-Mills (sYM) is that, because amplitudes scale as $\sim(\ell^2)^{-4}$ as $\ell\rightarrow\infty$, a natural choice of power-counting is `$4$-gon', which in four dimensions consists of scalar boxes and pentagons with a single loop-momentum numerator insertion: 
\eq{\mathfrak{B}_4\leftrightarrow\left\{\tikzBox{\oneLoopGraphElement[0]{4}}\,,\;\tikzBox{\oneLoopGraphElement[0]{5}\draw[markedEdgeR](a2)--(a3);}\;\right\}.}
The numerator space for the pentagons is conveniently described in a basis of the five inverse propagators of the graph, plus a single top-level numerator in the complement of that subspace. As every topology has a single {\color{topCount}top-level} degree of freedom, the contour prescriptions of this part of the basis are simple: every box may be normalized to be unit on the parity-even combination (the difference of residues) of its respective quad-cuts $f^{1,2}_{A,B,C,D}$, which we may denote diagrammatically as 
\eq{\frac{1}{2}\left(\,\tikzBox{\oneLoopGraphElementLS[0]{4};\node at (a1) [ordAmp] {};\node at (a2) [ordAmp] {};\node at (a3) [ordAmp] {};\node at (a4) [ordAmp] {};\node at (v0) {$1$};}+\tikzBox{\oneLoopGraphElementLS[0]{4};\node at (a1) [ordAmp] {};\node at (a2) [ordAmp] {};\node at (a3) [ordAmp] {};\node at (a4) [ordAmp] {};\node at (v0) {$2$};}\,\right)\leftrightarrow\tikzBox{\oneLoopGraphElement[0]{4}}\;,}
while each pentagon can be normalized on a parity-odd contour which results in integrands which integrate to zero. As has been discussed at length in \cite{Bourjaily:2017wjl}, this basis is unfortunately over-complete because the parity-odd pentagons are not all independent: the $\binom{n}{5}$ pentagons satisfy $\binom{n{-}1}{5}$ linear relations. Eliminating the resulting redundancy requires an (arbitrary) choice of an independent subset of pentagons. This choice, while not a significant roadblock at one loop---especially if one were only interested in integrated expressions---becomes substantially more problematic at higher loops. Fortunately, considering a basis with triangle power-counting is an elegant way of resolving this issue, at the cost of losing manifest dual conformal invariance in the planar part of the basis. 

\vspace{4pt}%
\subsection{Possible Choices of Triangle Power-Counting Bases at One Loop}\label{subsec:triangle_power_counting}\vspace{0pt}
The box power-counting basis of the previous subsection is perfectly natural for representing all one-loop amplitudes in sYM \emph{post-integration}. However, integrand-level representations of these amplitudes require an arbitrary choice of an independent set of parity-odd pentagons. This led to the development of the `upgraded' basis of \cite{Bourjaily:2013mma} resulting in the `chiral box' expansion, which made use of a particular, prescriptive basis of triangle power-counting integrands, cleanly separated into infrared finite and divergent subsets. These integrands were designed to have support on one solution to the box-cut, and vanish on the other. In addition, they were designed to vanish on the \emph{parity-even} contour (including infinity) of any \emph{three-mass} triangle sub-topology and on any soft and/or collinear region of any triangle with a massless corner. Thus, the `chiral box' basis described in \cite{Bourjaily:2013mma}, while not adequately described in these terms, corresponded to a prescriptive basis with the following choice of contours: chiral box-cut contours for each box integrand; soft and/or collinear regions and parity-even at infinity for all divergent or finite triangle integrals, respectively.\\

To clarify the role of the contour choices involved, it is worthwhile to revisit the question of triangle power-counting at one loop and derive two new prescriptive bases based upon slightly different contour choices. 

\newpage
\subsubsection{\textbf{Option 1}: Exploiting Residues on Poles at Infinity}\label{subsubsec:triangle_power_counting_at_infinity}\vspace{0pt}

In a basis with triangle power-counting in four dimensions, every scalar triangle integral has one simple pole at infinite loop momentum along each triple-cut. Thus, a natural choice for their contours would be the parity-even combination of these (which is the difference of the two contours), resulting in a normalization of the integrands given by 
\eq{\raisebox{-5pt}{\scalebox{1}{\triangle{\draw[int](v1)--(v2);\draw[int](v2)--(v3);\draw[int](v3)--(v1);\legAltLabelled{(v1)}{210}{$A$}\legAltLabelled{(v2)}{90}{$B$}\legAltLabelled{(v3)}{-30}{$\,\,C$}}}}\hspace{-10pt}\leftrightarrow\mathfrak{n}_{A,B,C}\equivR\frac{1}{2}\sqrt{(p_C^2{-}p_A^2{-}p_B^2)^2{-}4\,p_A^2\,p_B^2}\,.}
This normalization smoothly degenerates to all cases involving massless external momenta (the argument of the square root becoming a perfect square for all such degenerations).

As amplitudes in sYM (and SUGRA) have no support on poles at infinity at one loop, the coefficient of any all scalar triangle integrand for amplitudes in sYM will be zero. (This is much more in line with the original observation leading to the `no triangle' hypothesis for these theories \cite{Green:1982sw,Bern:1991aq,Bern:1992cz,Bern:1998sv,Bern:2005bb,Bjerrum-Bohr:2006xbk}.)

What about the box integral contours? Rather than taking the chiral solutions to the quad-cuts, let's use the parity-even and parity-odd combinations of contours---that is, the difference and sum of the chiral contours, respectively. Even before we discuss the numerators that result from this choice of contours, it is interesting to note that in the representation of a loop integrand, we would have coefficients arranged according to
\eqs{\label{eq:box_contours}&\frac{1}{2}\left(\;\tikzBox{\oneLoopGraphElementLS[0]{4};\node at (a1) [ordAmp] {};\node at (a2) [ordAmp] {};\node at (a3) [ordAmp] {};\node at (a4) [ordAmp] {};\node at (v0) {$1$};}+\tikzBox{\oneLoopGraphElementLS[0]{4};\node at (a1) [ordAmp] {};\node at (a2) [ordAmp] {};\node at (a3) [ordAmp] {};\node at (a4) [ordAmp] {};\node at (v0) {$2$};}\;\right)\leftrightarrow\tikzBox{\oneLoopGraphElement[0]{4};\node at (v0) {e};},\\
&\frac{1}{2}\left(\;\tikzBox{\oneLoopGraphElementLS[0]{4};\node at (a1) [ordAmp] {};\node at (a2) [ordAmp] {};\node at (a3) [ordAmp] {};\node at (a4) [ordAmp] {};\node at (v0) {$1$};}-\tikzBox{\oneLoopGraphElementLS[0]{4};\node at (a1) [ordAmp] {};\node at (a2) [ordAmp] {};\node at (a3) [ordAmp] {};\node at (a4) [ordAmp] {};\node at (v0) {$2$};}\;\right)\leftrightarrow\tikzBox{\oneLoopGraphElement[0]{4};\node at (v0) {o};}.}

For the parity-even contour, a loop-independent `scalar' box is an obvious candidate numerator---and obviously an element of $\mathfrak{B}_3$ (as $\mathfrak{B}_3\!\supset\!\mathfrak{B}_4$). Moreover, the scalar box integral is free of any poles at infinite loop momentum, making them automatically vanish on the contours defining the scalar triangle integrands. 

For the parity-odd contours for the boxes, it is clear that we must choose integrands that have equal residues on the two box-cuts so that these integrals vanish on the even-contours (so that the $2\!\times\!2$ system of top-level numerators for each box is diagonal). It turns out to be fairly easy to guess such numerators---which turn out to vanish automatically on all parity-even contours defining all triangle daughter topologies (contact terms).

\newpage
The final set of numerators will be
\eq{\fwbox{0pt}{\mathcal{I}_{A,B,C,D}^{e,o}\leftrightarrow\hspace{-10pt}\boxOL{\draw[int](v1)--(v2);\draw[int](v2)--(v3);\draw[int](v3)--(v4);\draw[int](v4)--(v1);\draw[directedEdge](v1)--(v2);\node[inner sep=2.5pt,anchor=east] at (connode) [] {$b$};\draw[directedEdge](v2)--(v3);\node[inner sep=2.5pt,anchor=south] at (connode) [] {$c$};\draw[directedEdge](v3)--(v4);\node[inner sep=2.5pt,anchor=west] at (connode) [] {$d$};\draw[directedEdge](v4)--(v1);\node[inner sep=2.5pt,anchor=north] at (connode) [] {$a$};\legAltLabelled{(v1)}{225}{$A$}\legAltLabelled{(v2)}{135}{$B$}\legAltLabelled{(v3)}{45}{$C$}\legAltLabelled{(v4)}{-45}{$D$}}\hspace{-10pt}\leftrightarrow\left\{\begin{array}{@{}l@{$\,$}l@{}}\\[-20pt]\mathfrak{n}_{A,B,C,D}^e\equivR&\frac{1}{2}\sqrt{\left(p_{AB}^2p_{BC}^2{-}p_A^2p_C^2{-}p_B^2p_D^2\right)^2{-}4p_A^2\,p_B^2\,p_C^2\,p_D^2}\\[10pt]
\mathfrak{n}_{A,B,C,D}^o\equivR&\frac{1}{2}\left(\br{p_A,b,c,p_C}-\br{b,c,p_C,p_A}\right)\end{array}\right.\;}\label{eq:one_loop_boxes1}}
(Here we use the kinematic bracket conventions from \cite{Bourjaily:2019iqr,Bourjaily:2019gqu} to denote contractions of momenta
\eq{\br{a_1,a_2,\cdots,c_1,c_2}\equivR\!\Big[(a_1\!\cdot\!a_2)^{\alpha}_{\phantom{\alpha}\beta}\cdots(c_1\!\cdot\!c_2)^{\gamma}_{\phantom{\gamma}\alpha}\!\Big]\,,\label{definition_of_br}}
where $(a_1\!\cdot\!a_2)^{\alpha}_{\phantom{\alpha}\beta}\equivR a_1^{\alpha\,\dot{\alpha}}\epsilon_{\dot{\alpha}\dot{\gamma}}a_2^{\dot{\gamma}\gamma}\epsilon_{\gamma\beta}$ and $a^{\alpha\dot\alpha}\equivR a^{\mu}\sigma_{\mu}^{\alpha\dot{\alpha}}$ are `$2\!\times\!2$' four-momenta, defined via the Pauli matrices. The `$\br{\cdots}$' object may be more familiar to some readers if written equivalently as `$\mathrm{tr}_+[\cdots]$'.) Both of these numerators smoothly degenerate under all massless limits of the corners. It is not hard to check that the entire basis is diagonal in the choice of contours described---and is hence \emph{prescriptive}. 

Notice that this basis is complete and not over-complete. Thus, the odd numerators for the box integrands are full-rank. Thus, considering that $\mathfrak{B}_3\!\supset\!\mathfrak{B}_4$, although each of these integrands scale like scalar triangle integral at infinite loop momentum, it is interesting to note that all parity-odd pentagon integrands are in fact spanned by these `parity-odd' boxes. This fact is further emphasized by the form that amplitudes take in this basis. For theories with maximal supersymmetry (which have box power-counting, and hence are expressible within $\mathfrak{B}_4$), amplitudes will take the form
\eqs{\mathcal{A}&=\sum_{A,B,C,D}\frac{1}{2}\left(f^1_{A,B,C,D}{+}f^2_{A,B,C,D}\right)\mathcal{I}^\text{e}_{A,B,C,D}+\frac{1}{2}\left(f^1_{A,B,C,D}{-}f^2_{A,B,C,D}\right)\mathcal{I}^\text{o}_{A,B,C,D}\\
&=\sum_{A,B,C,D}\frac{1}{2}\left(\;\tikzBox{\oneLoopGraphElementLS[0]{4};\node at (a1) [ordAmp] {};\node at (a2) [ordAmp] {};\node at (a3) [ordAmp] {};\node at (a4) [ordAmp] {};\node at (v0) {$1$};}+\tikzBox{\oneLoopGraphElementLS[0]{4};\node at (a1) [ordAmp] {};\node at (a2) [ordAmp] {};\node at (a3) [ordAmp] {};\node at (a4) [ordAmp] {};\node at (v0) {$2$};}\;\right)\times\tikzBox{\oneLoopGraphElement[0]{4};\node at (v0) {e};}+\frac{1}{2}\left(\;\tikzBox{\oneLoopGraphElementLS[0]{4};\node at (a1) [ordAmp] {};\node at (a2) [ordAmp] {};\node at (a3) [ordAmp] {};\node at (a4) [ordAmp] {};\node at (v0) {$1$};}-\tikzBox{\oneLoopGraphElementLS[0]{4};\node at (a1) [ordAmp] {};\node at (a2) [ordAmp] {};\node at (a3) [ordAmp] {};\node at (a4) [ordAmp] {};\node at (v0) {$2$};}\;\right)\times\tikzBox{\oneLoopGraphElement[0]{4};\node at (v0) {o};}\;.}
As far as integrals are concerned, this is essentially identical to the representation in $\mathfrak{B}_4$---namely, the only integrands that contribute upon integration are the scalar boxes $\mathcal{I}_{A,B,C,D}^e$. 

One disadvantage of this choice of contours, however, is that the IR structure of amplitudes is far from manifest: any scalar box with a massless leg will be IR divergent, and it is a non-trivial fact (an identity that led to the discovery of tree-level recursion relations for amplitudes, in fact \cite{BCF}) that the total IR divergence of an amplitude would be proportional to the tree. 

\vspace{4pt}%
\subsubsection{\textbf{Option 2}: Contours Supported in Regions of IR Divergence}\label{subsubsec:triangle_power_counting_in_IR}\vspace{0pt}

Rather than choosing contours at infinity for all the triangles, let us now consider choosing contours that (in addition to cutting all three propagators) enclose any soft and/or collinear region that may exist. These are the regions in loop-momentum space responsible for IR-divergences \cite{Bourjaily:2021ewt}.

This difference has very little effect on the triangle integrands (whose normalization changes merely by a sign---and by a factor of 2 if there are two branches for a triple-cut). However, it has a very strong effect on the box integrands in our basis: they must be altered so that each vanishes in all such regions of soft and/or collinear divergence. 

Interestingly, the odd box integrands $\mathcal{I}_{A,B,C,D}^o$ in (\ref{eq:one_loop_boxes1}) already vanish in every collinear region by virtue of the fact that these regions are parity-even. However, the scalar boxes, $\mathcal{I}^e_{A,B,C,D}$ in (\ref{eq:one_loop_boxes1}) must be modified accordingly. Moreover, this modification will \emph{change} which if any of the inflowing momenta $\{p_A,p_B,p_C,p_D\}$ are massless. 

Let us use (lower-case) Greek letters to denote a massless momentum; thus, we'll write `$p_\alpha$' for `$p_A$' if $p_A^2\!=\!0$ and similarly for $p_\beta,p_\gamma,p_\delta$. Using this, there are three cases to consider, each with different contact terms added to remove their regions of collinear divergence. The result is a basis of numerators for the boxes given by
\eqs{\label{eq:one_loop_boxes2}\hat{\mathfrak{n}}^e_{A,B,C,D}&\equivR\mathfrak{n}^e_{A,B,C,D}\\
\hat{\mathfrak{n}}^e_{\alpha,B,C,D}&\equivR\frac{1}{2}\left(\br{p_\alpha, b, c, C}{+}\br{b,c,p_C,p_\alpha}{+}\br{p_B,p_C}a^2{-}\br{p_{\alpha B},p_C}b^2\right),\\
\hat{\mathfrak{n}}^e_{\alpha,\beta,C,D}&\equivR\frac{1}{2}\left(\br{p_\alpha,b,c,p_C}{+}\br{b,c,p_C,p_\alpha}{-}\br{p_{\alpha B},p_C}b^2\right),\\
\hat{\mathfrak{n}}^e_{\alpha,B,\gamma,D}&\equivR\frac{1}{2}\left(\br{p_\alpha, b, c, p_\gamma}{+}\br{b,c,p_\gamma,p_\alpha}\right).}
Notice that we have used `$\,\hat{~}\,$'s' to disambiguate these numerators from those constructed in (\ref{eq:one_loop_boxes1}). 

By virtue of diagonalization implicit in the prescriptivity condition (\ref{prescriptivity_condition}), every even box integrand now automatically vanishes in all soft and/or collinear region of loop-momenta and is therefore IR-finite(!). Where do the divergences of amplitudes now go? The answer is that amplitudes in maximally supersymmetric theories, while vanishing on any contour taken at infinity, no longer vanish on the contours chosen for the triangle integrals. In particular, amplitudes always have support on the one-mass triangles' contours (the only ones responsible soft \emph{and} collinear  IR divergences)---with their leading singularities on these contours being simply the tree amplitude. 

Thus, amplitudes in maximally supersymmetric theories would take the form
\eq{\begin{split}
&\fwboxL{20pt}{\mathcal{A}\hspace{3pt}}\equivL\,\,\mathcal{A}^{\text{fin}}+\mathcal{A}^{\text{div}},\quad\text{where}\\
&\hspace{-00pt}\fwboxL{20pt}{\mathcal{A}^{\text{fin}}}\equivR\fwboxL{350pt}{\hspace{-5pt}\sum_{A,B,C,D}\frac{1}{2}\left(f^1_{A,B,C,D}{+}f^2_{A,B,C,D}\right)\hat{\mathcal{I}}^\text{e}_{A,B,C,D}+\frac{1}{2}\left(f^1_{A,B,C,D}{-}f^2_{A,B,C,D}\right)\hat{\mathcal{I}}^\text{o}_{A,B,C,D}\,,}\\
&\fwboxL{20pt}{\mathcal{A}^{\text{div}}}\equivR\sum_{\alpha,\beta,C}\mathcal{A}^{\text{tree}}\hat{\mathcal{I}}_{\alpha,\beta,C}\,\,.
\end{split}}
Notice that this is remarkably similar to the form of the `chiral box expansion' described in \cite{Bourjaily:2013mma}; in fact, the only distinction between the basis here and that of \cite{Bourjaily:2013mma} is that we have taken the even and odd combinations of chiral numerators.

\newpage\vspace{0pt}%
\section{Building a Non-Planar Integrand Basis at Two Loops}\label{sec:six_point_basis}\vspace{0pt}

To choose particular set of prescriptive integrand basis elements---for any power-counting, for any multiplicity---requires a spanning set of contours. As enumerated in \mbox{table \ref{enumeration_of_6pt_basis}} above, the complete, non-planar triangle power-counting basis for six external particles requires that we specify $373$ contours of integration. From only these, the integrand basis would be uniquely specified by the prescriptivity requirement (\ref{prescriptivity_condition}). However, this is easier said than done. 

To help illustrate how this can be done iteratively, with only minimal cleverness (or headache), it is worthwhile to describe how this can be done in stages:
\begin{enumerate}
\item For each integrand topology $\Gamma_I$, 
\begin{enumerate}
\item choose a spanning set of maximal-dimensional contours $\{\Omega_I^i\}$ to define its top-level numerators---each of which encircles all propagators of the given integrand topology.\\
The number of such numerators must equal the number of {\color{topCount}top-level} degrees of freedom in the integrand basis.
\begin{enumerate}
\item[\textbf{n.b.}] ensure the \emph{set} of contours is invariant under all graph isomorphisms.
\end{enumerate}
\item choose a \emph{spanning set} of \emph{initial} numerators $\{\hat{\mathfrak{n}}_I^i\}$ in the chosen integrand basis (with the desired power-counting) which is full-rank on the chosen contours; that is, make sure the \emph{period matrix}
\eq{\oint\limits_{\Omega_I^i}\mathcal{I}_I^{j}\equivL\, \mathbf{M}_{i\,j}}
is full-rank. 
\item diagonalize these \emph{initial} numerators to give `block-diagonal' {\color{topCount} top-level} numerators $\{\tilde{\mathfrak{n}}_I^i\}$; specifically, define 
\eq{\tilde{\mathfrak{n}}_{I}^i\equivR\, (\mathbf{M}^{-1})_{j\,i}\hat{\mathfrak{n}}_I^j\,.}
\end{enumerate}
The set of these numerators for each topology are now `\emph{block-diagonal}'---meaning that they are diagonal on their defining contours.
\item Diagonalize each numerator against the entire basis.\\
Provided all propagators of each graph are cut as part of each of its defining contours, then the only contours which need be checked are those which involve subsets of a given graph's propagators---that is, its daughter topologies (or `contact terms'). Because graph inclusion is triangular, diagonalization is analogous to diagonalizing an upper-triangular matrix. 
\begin{enumerate}
\item for each daughter topology $\Gamma_J\!\prec\!\Gamma_I$ determine the period integrals
\eq{\oint\limits_{\Omega_J^j}\tilde{\mathcal{I}}_I^{i}\equivL\,(\mathbf{M}_{IJ})_{i\,j}\,,}
and \emph{remove} these `contact terms' according to 
\eq{\tilde{\mathcal{I}}_I^i\mapsto\tilde{\mathcal{I}}_I^i-(\mathbf{M}_{IJ})_{i\,j}\tilde{\mathcal{I}}_{J}^j\,.}
Notice that each of these subtractions involves terms \emph{proportional to inverse propagators} appearing in $\Gamma_I$; thus, these subtractions have no effect on the periods of the integrands involving their own defining contours.\\~\\
Provided this is done iteratively, starting with the graphs with the fewest daughters, this process is guaranteed to result in numerators which are globally diagonal---and hence, a basis which is \emph{prescriptive}. 
\end{enumerate}
\end{enumerate} 

Of course, once contours have been chosen for all topologies (step 1 above), the resulting basis according to the prescriptivity condition (\ref{prescriptivity_condition}) will be unique. However, the process above makes this much more manageable---with the most artful step being the choice of initial numerators. 

In this section, we'd like to walk the reader through how this was done in the case of $\mathfrak{B}_3$ for the case of six external particles at two loops.

\vspace{4pt}%
\subsection{Identifying Candidate Numerators and Contours}\label{subsec:outline}\vspace{0pt}

The starting point in our construction of a basis of two-loop, triangle power-counting integrands in four dimensions is the enumeration of graph topologies and the counting of the number of independent degrees of freedom for which contours must be specified. This data was generated in the recent work \cite{Bourjaily:2020qca} and has been summarized in \mbox{table \ref{4d_3_gon_pc_graphs}} of section~\ref{subsec:review_of_size_of_basis}. 

For the case of six-particles, varying all the particular distributions of external legs, we find a total of 87 integrand topologies as enumerated in \mbox{table \ref{enumeration_of_6pt_basis}}. Because each vector space of numerators can be spanned with (sums of) products of generalized inverse propagators, it is straightforward to construct a complete---albeit far from diagonal---initial basis of integrands. 

As a representative example, consider a typical non-planar double pentagon, or $\Gamma_{[\r{3},\g{2},\b{3}]}$ topology (integrand \#11 in our list):
\eq{\label{eg_double_pent}\fwbox{0pt}{\hspace{10pt}\fwbox{10pt}{\dPent{\dPentScalarEdges\dPentLegs{1}{1}{0}{1}{1}{1}{1}\intDots{7}}} \hspace{45pt}  \mathfrak{N}_3(\Gamma_{[\r{3},\g{2},\b{3}]})=\text{span}\left\{\r{[\ell_A]^3}{\oplus}\r{[\ell_A]^2}\b{[\ell_C]}{\oplus}\r{[\ell_A]}\b{[\ell_C]^2}{\oplus}\r{[\ell_A]}\g{[\ell_B]}\b{[\ell_C]}\right\}.}}
Here, $\r{[\ell_A]}\!=\!\r{[a]}\!=\!\r{[b]}\!=\!\r{[c]}$, $\g{[\ell_B]}\!=\!\g{[g]}\!=\!\g{[h]}$, and $\b{[\ell_C]}\!=\!\b{[d]}\!=\!\b{[e]}\!=\!\b{[f]}$. As indicated in \mbox{table \ref{enumeration_of_6pt_basis}}, this vector space of numerators for this topology has {\color{totalCount}229 total} degrees of freedom, of which ${\color{topCount}8}$ {\color{topCount} top-level} and the rest (221) are contact terms. A random choice of {\color{topCount}8} elements involving inverse-propagators \emph{not} manifestly included in the graph would likely span the space of top-level numerators (and therefore suffice), but would be very far from diagonal in the contours chosen (or in any illuminating form). 

Even before discussing the choices for {\color{topCount}8} contours required to specify these top-level degrees of freedom, it is worthwhile to build some intuition about what numerators may be \emph{close} to diagonal. Conveniently, for any eight-propagator graph at two loops ($\Gamma_{[\r{4},\g{0},\b{4}]},\Gamma_{[\r{4},\g{1},\b{3}]},\Gamma_{[\r{4},\g{2},\b{2}]}, \Gamma_{[\r{3},\g{2},\b{3}]}$ in \mbox{table \ref{enumeration_of_6pt_basis}}---topologies indexed by $I\!\in\!\{1,\ldots,12\}$ in our basis), the number of {\color{topCount}top-level}degrees of freedom \emph{exactly} matches the number of solutions to the maximal-cut equations which put all eight propagators on-shell. Thus, there is a one-to-one correspondence between these cut configurations and initial top-level numerators $\tilde{\mathfrak{n}}_{11}^{i}$ for $i\!\in\!\{1,\ldots,8\}$.  

To be slightly pedantic, it is worth remembering that our first step is merely to find \emph{initial} `block-diagonal' numerators---which give integrands which satisfy
\eq{\oint\limits_{\Omega_{11}^j}\!\tilde{\mathcal{I}}_{11}^{i}=\delta_{i\,j}\,.\label{top_level_diagonality}}
Such integrands are not yet \emph{prescriptive}---as they may have support on the contours involving subsets of propagators (those used to define the top-level degrees of freedom of daughter (`contact term') integrand topologies). It is obvious that (\ref{top_level_diagonality}) is unchanged by the addition of any such contact terms. And so we still have work to do before we have fully diagonal---prescriptive---integrands. 

\vspace{4pt}%
\subsection{\textbf{Step 1}. Guessing a Spanning Set of Top-Level Numerators}\label{subsec:choosing_initial_numerators}\vspace{0pt}

As emphasized in section~\ref{sec:one_loop_review}, once a spanning set of contours has been chosen, this uniquely fixes every integrand in the basis. This means that in principle the choice of an initial set of spanning numerators for a given topology is unimportant, as the diagonalized numerators are entirely subservient to the contours. In practice, however, carrying out the diagonalization procedure is significantly easier if some thought is given to the initial, pre-diagonal basis of numerators. In particular, the chiral numerators appearing throughout \cite{Bourjaily:2019iqr,Bourjaily:2019gqu} often serve as excellent starting points, even when they require contact term corrections to be rendered diagonal with respect to subtopologies. 

There is little subtlety (or mystery) in the construction of `nice' chiral numerators for the eight-propagator integrands in our basis. To illustrate this, let us continue the double-pentagon example of (\ref{eg_double_pent}). In the notation of the ancillary files, this graph is numbered $\#11$, and has the topology shown above:
\eq{\label{eg_double_pent_6pt}\fwboxR{0pt}{\mathcal{I}_{11}\leftrightarrow}\hspace{0pt}\fwbox{90pt}{\dPent{\dPentScalarEdges\dPentLegs{1}{1}{0}{1}{1}{1}{1}\intDots{7}}}}
As mentioned above, the eight solutions to the cut equations correspond to the eight leading singularities associated to this graph. A block-diagonal set of numerators can be systematically constructed by the subspace of $\mathfrak{N}_3(\Gamma_{[\r{3},\g{2},\b{3}]})$ \emph{not} spanned by this graph's contact terms. In practice, however, it is often easier to guess a representative numerator by considering the kinematic conditions imposed on the on-shell loop momenta for each. For the topology (\ref{eg_double_pent_6pt}), the top-level degrees of freedom $\{\tilde{\mathfrak{n}}_{11}^i\}$---which, we stress, are only block-diagonal, and \emph{not} yet diagonal with respect to this topology's daughter graphs---can be chosen by inspection of the solutions to the cut equations.

For most cases of interest, the particular solutions to the cut equations can be identified by the parity of its three-point vertices---with {\color{mhvblue2}blue} for the `mhv' solution (all $\tilde\lambda$'s proportional) and white for the `$\overline{\text{mhv}}$' solution (all $\lambda$'s proportional). When a vertex has multiplicity $\!>\!3$, no such parity can be indicated---so black vertices are used in our contour diagrams. 

For example, we may identify and number the 8 solutions to the maximal-cut equations for the topology $\mathcal{I}_{11}$ in (\ref{eg_double_pent_6pt}) as follows:
\vspace{0pt}\eq{\fwbox{0pt}{\hspace{-22pt}\begin{array}{@{}c@{}c@{}c@{}c@{}}\fwboxR{0pt}{\left\{\rule[-10pt]{0pt}{50pt}\right.\hspace{-4pt}}\dPent{\draw[int](v1)--(v2);\draw[int](v2)--(v3);\draw[int](v3)--(v4);\draw[int](v4)--(v5);\draw[int](v5)--(v6);\draw[int](v6)--(v1);\draw[int](v6)--(v7);\draw[int](v7)--(v3);\dPentLegs{1}{1}{0}{1}{1}{1}{1}\contourVerts{1}{2}{1}{1}{2}{4}{2}},
&\dPent{\draw[int](v1)--(v2);\draw[int](v2)--(v3);\draw[int](v3)--(v4);\draw[int](v4)--(v5);\draw[int](v5)--(v6);\draw[int](v6)--(v1);\draw[int](v6)--(v7);\draw[int](v7)--(v3);\dPentLegs{1}{1}{0}{1}{1}{1}{1}\contourVerts{1}{2}{1}{2}{1}{4}{2}},
&\dPent{\draw[int](v1)--(v2);\draw[int](v2)--(v3);\draw[int](v3)--(v4);\draw[int](v4)--(v5);\draw[int](v5)--(v6);\draw[int](v6)--(v1);\draw[int](v6)--(v7);\draw[int](v7)--(v3);\dPentLegs{1}{1}{0}{1}{1}{1}{1}\contourVerts{1}{2}{2}{1}{2}{4}{1}},
&\dPent{\draw[int](v1)--(v2);\draw[int](v2)--(v3);\draw[int](v3)--(v4);\draw[int](v4)--(v5);\draw[int](v5)--(v6);\draw[int](v6)--(v1);\draw[int](v6)--(v7);\draw[int](v7)--(v3);\dPentLegs{1}{1}{0}{1}{1}{1}{1}\contourVerts{1}{2}{1}{2}{1}{4}{1}},\\
\dPent{\draw[int](v1)--(v2);\draw[int](v2)--(v3);\draw[int](v3)--(v4);\draw[int](v4)--(v5);\draw[int](v5)--(v6);\draw[int](v6)--(v1);\draw[int](v6)--(v7);\draw[int](v7)--(v3);\dPentLegs{1}{1}{0}{1}{1}{1}{1}\contourVerts{2}{1}{2}{1}{2}{4}{2}},
&\dPent{\draw[int](v1)--(v2);\draw[int](v2)--(v3);\draw[int](v3)--(v4);\draw[int](v4)--(v5);\draw[int](v5)--(v6);\draw[int](v6)--(v1);\draw[int](v6)--(v7);\draw[int](v7)--(v3);\dPentLegs{1}{1}{0}{1}{1}{1}{1}\contourVerts{2}{1}{1}{2}{1}{4}{2}},
&\dPent{\draw[int](v1)--(v2);\draw[int](v2)--(v3);\draw[int](v3)--(v4);\draw[int](v4)--(v5);\draw[int](v5)--(v6);\draw[int](v6)--(v1);\draw[int](v6)--(v7);\draw[int](v7)--(v3);\dPentLegs{1}{1}{0}{1}{1}{1}{1}\contourVerts{2}{1}{2}{1}{2}{4}{1}},
&\dPent{\draw[int](v1)--(v2);\draw[int](v2)--(v3);\draw[int](v3)--(v4);\draw[int](v4)--(v5);\draw[int](v5)--(v6);\draw[int](v6)--(v1);\draw[int](v6)--(v7);\draw[int](v7)--(v3);\dPentLegs{1}{1}{0}{1}{1}{1}{1}\contourVerts{2}{1}{2}{2}{1}{4}{1}}\fwboxL{0pt}{\hspace{-4pt}\left.\rule[-10pt]{0pt}{50pt}\right\}}
\end{array}}\label{omega11_contours}}
These figures represent contours $\{\Omega_{11}^{1},\ldots,\Omega_{11}^{8}\}$. 

It is easier than it may at first appear to construct a numerator which has support on each of the corresponding contours. For example, to construct a numerator which vanishes on all but the first contour, one need only require that it vanish whenever the vertices have the wrong parity. For example, consider the tentative numerator 
\eq{\tilde{\mathfrak{n}}_{11}^1\equivR\br{p_1,b,c,h,g,p_4,e,d}\,;}
this numerator vanishes whenever $\lambda_1\!\propto\!\lambda_b$ or $\lambda_c\!\propto\!\lambda_h$ or when $\lambda_e\!\propto\!\lambda_d$---that is, on all cases where the vertex involving $p_1$, $p_3$ or the top-middle vertex is colored blue, at least one of which is the case for every contour \emph{except} $\Omega_{11}^1$. (To see this, recall the definition of ``$\br{\cdots}$'' in (\ref{definition_of_br}).) Moreover, it is not hard to verify that this numerator gives rise to an integrand which integrates to 1 on $\Omega_{11}^1$. That is, 
\eq{\oint\limits_{\Omega_{11}^i}\tilde{\mathcal{I}}_{11}^1\equivR\oint\limits_{\Omega_{11}^i}\!\dbar^4\ell_1\,\dbar^4\ell_2\,\frac{\tilde{\mathfrak{n}}_{11}^1}{a^2\,b^2\,c^2\,d^2\,e^2\,f^2\,g^2\,h^2}=\left\{\begin{array}{@{}l@{$\;\;$}r@{}}1&i=1\\0&i\neq 1\end{array}\right.\,.}

Continuing in this way results in an initial `block-diagonal' set of numerators for topology \#11. Specifically, we find that the following numerators are diagonal on the contours for $\Omega_{11}$:
\eq{\begin{array}{@{}l@{}l@{}l@{$\;\;\;\;\;\;\;$}l@{}l@{}l@{}l@{}}\tilde{\mathfrak{n}}_{11}^1&\equivR&\phantom{{-}}\br{p_1,b,c,h,g,p_4,e,d}\,,
&\tilde{\mathfrak{n}}_{11}^{5}&\equivR&\frac{1}{2}\big(\br{e,d,c,b,p_1,g,h,p_4}\\
\tilde{\mathfrak{n}}_{11}^2&\equivR&\frac{1}{2}\big(\br{p_1,b,c,h,g,p_3,e,f}
&&&{-}\br{b,c,d,e,p_4,g,h,p_1}\big)\,,\\
&&{-}\br{p_4,e,d,h,g,p_2,b,a}\big)\,,&\tilde{\mathfrak{n}}_{11}^6&\equivR&{-}\br{p_4,e,d,h,g,p_1,b,c}\,,\\
\tilde{\mathfrak{n}}_{11}^3&\equivR&\phantom{{-}}\br{e,d,h,g,a,b,p_2,p_4}\,,&\tilde{\mathfrak{n}}_{11}^7&\equivR&\frac{1}{2}\big(\br{f,p_1,b,c,h,g,p_3,e}\\
\tilde{\mathfrak{n}}_{11}^4&\equivR&\frac{1}{2}\big(\br{p_4,e,d,c,b,p_1,g,h}
&&&{-}\br{a,p_4,e,d,h,g,p_2,b}\big)\,,\\
&&{-}\br{p_1,b,c,d,e,p_4,g,h}\big)\,,&\tilde{\mathfrak{n}}_{11}^8&\equivR&{-}\br{b,c,h,g,f,e,p_3,p_1}\,.
\end{array}\label{omega11_numerators}}
In cases where these numerators have factors of 2, this reflects contour-graph isomorphisms in (\ref{omega11_contours}). It is not hard to verify that these numerators are all unit on their corresponding contour, and vanish on all the others. Thus, these numerators provide good starting points for further diagonalization (against daughter-topology contours).\\[-5pt]

Since all eight-propagator integrands have exactly the same number of solutions to the cut equations as numerator degrees of freedom, this procedure can be repeated without any subtlety whatsoever for every such topology. 

There is a single case at six particles, however, where coloring of the vertices alone does not suffice to distinguish all solutions to the maximal cut equations. The exceptional case involves topology \#10 in our basis: 
\eq{\label{eg_double_pent_6pt_special}\fwboxR{0pt}{\mathcal{I}_{10}\leftrightarrow}\hspace{00pt}\fwbox{90pt}{\dPent{\dPentScalarEdges\dPentLegs{1}{1}{0}{1}{1}{0}{2}\intDots{7}}}.}
For two sets of `colorings' of this graph (contours encoded by the parity of the three-point vertices), there are two pairs of solutions to the cut equations related to the choice of sign in front of the square root of $\Delta^2\equivR\br{p_{12},p_{34}}^2-4s_{12}s_{34}$. In each case, we may match these two leading singularities with numerators that are chiral, or parity-even and parity-odd; throughout this work, we always choose the latter.  Thus, for example, we choose the following pair of contours:
\eq{\begin{split}&\hspace{-85pt}\Omega_{10}^2\equivR\hspace{-5pt}\dPent{\draw[int](v1)--(v2);\draw[int](v2)--(v3);\draw[int](v3)--(v4);\draw[int](v4)--(v5);\draw[int](v5)--(v6);\draw[int](v6)--(v1);\draw[int](v6)--(v7);\draw[int](v7)--(v3);\dPentLegs{1}{1}{0}{1}{1}{0}{2}\contourVerts{1}{2}{1}{2}{1}{2}{4}\node at (0,-1.15) {\footnotesize{(odd)}};}\hspace{-5pt}\leftrightarrow\tilde{\mathfrak{n}}_{10}^2\equivR{-}\frac{1}{2}\br{e,p_4,p_1,b,c,d}\Delta\,;\\
&\hspace{-85pt}\Omega_{10}^4\equivR\hspace{-5pt}\dPent{\draw[int](v1)--(v2);\draw[int](v2)--(v3);\draw[int](v3)--(v4);\draw[int](v4)--(v5);\draw[int](v5)--(v6);\draw[int](v6)--(v1);\draw[int](v6)--(v7);\draw[int](v7)--(v3);\dPentLegs{1}{1}{0}{1}{1}{0}{2}\contourVerts{1}{2}{1}{2}{1}{2}{4}\node at (0,-1.15) {\footnotesize{(even)}};}\hspace{-5pt}\leftrightarrow\,\tilde{\mathfrak{n}}_{10}^4\equivR\begin{array}{@{}l@{}}~\\\frac{1}{4}\big(\br{g,p_2,b,a,f,e,p_3,h}{-}\br{p_1,b,c,d,e,p_4,g,h}\\
\hspace{2.75pt}{+}\br{p_4,e,d,c,b,p_1,g,h}{-}\br{g,p_3,e,f,a,b,p_2,h}\big)\,.\end{array}\hspace{-100pt}
\end{split}\label{even_odd_examples}}
\textbf{Nota bene}: the labels `even' and `odd' do \emph{not} refer to parity: they refer to the \emph{even} or \emph{odd} sum of the multiple solutions to the corresponding contour. In addition to these, there are two more cases which have been split into even/odd combinations; these are related to those of (\ref{even_odd_examples}) by an up/down flip of the graphs.

\vspace{4pt}\subsection{\textbf{Step 2}. Block-Diagonalization with Respect to a Choice of Contours}\label{subsec:block_diagonalization}\vspace{0pt}

While the contour specification for eight-propagator integrand topologies is quite rigid, for integrands with fewer propagators there is significantly more freedom. By virtue of the requirement that the basis be diagonal in contours, these choices have significant effects on the rest of the basis, including the contact terms of the eight-propagator integrands. A simple one-loop analogue of this freedom is in the normalization of the (scalar) triangle integrand. In sections \ref{subsubsec:triangle_power_counting_at_infinity} and \ref{subsubsec:triangle_power_counting_in_IR} we emphasized that the form of the triangle power-counting basis, as well as what features were made manifest, depended heavily on whether the (one and two-mass) triangles were normalized at infinite loop momentum or in the soft and/or collinear regions. This phenomenon is even more prominent at two loops, where every integrand topology with fewer than eight propagators requires a choice of contours which has trickle-up (and down) effects on the rest of the basis.\\[-5pt]

The conventions for our contour choices are as follows. Whenever an integrand topology is such that soft and/or collinear residues are accessible, we choose a corresponding contour as part of our spanning set. When there are no such contours (analogous to the three-mass triangle at one loop) we choose contours involving (at least) one loop-momentum cycle in the graph being sent to infinity. In addition, when the scalar graph topology has non-trivial graph automorphisms, our contour choices are such that the entire \emph{set} (but not necessarily the individual contours themselves) is invariant up to overall signs. We shall illustrate the r\^{o}le of graph symmetries in greater detail below. 

The motivation for our preference for soft and/or collinear contours is simple: we aim to construct a basis of integrands maximally stratified according to IR finiteness or divergence. We expect that every integrand \emph{not} normalized on a soft or collinear contour will vanish entirely in every such region of loop momentum space, which is highly suggestive of it being infrared finite upon integration. In the representation of an amplitude, our basis is designed to match the IR-divergence of the amplitude manifestly, while every other integrand with a non-vanishing coefficient vanishing (by construction) in all collinear regions. 

\vspace{-1pt}\subsubsection{Requiring the Set of Contours be Graph Symmetric}\label{subsubsec:graph_symmetries}
\vspace{0pt}
The attentive reader may have noted that in the top-level numerators of the non-planar double pentagon example (\ref{omega11_numerators}), several of the numerators had `symmetry factors' in their definition. In this case, it is natural---though not essential---to use numerators which respect the symmetries of the leading singularity graphs associated with the contours. For contours involving composite conditions, wherein at least one set of momenta are either soft and/or collinear, our prescriptive basis of integrands always respects the symmetries of the composite leading singularities. Provided this is done throughout the basis for leading singularities with amplitude support in sYM, any such amplitude integrand's representation is simple to write down---namely, in the diagonal basis, the integrand will be a sum of inequivalent leading singularities, each decorated by the corresponding basis element \cite{Bourjaily:2017wjl}.

In addition to contours which have amplitude-support in sYM, there are many degrees of freedom in our basis normalized on contours involving infinite loop momenta. In these cases, our convention is to impose a slightly less restrictive symmetry constraint: we require only that the \emph{set} of numerators associated with contours defined at infinity to be closed under the automorphism group of the graph.

To explain our conventions in further detail, let us consider an example in which the automorphism group is non-trivial. Among others, we have a double-box topology $\#17$ with eight degrees of freedom
\vspace{-4pt}\eq{\label{eq:double_box_example}\fwbox{0pt}{\fwboxR{0pt}{\mathcal{I}_{17}\leftrightarrow}\dBox{\dBoxScalarEdges\dBoxLegs{1}{1}{0}{1}{1}{2}\intDots{6}}\fwboxL{0pt}{,}}\vspace{-4pt}}
two contours of which are defined as even combinations of contours taken by starting from a heptacut and sending one loop-momentum cycle to infinity:
\eqs{\label{eq:contours_at_infinity_example}&\fwbox{0pt}{\hspace{-10pt}\Omega_{17}^4\equivR\left\{\dBox{\dBoxInfB\draw[int](v1)--(v2);\draw[int](v2)--(v3);\draw[int](v3)--(v4);\draw[int](v4)--(v5);\draw[int](v5)--(v6);\draw[int](v6)--(v1);\draw[int](v6)--(v3);\dBoxLegs{1}{1}{0}{1}{1}{2}\contourVerts{1}{2}{1}{2}{1}{4}{5}}+\dBox{\dBoxInfA\draw[int](v1)--(v2);\draw[int](v2)--(v3);\draw[int](v3)--(v4);\draw[int](v4)--(v5);\draw[int](v5)--(v6);\draw[int](v6)--(v1);\draw[int](v6)--(v3);\dBoxLegs{1}{1}{0}{1}{1}{2}\contourVerts{1}{2}{1}{2}{1}{4}{5}}\right\}\leftrightarrow \textcolor{purple}{\tilde{\mathfrak{n}}_{17}^4\equivR\frac{1}{2}\br{c,d,e,p_4,p_1,b}}\,,}\\
&\fwbox{0pt}{\hspace{-10pt}\Omega_{17}^8\equivR\left\{\dBox{\dBoxInfB\draw[int](v1)--(v2);\draw[int](v2)--(v3);\draw[int](v3)--(v4);\draw[int](v4)--(v5);\draw[int](v5)--(v6);\draw[int](v6)--(v1);\draw[int](v6)--(v3);\dBoxLegs{1}{1}{0}{1}{1}{2}\contourVerts{2}{1}{2}{1}{2}{4}{5}}+\dBox{\dBoxInfA\draw[int](v1)--(v2);\draw[int](v2)--(v3);\draw[int](v3)--(v4);\draw[int](v4)--(v5);\draw[int](v5)--(v6);\draw[int](v6)--(v1);\draw[int](v6)--(v3);\dBoxLegs{1}{1}{0}{1}{1}{2}\contourVerts{2}{1}{2}{1}{2}{4}{5}}\right\}\leftrightarrow \textcolor{hblue}{\tilde{\mathfrak{n}}_{17}^8\equivR\frac{1}{2}\br{b,c,d,e,p_4,p_1}}\,.}
}
In addition to these, there are four degrees of freedom normalized on soft contours associated with either the momentum through edges $b$ or $e$ to zero. These contours---and corresponding numerators which are unit on them---are easy to identify:
\eqs{\label{eq:block_diagonal_nums_example}
&\fwboxR{10pt}{\Omega_{17}^1\equivR}\dBox{\draw[int](v2)--(v3);\draw[int](v3)--(v4);\draw[int](v4)--(v5);\draw[int](v5)--(v6);\draw[int](v6)--(v1);\draw[int](v6)--(v3);\draw[dashed](v1)--(v2);\dBoxLegs{1}{1}{0}{1}{1}{2}\contourVerts{0}{0}{2}{1}{2}{4}{5}}\leftrightarrow \tilde{\mathfrak{n}}_{17}^1\equivR s_{12}\br{e,d,c,p_4}\textcolor{hblue}{+\frac{1}{2}\br{b,c,d,e,p_4,p_1}}\,,\\
&\fwboxR{10pt}{\Omega_{17}^5\equivR}\dBox{\draw[int](v2)--(v3);\draw[int](v3)--(v4);\draw[int](v4)--(v5);\draw[int](v5)--(v6);\draw[int](v6)--(v1);\draw[int](v6)--(v3);\draw[dashed](v1)--(v2);\dBoxLegs{1}{1}{0}{1}{1}{2}\contourVerts{0}{0}{1}{2}{1}{4}{5}}\leftrightarrow \tilde{\mathfrak{n}}_{17}^5\equivR s_{12}\br{d,c,p_4,e}\textcolor{purple}{+\frac{1}{2}\br{c,d,e,p_4,p_1,b}}\,,\\
&\fwboxR{10pt}{\Omega_{17}^3\equivR}\dBox{\draw[int](v1)--(v2);\draw[int](v2)--(v3);\draw[int](v3)--(v4);\draw[int](v5)--(v6);\draw[int](v6)--(v1);\draw[int](v6)--(v3);\draw[dashed](v4)--(v5);\dBoxLegs{1}{1}{0}{1}{1}{2}\contourVerts{1}{2}{1}{0}{0}{4}{5}}\leftrightarrow \tilde{\mathfrak{n}}_{17}^3\equivR s_{34}\br{c,d,p_1,b}\textcolor{purple}{-\frac{1}{2}\br{d,c,b,p_1,p_4,e}}\,,\\
&\fwboxR{10pt}{\Omega_{17}^7\equivR}\dBox{\draw[int](v1)--(v2);\draw[int](v2)--(v3);\draw[int](v3)--(v4);\draw[int](v5)--(v6);\draw[int](v6)--(v1);\draw[int](v6)--(v3);\draw[dashed](v4)--(v5);\dBoxLegs{1}{1}{0}{1}{1}{2}\contourVerts{2}{1}{2}{0}{0}{4}{5}}\leftrightarrow \tilde{\mathfrak{n}}_{17}^7\equivR s_{34}\br{b,c,d,p_1}\textcolor{hblue}{-\frac{1}{2}\br{e,d,c,b,p_1,p_4}}\,.
}
The appearance of the colored numerators in the expressions (\ref{eq:block_diagonal_nums_example}) are needed in order for the combined numerators to vanish on the defining contours in (\ref{eq:contours_at_infinity_example}). The rest of the eight-dimensional basis of numerators is furnished by contours defined by starting from a hepta-cut and imposing collinearity at the vertex involving edges $\{c,d,g\}$,
\eqs{
&\hspace{-100.5pt}\fwboxR{10pt}{\Omega_{17}^2\equivR}\dBox{\draw[int](v1)--(v2);\draw[int](v2)--(v3);\draw[int](v3)--(v4);\draw[int](v4)--(v5);\draw[int](v5)--(v6);\draw[int](v6)--(v1);\draw[int](v6)--(v3);\dBoxLegs{1}{1}{0}{1}{1}{2}\contourVerts{1}{2}{3}{1}{2}{4}{5}}\leftrightarrow \tilde{\mathfrak{n}}_{17}^2\equivR-\br{e,d,a,b,p_2,p_4}\,,\\
&\hspace{-100.5pt}\fwboxR{10pt}{\Omega_{17}^6\equivR}\dBox{\draw[int](v1)--(v2);\draw[int](v2)--(v3);\draw[int](v3)--(v4);\draw[int](v4)--(v5);\draw[int](v5)--(v6);\draw[int](v6)--(v1);\draw[int](v6)--(v3);\dBoxLegs{1}{1}{0}{1}{1}{2}\contourVerts{2}{1}{3}{2}{1}{4}{5}}\leftrightarrow \tilde{\mathfrak{n}}_{17}^6\equivR-\br{p_4,e,d,a,b,p_2}\,.\\
}

\newpage\subsubsection{Contours Involving Double-Poles}\label{subsubsec:double_pole_discussion}
\vspace{0pt}
There is one additional complication regarding contour choices: for triangle power-counting, there are an \emph{insufficient} number of maximal co-dimension residues of logarithmic type to fill out the basis. Said differently, the basis of integrands cannot be spanned by polylogarithmic integrals of uniform and \emph{maximal} transcendental weight. Instead, there are some degrees of freedom must be normalized on contours involving double-poles. This phenomenon first appears at the six-propagator level, and is exemplified by considering the following example, $\#47$ in our basis:
\eq{\fwboxR{0pt}{\mathcal{I}_{47}\leftrightarrow}\hspace{0pt}\fwbox{90pt}{\bT{\bTScalarEdges\bTLegs{1}{1}{0}{4}{0}\intDots{5}}\fwboxL{0pt}{\hspace{10pt}.}}}
For triangle power-counting, this topology requires a choice of three contours. However, in this case, due to the degenerate kinematics where only a single vertex is massive, there is only a single independent contour which is logarithmic. We can see detect the presence of double-poles in this topology's cut structure by considering a hexa-cut where all six propagators are on-shell. One solution to the cut equations may be parametrized in a basis of spinors $\{\lambda_1,\tilde{\lambda}_1,\lambda_2,\tilde{\lambda}_2\}$ as,
\eq{\label{eq:double_pole_example}\hspace{10pt}\fwbox{20pt}{\bT{\bTScalarEdges\bTLegs{1}{1}{0}{4}{0}\intDots{5}\contourVerts{2}{1}{2}{4}{1}{5}{5}}}\hspace{40pt} \leftrightarrow b^\star=\alpha \lambda_2\tilde{\lambda}_1\,,\quad d^\star=\left[\beta\lambda_1+\alpha(1-\beta)\lambda_2\right]\left(\tilde{\lambda}_1+\frac{1}{\alpha}\tilde{\lambda}_2\right)\,,\hspace{-20pt}}
where the Jacobian of the hexa-cut is $J{=}s_{12}^2\alpha\beta$. Trivially, we see that the scalar numerator $\tilde{\mathfrak{n}}_{47}^1\!=\!s_{12}^2$ yields a logarithmic two-form $d\log \alpha \,d\log \beta$, and the corresponding contour may be normalized at some combination of $\alpha,\beta\!\rightarrow\!\{0,\infty\}$. However, it is also easy to see that there is no other independent contour (accessible using a loop-momentum dependent numerator) which does not involve a double-pole (at infinity) in either $\alpha$ or $\beta$. Indeed, to span the full space of triangle power-counting numerators for this topology, we are forced to introduce two double-pole numerators $\tilde{\mathfrak{n}}_{47}^{2,3}$. Since the power-counting allows an $\ell_1$-dependent numerator, we see that any integrand proportional to $\alpha$ on the cut (\ref{eq:double_pole_example}) will have a double-pole in the (ordered) limit $\beta,\alpha\!\rightarrow\!\infty$; these two conditions, together with the hexacut, define the double-pole contour. The parity conjugate of this contour fills out the rest of the basis, and a block-diagonal set of numerators is:
\eqs{&\fwboxR{10pt}{\Omega_{47}^1\equivR}\bT{\draw[int](v2)--(v3);\draw[int](v3)--(v4);\draw[int](v4)--(v5);\draw[int](v5)--(v1);\draw[dashed](v1)--(v2);\draw[dashed](v5)--(v3);\bTLegs{1}{1}{0}{4}{0}\contourVerts{0}{0}{0}{4}{0}{5}{5}}\leftrightarrow \tilde{\mathfrak{n}}_{47}^1\equivR s_{12}\br{c,p_{12}}\,,\\[5pt]
&\fwboxR{10pt}{\Omega_{47}^2\equivR}\bT{\bTInfAB\draw[int](v1)--(v2);\draw[int](v2)--(v3);\draw[int](v3)--(v4);\draw[int](v4)--(v5);\draw[int](v5)--(v1);\draw[int](v5)--(v3);\bTLegs{1}{1}{0}{4}{0}\contourVerts{1}{2}{1}{4}{2}{5}{5}\node at (0,-1.15) {\footnotesize{(double-pole)}};} \leftrightarrow \tilde{\mathfrak{n}}_{47}^2\equivR\frac{1}{2}\left(\br{a,p_1,p_2,p_{45}}+\br{p_{2},p_1,p_{45},c}\right)\,,\\[5pt]
&\fwboxR{10pt}{\Omega_{47}^3\equivR}\bT{\bTInfAB\draw[int](v1)--(v2);\draw[int](v2)--(v3);\draw[int](v3)--(v4);\draw[int](v4)--(v5);\draw[int](v5)--(v1);\draw[int](v5)--(v3);\bTLegs{1}{1}{0}{4}{0}\contourVerts{2}{1}{2}{4}{1}{5}{5}\node at (0,-1.15) {\footnotesize{(double-pole)}};}\leftrightarrow \tilde{\mathfrak{n}}_{47}^3\equivR\frac{1}{2}\left(\br{c,p_2,p_1,p_{45}}+\br{p_1,p_2,p_{45},a}\right)\,.
}
While the appearance of double-pole degrees of freedom may seem avoidable in the previous example, at the five propagator level there are, in fact, topologies where there exists \emph{no} logarithmic contours whatsoever. An example of this is $\#85$ in our basis, the $\Gamma_{[\r{2},\g{1},\b{2}]}$ topology:
\eq{\label{eq:double_pole_dt}\fwboxR{0pt}{\mathcal{I}_{85}\leftrightarrow}\fwbox{90pt}{\dT{\dTScalarEdges\dTLegs{2}{0}{4}{0}\intDots{4}}\fwboxL{0pt}{.}}}
The only allowed numerator for triangle power-counting is a scalar and---regardless of its normalization---the resulting integrand has double-poles in its cut structure (indicative of a drop in transcendental weight upon integration). In particular, a contour defined by imposing collinearity at both three-point vertices in (\ref{eq:double_pole_dt}) and sending the loop momenta to infinity may be accessed by starting from the co-dimension six configuration
\eq{\label{eq:double_pole_dt_contour}a=d=\left(\alpha \lambda_1+\beta\lambda_2\right)\left[\tilde{\lambda}_1-\left(\frac{\alpha+1}{\beta}\right)\tilde{\lambda}_2\right],\quad J=-s_{12}\beta\,.}
Upon taking a residue at $\beta\rightarrow\infty$, the integrand with scalar numerator $s_{12}$ evaluates on (\ref{eq:double_pole_dt_contour}) to 
\eq{\underset{(\ref{eq:double_pole_dt_contour})}{\mathrm{Res}}\left(\mathcal{I}_{85}^1\right)=\frac{-d\alpha d\beta}{\beta}\,.}
After taking an additional residue at $\beta\!\rightarrow\!\infty$, we may choose to normalize this degree of freedom to be $1$ at the double-pole at $\alpha\!\rightarrow\!\infty$. In fact, however, this contour prescription does not enjoy the up-down flip symmetry of the scalar graph in (\ref{eq:double_pole_dt}). Using a contour which is compatible with the symmetries of the scalar graph produces a symmetry factor in the corresponding numerator, which is $\mathfrak{n}_{85}^1\equivR\frac{1}{2}s_{12}$. (Note that this is indeed the globally diagonal numerator because this integrand has no contact terms whatsoever, which explains why we have omitted the $\sim$ label above $\mathfrak{n}_{85}^1$.) Diagrammatically, we indicate this contour as: 
\eq{\label{eq:int_85_contour_picture}\fwboxR{0pt}{\Omega_{85}^1\equivR}\dT{\dTInfAB\draw[int](v1)--(v2);\draw[int](v2)--(v3);\draw[int](v3)--(v4);\draw[int](v4)--(v1);\draw[dashed](v4)--(v2);\dTLegs{2}{0}{4}{0}\contourVerts{4}{0}{4}{0}{5}{5}{5}\node at (0,-1.15) {\footnotesize{(odd) double-pole}};} \,\,.}

\vspace{4pt}\subsection{\textbf{Step 3}. Global Diagonalization of the Basis}\label{subsec:complete_diagonalization}\vspace{0pt}
The numerators obtained as described in the previous subsection~\ref{subsec:choosing_initial_numerators} are locally-diagonal, and provide an excellent starting point in the construction of a complete, globally-diagonal basis. If we denote the set of block-diagonal integrands for topology $I$ as $\{\tilde{\mathcal{I}}_I^i\}$ and the corresponding contours as $\{\Omega_I^i\}$, where $i$ is an index running over the number of {\color{topCount}top-level} degrees of freedom in triangle power-counting basis for the $I$th topology, then by construction we have
\eq{\oint\limits_{\Omega_I^j}\tilde{\mathcal{I}}_I^i=\delta_{i,j}.}
However, generically the integrands of the parent topologies will still have support on the contours of their daughters (the set of graphs generated by contracting some number of internal edges). That is, the complete top-level and contact term period matrix is of the form
\eq{\label{eq:partially_diagonal}\oint\limits_{\Omega_J^j}\tilde{\mathcal{I}}_I^i\equivL\left\{\begin{array}{lr} \delta_{i\,j} & I=J, \\
(\mathbf{M}_{I\,J})_{i\,j} & I\neq J. \end{array}\right.}
By contrast, in a globally diagonal basis the contact-term matrices $(\mathbf{M}_{IJ})$ \emph{must} vanish identically for all pairs $I,J$. 

Notice that the set of period matrices $\mathbf{M}_{I\,J}$ above is automatically \emph{upper-triangular} in its indices $\{I,J\}$: it is a consequence of the triangular-nature of graph inclusion: $\mathbf{M}_{I\,J}$ vanishes for any graph $\Gamma_J\!\!\nprec\!\Gamma_I$---that is for any set of contours $\{\Omega_J\}$ \emph{not} corresponding to a daughter of the graph $\Gamma_I$. This makes it extremely easy to diagonalize the entire matrix. 

To diagonalize the partially-diagonal basis (\ref{eq:partially_diagonal}), it is useful to take a `bottom-up' approach. To begin, we partition the $87$ integrand topologies relevant for two-loop, triangle power-counting at six particles according to the number of propagators. Thus, at the `bottom' of the list are the five-propagator, $\Gamma_{[\r{2},\g{1},\b{2}]}$ topologies, each with a single degree of freedom. Trivially, these basis elements already vanish on every other defining contour by virtue of the fact that they lack the additional propagators necessary to access the six-propagator and higher contours. That is, in a slight abuse of notation we have schematically
\eq{\oint\limits_{\Omega_{\text{6,7, or 8 props}}}\mathcal{I}_{[\r{2},\g{1},\b{2}]}=0,}
where $\Omega_{\text{6,7, or 8 props}}$ denotes any six-, seven- or eight-propagator-cutting contour, and $\mathcal{I}_{[\r{2},\g{1},\b{2}]}$ any five-propagator integrand in the basis. 

The next step is to diagonalize each of the six-propagator integrands, which are of either $\Gamma_{[\r{3},\g{0},\b{3}]}$, $\Gamma_{[\r{2},\g{2},\b{2}]}$ or $\Gamma_{[\r{3},\g{1},\b{2}]}$ type, with respect to every five-propagator graph obtained by single internal edge contractions. Let us illustrate how this works in practice with a particular example: in the notation of the ancillary files of this paper, one of the degrees of freedom of topology $\#61$,
\eq{\label{eq:global_diag_example}\fwboxR{0pt}{\mathcal{I}_{61}\leftrightarrow}\fwbox{110pt}{\raisebox{-5pt}{\tardi{\tardiScalarEdges\tardiLegs{1}{0}{4}{0}{1}\intDots{5}}\fwboxL{0pt}{\hspace{10pt},}}}\vspace{-12pt}}
is naturally associated with a contour where edges $a,f\rightarrow0$ are soft. A corresponding numerator which is unit on this contour (and which is already diagonal with respect to the other five top-level contours of this topology) is,
\eq{\label{eq:top_level_num_eg}\raisebox{-5pt}{\tardi{\draw[int](v1)--(v2);\draw[int](v2)--(v3);\draw[int](v3)--(v4);\draw[int](v4)--(v5);\draw[dashed](v4)--(v1);\draw[dashed](v5)--(v2);\tardiLegs{1}{0}{4}{0}{1}\contourVerts{0}{0}{4}{0}{0}{5}{5}}}\leftrightarrow \tilde{\mathfrak{n}}_{61}^1=\br{c,d,p_1,p_6}.\vspace{-8pt}}
The six contact-term degrees of freedom of this integrand are proportional to the inverse propagators of the graph; thus, the final, diagonalized numerator $\mathfrak{n}_{61}^1$ will take the form 
\eq{\mathfrak{n}_{61}^1\equivR\tilde{\mathfrak{n}}_{61}^1{+}c_1 a^2+\cdots+c_6 f^2\,.}
Consider the term proportional to $e^2$. Upon collapsing this edge in (\ref{eq:global_diag_example}), we obtain (a re-labelled version of) the five-propagator topology $\#78$,
\vspace{-10pt}\eq{\fwboxR{0pt}{\mathcal{I}_{78}\leftrightarrow}\fwbox{100pt}{\dT{\dTScalarEdges\dTLegs{1}{0}{4}{1}\intDots{4}}\fwboxL{0pt}{\hspace{10pt},}}}
whose scalar degree of freedom is normalized on the contour
\vspace{-10pt}\eq{\fwboxR{10pt}{\Omega_{78}^1\equivR}\dT{\dTInfB\draw[int](v1)--(v2);\draw[int](v2)--(v3);\draw[int](v3)--(v4);\draw[int](v4)--(v1);\draw[int](v4)--(v2);\node at (0,-1.7) {{\footnotesize(odd)}};\dTLegs{1}{0}{4}{1}\contourVerts{3}{3}{4}{4}{5}{5}{5}}\leftrightarrow\mathfrak{n}_{78}^1\equivR{-}\frac{1}{2}s_{16}\,.\vspace{16pt}}
While this contour is an odd combination of two residues, the top-level numerator (\ref{eq:top_level_num_eg}) happens to have support on only one of them. An analytic representation of the relevant part of this contour is given by parametrizing $b\!=\!0$ and $c\!=\!(\lambda_1{+}\alpha \lambda_6)\tilde{\lambda}_1$ with Jacobian $J\!=\!\alpha s_{16}$, and where the final residue is taken at $\alpha\!\rightarrow\!\infty$. The residue of the pre-diagonal integrand $\#61$, degree of freedom $i\!=\!1$ may be computed by evaluating the numerator and dividing by both the Jacobian and the uncut propagator, and we find
 \eq{\oint\limits_{\Omega_{78}^1}\tilde{\mathcal{I}}_{61}^1=\underset{\alpha\rightarrow\infty}{\mathrm{Res}}\left(\frac{\br{c,d,p_1,p_6}d\alpha}{J(d-a)^2}\right)=\underset{\alpha\rightarrow\infty}{\mathrm{Res}}\left(\frac{d\alpha}{\alpha}\right)={-}1\,.}
 To remove the integrand's support on this contour fixes the contact term involving $e^2$ to have the coefficient $c_5\!=\!2\,\mathfrak{n}_{78}^1\equivR{-}s_{16}$. Similar calculations for the remaining five contact terms of integrand $\#61$ lead to the final expression for the diagonalized numerator, 
 \eq{\mathfrak{n}_{61}^1\equivR\br{c,d,p_1,p_6}-\frac{s_{16}}{2}\left[c^2+d^2+2(b^2+e^2)\right]\,.}
 This procedure must be repeated for every degree of freedom and every topology with six propagators. Once this is done, the resulting basis is diagonal and fully fixed at both the five- and six-propagator levels, and we have
 \eq{\oint\limits_{\Omega_{[\r{2},\g{1},\b{2}]}}\mathcal{I}_{\gamma}=0\,,\vspace{-5pt}}
 where $\gamma\!\in\!\{[\r{3},\g{0},\b{3}],[\r{2},\g{2},\b{2}],[\r{3},\g{1},\b{2}]\}$ and $\Omega_{[\r{2},\g{1},\b{2}]}$ denotes any five-cut contour.
 
 This same procedure must be iteratively repeated to fix the contact terms of the seven-propagator integrands by demanding orthogonality with respect to both the five- and six-propagator basis elements. For example, the $45$ contact term degrees of freedom for topology $\#38$ are fixed by enumerating the daughter topologies (and counting with the appropriate multiplicity)
\definecolor{labelcolor}{rgb}{1,1,1}
\vspace{-10pt}\eq{\begin{split}\\[-26pt]&\fwbox{200pt}{\hspace{-2pt}\npPbox{\draw[int](v1)--(v2)--(v3)--(v4)--(v5)--(v6)--(v3);\draw[int](v1)--(v5);\npPboxLegs{1}{1}{1}{1}{1}{1}\intDots{6}}\hspace{-15pt}\succ\!\!\Bigg\{ 2\!\times\!\Bigg(\raisebox{-5.pt}{\fwbox{75pt}{\tardi{\draw[int](v1)--(v2)--(v3)--(v4)--(v5)--(v2);\draw[int](v4)--(v1);\tardiLegs{1}{1}{1}{2}{1}\intDots{5}}}}\Bigg),1\!\times\!\Bigg(\raisebox{-5pt}{\fwbox{75pt}{\tardi{\draw[int](v1)--(v2)--(v3)--(v4)--(v5)--(v2);\draw[int](v4)--(v1);\tardiLegs{1}{1}{2}{1}{1}\intDots{5}}}}\Bigg), 4\!\times\!\Bigg(\fwbox{75pt}{\bT{\draw[int](v1)--(v2)--(v3)--(v4)--(v5)--(v1);\draw[int](v5)--(v3);\bTLegs{1}{1}{1}{1}{2}\intDots{5}}}\Bigg),}\\[-17pt]
 &\fwbox{200pt}{\hspace{104pt}\hspace{4pt}6\!\times\!\Bigg(\fwbox{67pt}{\dT{\draw[int](v1)--(v2)--(v3)--(v4)--(v1);\draw[int](v4)--(v2);\dTLegs{1}{1}{1}{3}\intDots{4}}}\Bigg),\hspace{8pt}5\!\times\!\Bigg(\fwbox{67pt}{\dT{\draw[int](v1)--(v2)--(v3)--(v4)--(v1);\draw[int](v4)--(v2);\dTLegs{1}{2}{1}{2}\intDots{4}}}\Bigg),\hspace{4pt}4\!\times\!\Bigg(\fwbox{67pt}{\dT{\draw[int](v1)--(v2)--(v3)--(v4)--(v1);\draw[int](v4)--(v2);\dTLegs{1}{1}{2}{2}\intDots{4}}}\Bigg)\Bigg\}.}\end{split}\vspace{-20pt}}
\definecolor{labelcolor}{rgb}{0,0,0}
~\\[-20pt]
A final, prescriptive and globally-diagonal basis is obtained by removing all relevant contact terms from the eight-propagator integrands. 

The end result of the above procedure is a fully diagonal basis of integrands and contours $\{\mathcal{I}_I^i,\Omega_I^i\}$,
\eq{\label{eq:fully_diagonal}\oint\limits_{\Omega_J^j}\mathcal{I}_I^i=\left\{\begin{array}{lr} \delta_{i\,j} & I=J, \\
0\, & I\neq J. \end{array}\right.}
The complete list of integrands and contour definitions which satisfy (\ref{eq:fully_diagonal}) are given in the ancillary files of this paper and described in greater detail in appendix~\ref{ancillary_files}. 

\vspace{4pt}%
\section{Features of the Resulting Basis of Integrands}\label{sec:discussion_of_our_basis}\vspace{0pt}
The triangle power-counting basis of local integrands constructed according to the procedure of section~\ref{sec:six_point_basis} has several noteworthy properties worth emphasizing. First, this basis is prescriptive: every integrand is normalized to be unit on a single co-dimension eight contour, and vanishes on every other contour in the spanning set. Producing a unique representation of an amplitude integrand $\mathcal{A}$ which is of box or triangle power-counting amounts to computing a set of contour integrals
\eq{\label{eq:amp_in_basis}\mathcal{A}=\sum_{I,i}\mathfrak{a}_I^i\mathcal{I}^i,\quad\text{where}\quad\mathfrak{a}_I^i\equivR\oint\limits_{\Omega_I^i}\mathcal{A}\,.\vspace{-6pt}}
The non-zero coefficients $\mathfrak{a}_i^i$ are the non-vanishing leading singularities of the amplitude in question. To be clear, this procedure does not require computing \emph{every} leading singularity of an amplitude, but only those in the spanning set of contours $\{\Omega_I^i\}$; all other contours are matched automatically in (\ref{eq:amp_in_basis}) by the completeness of the basis. Furthermore, for all contours not manifestly matched in (\ref{eq:amp_in_basis}), each integrand $\{\mathcal{I}^i\}$ which is \emph{not} normalized on a double-pole at infinity evaluates to either $\pm1$ or $\pm\frac{1}{2}$. Our basis therefore splits into pure, unit-leading-singularity polylogarithmic integrals as well as integrals which have simple transcendental weight drops upon integration. As amplitude integrands in sYM are free of double-poles, we expect the representation (\ref{eq:amp_in_basis}) is entirely free of transcendental weight drops. 

Our basis is also partitioned according to infrared structure. From the complete list of contours (which can be found in the ancillary files), every integrand normalized on a contour in the collinear region of loop momentum space may be identified; these integrands generate infrared divergences upon integration, while every other basis element are explicitly infrared finite. The partitioning according to IR finiteness/divergence will obviously descend to the expression for amplitude integrands (\ref{eq:amp_in_basis}), where every soft-collinear divergence of the amplitude is matched manifestly by individual elements of basis designed to have support on the corresponding soft-collinear contours.

As indicated in the ancillary files for this work, the final basis includes $183$ IR-finite integrands and 190 IR-divergent integrands; and of these only 24 have support on double-poles---which should be the only integrands with less than maximal weight.

\vspace{4pt}
\section{Conclusions and Future Directions}\label{sec:conclusions}\vspace{0pt}
In this work, we have constructed a diagonal basis of four-dimensional local loop integrands with triangle power-counting at two loops and six particles. As we have emphasized, the properties of this basis are entirely dependent on the choice of eight-dimensional compact contours on which the integrands are normalized to be unit. The specific choices in this paper amount to making infrared structure, polylogarithmicity, and purity as manifest as possible; this is important for practical applications including carrying out the loop integrals, since uniformly transcendental integrals obey particularly simple differential equations \cite{Kotikov:1990kg,Kotikov:2000vn,Remiddi:1997ny,Henn:2013pwa}, among other nice properties \cite{ArkaniHamed:2010gh}. Of course, integrating the basis of this paper in dimensional regularization requires (at least for those elements with infrared divergences) the addition of the extra-dimensional terms relevant for $d\!=\!4{-}2\epsilon$ dimensions. In addition to this, we expect the finite part of our basis to be an ideal testing ground for testing the methods of direct integration \cite{Bourjaily:2019jrk,Bourjaily:2019vby} in the non-planar sector. 

 This basis is sufficient to express the six-particle MHV and NMHV amplitudes in both sYM and SUGRA, and doing so requires only the computation of the physical leading singularities which are among the spanning set chosen in this work. It should be relatively straightforward to compute the relevant fully color-dressed leading singularities in sYM, and we hope to do so in a forthcoming work \cite{6ptAmps}.

Because the basis of integrands is identical for both MHV and NMHV helicity configurations, having representations of both amplitude integrands in hand immediately suggests an obvious application: namely, the study of non-planar versions of the two-loop ratio function. The universal nature of the infrared divergence of non-planar amplitudes strongly suggests that the ratio of the NMHV and MHV amplitudes should define a \emph{locally-finite} quantity according to the definition of \cite{Bourjaily:2021ewt}, but due to the presence of color factors, the details of how this works remain unclear. 

Another natural extension of this work would be to consider bubble or worse degrees of power-counting. Although this introduces new degrees of freedom and higher poles at infinity, because $\mathfrak{B}_{p}\!\subset\!\mathfrak{B}_{q}$ for any $p\!>\!q$, \emph{every} contour used in the basis constructed in this work can be recycled and used to fill out a large subset of the spanning set of contours for bubble power-counting (and beyond). Note that while the integrands will be modified by both additional top-level and contact term numerators, the contours themselves require no modification whatsoever. 

Instead of boosting the power-counting, implementing the diagonalization procedure of this work at higher multiplicities and higher loops offers an additional direction for future investigations. Generating diagonal bases beyond this work will inevitably require non-polylogarithmic contours and integrands \cite{Bourjaily:2017bsb,Bourjaily:2018ycu}; here, we expect the extension of prescriptive unitarity to elliptic (and beyond) leading singularities outlined in \cite{Bourjaily:2021vyj} to prove quite useful.

\vspace{\fill}\vspace{-4pt}
\section*{Acknowledgements}%
\vspace{-4pt}
\noindent The authors gratefully acknowledge fruitful conversations with Enrico Herrmann and Jaroslav Trnka. This project has been supported by an ERC Starting Grant \mbox{(No.\ 757978)}, a grant from the Villum Fonden \mbox{(No.\ 15369)}, by a grant from the US Department of Energy \mbox{(No.\ DE-SC00019066)}.

\newpage
\appendix
\section{Organization of Ancillary Files}\label{ancillary_files}
The complete details of the basis constructed for triangle power-counting at two loops for six external particles are included as ancillary files for this work's submission to the \texttt{arXiv}. These files are available directly on this work's abstract page on the \texttt{arXiv} (under `Download' on the right-hand pane). In particular, these files include:\\[-12pt]
\vspace{-5pt}\begin{itemize}
\item \textbf{{integrand\uscore basis\uscore data.m}}: a \emph{plain-text} enumeration of:
\begin{itemize}
\item all 87 non-isomorphic two-loop topologies described as 
\begin{itemize}
\item \fun{edgeList}{\tt[\var{I}]}: lists of oriented edges given as ordered pairs between vertices (the third item in each edge is its label);
\item \fun{propRules}{\tt[\var{I}]}: example graph routing for these edge variables in terms of $\ell_1,\ell_2$ (for the sake of numerical evaluations);
\end{itemize}
\item all 373 integrand numerators expressed as \fun{num}{\tt[\var{I},\var{i}]}$\equivR\mathfrak{n}_{I}^i$;
\item all 373 contours used to diagonalize the basis expressed as \fun{contourLists}{\tt[\var{I}]}, returning the \texttt{List} of contours for the $I$th integrand topology; for each contour, we encode this data as:
\begin{itemize}
\item a \fun{pt}, \fun{pl}, or \fun{seq} if it is given by a `point' (a single contour picture---possibly representing an even/odd combination of multiple solutions for the given picture), a `sum' of points (a sum of single-contour pictures), or a `sequence' of constraints, respectively; for each, 
\begin{itemize}
\item color assignments for each vertex---which can be `\fun{b}', `\fun{w}', `\fun{c}', or `\fun{k}' for vertices which are taken in the `blue' (mhv),
`white' ($\overline{\text{mhv}}$), `composite' (collinear), or generic configurations, respectively;
\item a list of any soft edges (denoted as `\fun{s}{\tt[\var{i},\var{j}]}' for a soft edge connecting vertices \fun{v}{\tt[\var{i}]} to \fun{v}{\tt[\var{j}]})
\item a list of any cycles taken to infinity;
\item extra information about the contour---such as \fun{dp} for `double-pole', or \fun{odd} to indicate the \emph{difference} of different solutions to the cuts.
\end{itemize}
\end{itemize}
This information is arguably encoded somewhat idiosyncratically; it is correctly parsed using the \fun{drawContour} function in the \textsc{Mathematica} package.
\end{itemize}
\item \textbf{integrand\uscore basis\uscore tools.m}: a \textsc{Mathematica} package file that includes many evaluation and graph-drawing features for using, visualizing, and manipulating the data for this basis;
\item \textbf{integrand\uscore basis\uscore walkthrough.nb}: a \textsc{Mathematica} notebook that illustrates the main functionality provided by the codebase given in the code-base file \textbf{integrand\uscore basis\uscore tools.m}. This file should be viewed as self-contained documentation for the package and tools---and should be consulted for such.\\[-16pt]
\end{itemize}
Some of the tools and code are based on earlier packages made available as part of refs.~\cite{Bourjaily:2010wh,Bourjaily:2012gy,Bourjaily:2013mma,Bourjaily:2015jna,Bourjaily:2018omh,Bourjaily:2019gqu}.

\newpage

\providecommand{\href}[2]{#2}\begingroup\raggedright\endgroup

\end{document}